%% file: iris-hep-strategic-plan.tex
\documentclass[11pt,letterpaper,fleqn,x11names]{article}
\usepackage{graphicx}
\usepackage{tabulary}
\usepackage{lineno}
\usepackage{hyperref}
\usepackage{subfigure}
\usepackage{cite}
\usepackage{tcolorbox}
\usepackage{xcolor,colortbl}
\usepackage{wrapfig}
\usepackage{enumitem} 
\usepackage{booktabs}

\newtcolorbox{iriscolorbox}[1]{colback=LightBlue1,colframe=msdarkblue,fonttitle=\bfseries, title={#1}}

\selectcolormodel{rgb}
\definecolor{LHCb dark}{rgb}{0.0000,0.3412,0.6549}
\definecolor{UC red}{rgb}{0.8196,0.1176,0.2314} 
\definecolor{brickred}{rgb}{0.8, 0.25, 0.33}
\definecolor{LHCb dark}{rgb}{0.0000,0.3412,0.6549}
\definecolor{Gray}{gray}{0.85}
\definecolor{asparagus}{rgb}{0.53, 0.66, 0.42}
\definecolor{brightcerulean}{rgb}{0.11, 0.67, 0.84}
\definecolor{brightgreen}{rgb}{0.4, 1.0, 0.0}
\definecolor{candyapplered}{rgb}{1.0, 0.03, 0.0}

\definecolor{mslightblue}{HTML}{DFEAF7}
\definecolor{msdarkblue}{HTML}{1D2F5A}
\definecolor{verylightgray}{rgb}{0.88, 0.88, 0.88}

\usepackage{fancyhdr}
\newcommand{\changefont}{%
    \fontsize{9}{11}\selectfont
}
\let\oldenumerate\itemize
\renewcommand{\itemize}{
  \oldenumerate
  \setlength{\itemsep}{1pt}
  \setlength{\parskip}{1pt}
  \setlength{\parsep}{1pt}
}

\newcommand{\tempnewpage}[0]{{\newpage}}
\def\s2i2{$S^2 I^2$}

\title{{\bf IRIS-HEP Strategic Plan for the Next Phase \\
       of Software Upgrades for HL-LHC Physics}}
\date{\today~- Version 1.0}
\begin{document}


\vbox{
    \centering
    \includegraphics[width=1.0\textwidth]{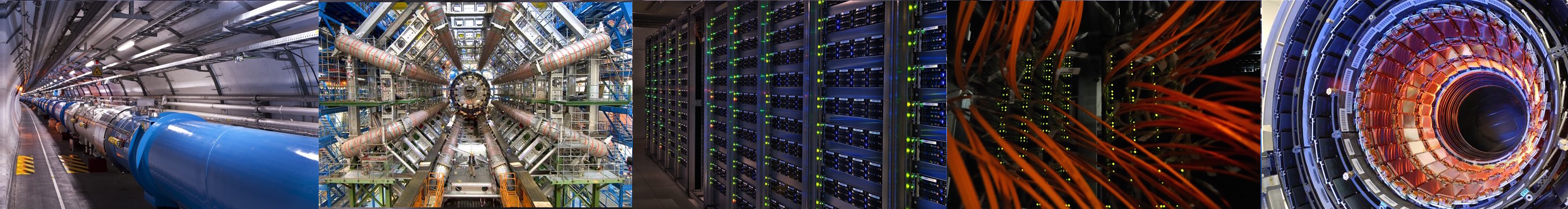}
    \maketitle 

        {\bf Editors:}
        Brian Bockelman (Morgridge Institute), Peter Elmer (Princeton University)\\
        Gordon Watts (University of Washington)\\ 
        ~~\\
        {\bf Contributors:}
        Kyle Cranmer (University of Wisconsin-Madison),
        Matthew Feickert (University of Wisconsin-Madison),
        Rob Gardner (University of Chicago),
        Heather Gray (UC Berkeley and Lawrence Berkeley Laboratory),
        Alexander Held (University of Wisconsin-Madison),
        Daniel S. Katz (University of Illinois at Urbana-Champaign),
        David Lange (Princeton University),
        Kilian Lieret (Princeton University),
        Mark Neubauer (University of Illinois at Urbana-Champaign),
        Oksana Shadura (University of Nebraska-Lincoln),
        Mike Sokoloff (University of Cincinnati),
        Robert Tuck (Princeton University),
        Frank Wuerthwein (University of California, San Diego)
        \\

\vskip 2.5in
\begin{tabulary}{1.0\textwidth}{lr}
\includegraphics[width=0.1\textwidth]{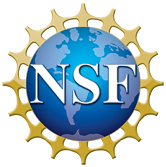} & \parbox{0.85\textwidth}{This report has been produced by the IRIS-HEP project (\url{http://iris-hep.org}) and supported by National Science Foundation grants OAC-1836650. Any opinions, findings, conclusions or recommendations expressed in this material are those of the project participants and do not necessarily reflect the views of the National Science Foundation.}
\end{tabulary}
}

\thispagestyle{empty}
\newpage
\pagenumbering{roman}
\input{005-executive-summary.tex}

\input{000-endorsers.tex}
\newpage
\input{002-revisions.tex}
\thispagestyle{empty}
\newpage
\tableofcontents
\thispagestyle{empty}


\newpage
\pagenumbering{arabic}
\pagestyle{fancy}
\setcounter{page}{1}
\input{010-introduction.tex}
\input{020-science-drivers.tex}
\input{025-computing-challenges.tex}
\input{026-computing-challenges-sustainability.tex}

\input{050-institute-role.tex}

\input{100-strategic-areas.tex}

\input{300-grand-challenges.tex}
\input{320-broadening-participation.tex}
\input{350-funding-levels.tex}

\appendix
\input{810-workshop-list.tex}

\input{820-institute-organization.tex}

\input{840-glossary.tex}

\newpage 
\bibliographystyle{unsrt}
\bibliography{iris-hep-bib-1,iris-hep-bib-2}

\end{document}

%% file: 005-executive-summary.tex
\section*{Executive Summary}
\pagestyle{plain}
The quest to understand the fundamental building blocks of nature
and their interactions is one of the oldest and most ambitious of
human scientific endeavors. CERN's Large Hadron
Collider (LHC) represents a huge step forward in this quest.  The
discovery of the Higgs boson, the observation of exceedingly rare
decays of $B$ mesons, and stringent constraints on many viable
theories of physics beyond the Standard Model (SM) demonstrate the
great scientific value of the LHC physics program. The next phase
of this global scientific project will be the High-Luminosity LHC
(HL-LHC) which will collect data starting circa 2029 and continue
through the 2030s.  The primary science goal is to search for physics
beyond the SM and, should it be discovered, to study its 
implications.  In the HL-LHC era, the ATLAS and CMS experiments
will record ${\sim}100$ times as many collisions as were used to
discover the Higgs boson (and at twice the energy). Both NSF and DOE 
are making large detector upgrade investments so the HL-LHC can
operate in this high-rate environment.  Similar investment in {\em software}
R\&D for acquiring, managing, processing and analyzing
HL-LHC data is critical to maximize the return-on-investment in the
upgraded accelerator and detectors.

This report presents a strategic plan for a possible second 
5-year funded phase (2023 through 2028) for the {\bf Institute for Research
and Innovation in Software 
for High Energy Physics (IRIS-HEP)}~\cite{iris-hep,iris-hep-nsf-award}
which will close remaining software and computing gaps to deliver HL-LHC
science.
IRIS-HEP was originally funded by the NSF in September 2018 and its 
current funding extends through August 2023. IRIS-HEP was the result 
of an international
community planning process in 2016-2017 that prepared a roadmap for
HEP software R\&D in the 2020s~\cite{CWPDOC} as well as the original
strategic plan~\cite{S2I2HEPSP} for how an NSF-funded Software
Institute for High Energy Physics (HEP) could play a key role in
meeting HL-LHC challenges. 
The mission of IRIS-HEP is two-fold: to serve as 
an active center for software R\&D {\em and} as an intellectual hub for 
the larger community software R\&D effort required to ensure the success 
of the HL-LHC scientific program. The ensemble of its activities are meant
to bridge 4 key gaps to that success: (G1) raw resource gaps, 
(G2) scalability of the distributed computing cyberinfrastructure, 
(G3) executing analyses at the HL-LHC scale and (G4) sustainability of the 
software and computing system through the lifetime of the HL-LHC and 
beyond.

Building on its successes over the past 5 years, and informed by recent 
community workshops, IRIS-HEP will continue to
pursue several high-impact R\&D areas identified as those required to 
close those gaps and enable HL-LHC science:
(1)~development of highly performant analysis systems that 
reduce `time-to-insight' and maximize the HL-LHC physics potential;
(2)~development of advanced innovative algorithms for data reconstruction 
and triggering; 
(3)~development of data organization, management and access systems
for the Exabyte era; 
and (4)~facilities R\&D to support the evolving 
requirements of the other R\&D areas. A potentially important emerging 
opportunity, (5) Translational AI, has also been identified. 
IRIS-HEP will sustain the investments
in the fabric for (6) distributed high-throughput computing which provides the LHC community's cyberinfrastructure with a route to the HL-LHC. Specific goals and objectives for these areas of activity include:

\input{101-strategic-area-summary-include.tex}

IRIS-HEP will continue to use the concept of a ``Grand Challenge" to 
focus activities on long-term, large-scale goals as part of the 
institute's overall vision. A Grand Challenge differs from a more 
traditional milestone or deliverable by its scale (often requiring 
cross-cutting teams working together), a multi-year timeframe, and 
the fact the entirety of the approach may not be known upfront.  
The challenges are executed through a series of increasingly difficult exercises coordinated throughout the community.  The currently defined
and planned grand challenges are:

\begin{itemize}
\item {\bf Analysis Grand Challenge (AGC)}: The AGC aims to execute realistic analyses at the scale and complexity envisioned by the HL-LHC using a set of tools, facilities, and services developed within IRIS-HEP as exemplars.  The AGC team coordinates an annual workshop to demonstrate current progress against the goals and to update the vision and approach as necessary.
\item {\bf Data Grand Challenge (DGC)}: The DGC for IRIS-HEP is realized as a set of global data challenges coordinated with the WLCG.  These challenges, occurring biennially, bring the entire global community together to demonstrate aggregate transfer data rates and compare what is currently achievable with the HL-LHC roadmap.  These challenges also provide an opportunity for integrating new technologies being worked on by DOMA into the production infrastructure.
\item {\bf Training Grand Challenge (TGC)}: To tackle the challenges of 
the HL-LHC, we need a workforce with broad software knowledge, spanning 
from basic programming skills to highly specialized training. The TGC 
defines a roadmap to efficiently scale up training activities and 
provide adequate training to create the software-skilled workforce
that will realize HL-LHC science.
\end{itemize}
\vskip -0.3cm
\noindent Additional Institute activities, such as data reconstruction 
and tracking, are more naturally coordinated by each experiment. 
IRIS-HEP will work with the experiments to integrate those deliverables 
with their timelines and production codes.
Building on the successful
R\&D efforts of the first five years of the project, the second phase of
IRIS-HEP will develop production-quality tools, facilities, and services
that the experiments will be able to rely on for their science programs.

\vskip 0.1cm
Lastly, as an intellectual hub, the Institute will continue to lead efforts 
to (1) bring together the IRIS-HEP team, key stakeholders and relevant domain experts 
to inform the Institute’s mission, (2) bring together U.S.\ and international R\&D efforts 
in ``blueprint'' workshops to ensure full alignment of the various efforts (3) develop partnerships between HEP and the cyberinfrastructure,
computer science, and data science communities for novel approaches to
meeting HL-LHC challenges, (4) bring in new effort from U.S.\
universities emphasizing professional development and training, and
(5) sustain HEP software and underlying algorithmic and implementation
knowledge over the two decades required.
HEP is a global, complex, scientific endeavor.  These activities
will ensure that the software developed and deployed by a
globally distributed community will extend the science reach of the
HL-LHC and will be sustained over its lifetime.

The strategic plan targeting HL-LHC physics presented in this report 
for the period 2023-2028 reflects a community vision.  It 
builds on the original planning
activities in 2016-2017 and is informed by ongoing IRIS-HEP 
activities which engage the community including a dedicated series of 
workshops and conferences.
The plan complements and is aligned with the experiments' HL-LHC 
planning activities~\cite{atlas2022reviewdoc,cms2022reviewdoc,LHCbCollaboration:2776420}.
IRIS-HEP is ready to deliver the software required to enable best possible HL-LHC science. 


\newpage

%% file: 101-strategic-area-summary-include.tex
\begin{enumerate}

\item \textbf{Analysis Systems}: Modernize and evolve tools and
techniques for analysis of HL-LHC data sets.  The Analysis Systems
area will concentrate on {\em G3 (Analysis at scale)} topics of
managing order-of-magnitude larger data sets, enabling more complex
techniques including use of modern machine learning, and the adoption
of data science tools toward {\em G4 (Sustainability)} goals. 
Deliverables will include the tools supporting a full pipeline for 
distributed columnar data analysis at scale that are interoperable with 
other elements of the HEP and broader data science ecosystems. 
Modern machine learning techniques and operations, including differentiable analysis pipelines,
will be integrated, as well as full support for analysis preservation 
and reinterpretation.

\item \textbf{Reconstruction and Trigger Algorithms}: Develop and
evolve pattern-recognition software able to exploit next-generation
detector technologies, computing platforms, and programming techniques
to accurately and efficiently identify charged particle trajectories.
Efforts in this area are key to eliminating {\em G1 (resource)}
gaps due to new, and more capable, experimental apparatus, larger
data rates, and evolving computing hardware. Algorithms must be
engineered to be adequately {\em G4 (sustainable)} over the course
of HL-LHC operations.
This area will deliver critical components of the tracking pipeline, prioritizing algorithmic interoperability, achieving high levels of parallelism, and implementing robust algorithms. Both traditional and novel approaches to tracking algorithms will be considered as tracking is a multifaceted problem and has optimization points that vary depending on the experimental apparatus design and computing technical design. 

%

\item \textbf{Translational AI}: Exploit Machine Learning approaches
to improve the physics reach of the HL-LHC.  The Translational AI
area will leverage fundamental research, such as the work done by
the NSF AI Institutes, and focus on helping the HL-LHC experiments
translate these capabilities into production.  Activities include
working to enable and use ML-based services, helping the field
connect to available infrastructure, and working on enabling the
``retraining'' of in-use models for new data.

\item \textbf{Data Organization, Management, and Access (DOMA)}:
Scale and modernize the bulk data transfer infrastructure including new
authorization schemes, transfer protocols, and network integration; provide
new data delivery services and techniques for use in analysis
facilities.  DOMA contributes to the {\em G2 (Scalability)} computing
gap to close the 20x difference in the wide-area data rates expected
between now and the start of the HL-LHC.  DOMA innovates new authorization 
schemes for inter-site bulk data transfer coordinates international
data challenges to mark progress toward the target data rates and
integrate new technologies.  For analysis, DOMA will develop services
to deliver columnar data and to provide modern data management
techniques from the database community to HL-LHC analysis environments.

\item \textbf{Facilities R\&D}: Innovates new approaches to building
facilities for the U.S. LHC and aligns the community with approaches
in the larger NSF coordinated cyberinfrastructure.  The area works
with Kubernetes as a ``substrate" to orchestrate portable services,
provides testbed facilities to projects within the Institute,
and investigates the use of multi-site Kubernetes clusters to provide
agility to the operation of distributed services.  Facilities R\&D
contributes to the {\em G4 (Sustainability)} goal by reducing the
operational complexity and costs and to {\em G3 (Analysis at scale)}
by applying agile techniques for future analysis facilities.

\item \textbf{Fabric of distributed high-throughput computing
services (OSG)}: Operates a fabric of services specifically to meet
the needs of the LHC and provides a stable route to their evolution
for the HL-LHC.  The OSG-LHC group ensures the needs of the LHC
experiments while making contributions to the larger consortium,
allowing the LHC to benefit from broader common services as well.

\item \textbf{Training, Workforce Development, and Outreach}:
Executes and coordinates a broad range of events to help with the
training, workforce development, and outreach needs of the HL-LHC
community.  The area in the Institute will ensure there is training
available at multiple levels of need (undergraduate, graduate,
postdoc, professional) and will run the IRIS-HEP Fellows program,
a in-depth virtual mentoring program targeting senior undergraduates
and junior graduate students. This is key to bridging the 
{\em G4 (Sustainability)} gap.
\end{enumerate}

%% file: 000-endorsers.tex
\section*{Endorsers}
\thispagestyle{empty}

This strategic plan has been explicitly endorsed by ATLAS
and CMS experiment representatives, as well as representatives
from the US-ATLAS and US-CMS Operations programs. 
Specific endorsers include:
Wolfgang Adam (HEPHY Vienna),
Lothar Bauerdick (Fermilab),
Brian Bockelman (Morgridge Institute for Research),
Tulika Bose (University of Wisconsin-Madison),
Kenneth Bloom (University of Nebraska-Lincoln),
Paolo Calafiura (Lawrence Berkeley National Laboratory),
Kyle Cranmer (University of Wisconsin - Madison),
Alessandro Di Girolamo (CERN),
Peter Elmer (Princeton University),
Matthew Feickert (University of Wisconsin - Madison),
Robert Gardner (University of Chicago),
Heather Gray (University of California, Berkeley and Lawrence Berkeley National Laboratory),
Oliver Gutsche (Fermilab),
Alexander Held (University of Wisconsin - Madison),
Michael Hildreth (University of Notre Dame),
Daniel S. Katz (University of Illinois at Urbana-Champaign),
Patrick Koppenburg (Nikhef),
David Lange (Princeton University),
James Letts (University of California, San Diego),
Kilian Lieret (Princeton University),
Carlos Maltzahn (University of California, Santa Cruz),
Zachary Marshall (Lawrence Berkeley National Laboratory),
Verena Martinez Outschoorn (University of Massachusetts Amherst),
Patricia McBride (FNAL),
Shawn McKee (University of Michigan),
Mark S Neubauer (University of Illinois at Urbana-Champaign),
Ianna Osborne (Princeton University),
Danilo Piparo (CERN),
Eduardo Rodrigues (University of Liverpool),
Elizabeth Sexton-Kennedy (Fermilab),
Oksana Shadura (University of Nebraska-Lincoln),
Lucia Silvestris (INFN-Bari),
Michael D Sokoloff (University of Cincinnati),
Matev\v{z} Tadel (University of California, San Diego),
Lauren Tompkins (Stanford University),
Robert Tuck (Princeton University),
Vassil Vassilev (Princeton University),
Gordon Watts (University of Washington),
Derek Weitzel (University of Nebraska-Lincoln),
Mike Williams (MIT), 
Peter Wittich (Cornell University), 
Frank Wuerthwein (University of California, San Diego) and
Avi Yagil (University of California, San Diego). 

\newpage

%% file: 002-revisions.tex
\section*{Document revisions}
\thispagestyle{empty}
The revision history is:
\begin{description}
\item[19 December 2022] Version 0.9 - Initial draft of the document circulated for feedback and endorsements.
\item[23 January 2023] Version 0.95 - Improved version of the document based on feedback and endorsements from version 0.9. More complete Executive Summary. Provisional version submitted to NSF.
\item[2 February 2023] Version 1.0 - submitted to arXiv
\end{description}
\newpage

%% file: 010-introduction.tex
\section{Introduction}
\label{sec:intro}

The High-Luminosity Large Hadron Collider (HL-LHC) is scheduled to
start producing data in 2029 and extend the LHC physics program
through the 2030s.  Its primary science goal is to search for Beyond
the Standard Model (BSM) physics and, should it be discovered, to study its
implications. Although the basic constituents of
ordinary matter and their interactions are extraordinarily well
described by the Standard Model (SM) of particle physics, a quantum
field theory built on top of simple but powerful symmetry principles,
it is incomplete.  For example, most of the gravitationally interacting
matter in the universe does not interact via electromagnetic or
strong nuclear interactions.  As it produces no directly visible
signals, it is called dark matter.  No particles or fields of the SM can account for its existence.  Equally as important, the SM does not
address fundamental questions related to the detailed properties
of its {\em own} constituent particles or the specific symmetries
governing their interactions.  To achieve this scientific program,
the HL-LHC will record data from 100 times as many proton-proton
collisions as did Run 1 of the LHC, with both more complex detectors and
where each collision will be far more complex to process.  

Realizing the full potential of the HL-LHC requires large investments
in upgraded hardware. The construction projects for these hardware
upgrades are now underway.  The two
general purpose detectors at the LHC, ATLAS and CMS, are operated
by collaborations of more than 3000 scientists each.  U.S.\ personnel
constitute about 30\% of the collaborators on these experiments.
Within the U.S., funding for the construction and operation of ATLAS
and CMS is jointly provided by the Department of Energy (DOE) and
the National Science Foundation (NSF).  Funding for U.S.\ participation
in the LHCb experiment is provided only by the NSF.  The NSF is
also playing a major role in the hardware upgrade of the ATLAS and
CMS detectors for the HL-LHC, through Major Research Equipment and 
Facilities Construction (MREFC) funding which began in 2020.

The HL-LHC {\bf also requires a commensurate investment in software
research and development} and deployment. Software is necessary to 
acquire, manage, process, and analyze the data.  Current estimates
of HL-LHC computing needs significantly exceed what will be possible
-- even assuming Moore's Law -- with constant operational
budgets~\cite{atlas2022reviewdoc,cms2022reviewdoc}. The underlying 
nature of computing hardware (processors, storage, networks) is
also evolving, the quantity of data to be processed is increasing
dramatically, its complexity is increasing, and more sophisticated
analyses will be required to maximize the HL-LHC physics yield.
The magnitude of the HL-LHC computing problems to be solved will
require different approaches.

The existing computing system of the LHC experiments is the result
of almost 20 years of effort and experience. In addition to addressing
the significant future challenges, sustaining the fundamental aspects
of what has been built to date is also critical.  Fortunately, the
collider nature of this physics program implies that essentially
all computational challenges are "pleasantly" parallel.  The large
LHC collaborations each produce tens of billions of events per year
through a mix of simulation and data triggers recorded by their
experiments, and all events are statistically independent of each other, and can be processed in parallel.
This intrinsic simplification from the science itself permits
aggregation of distributed computing resources and is well-matched
to the use of {\it high throughput computing} to meet LHC and HL-LHC
computing needs.  In addition, the LHC today requires more computing
resources than will be provided by funding agencies in any single
location (such as CERN).
Thus {\it distributed high-throughput
computing} (DHTC) will continue to be a fundamental characteristic
of the HL-LHC and evolving the DHTC is essential for the HEP
community.

\begin{figure}[htp]
\begin{center}
\includegraphics[width=0.80\textwidth]{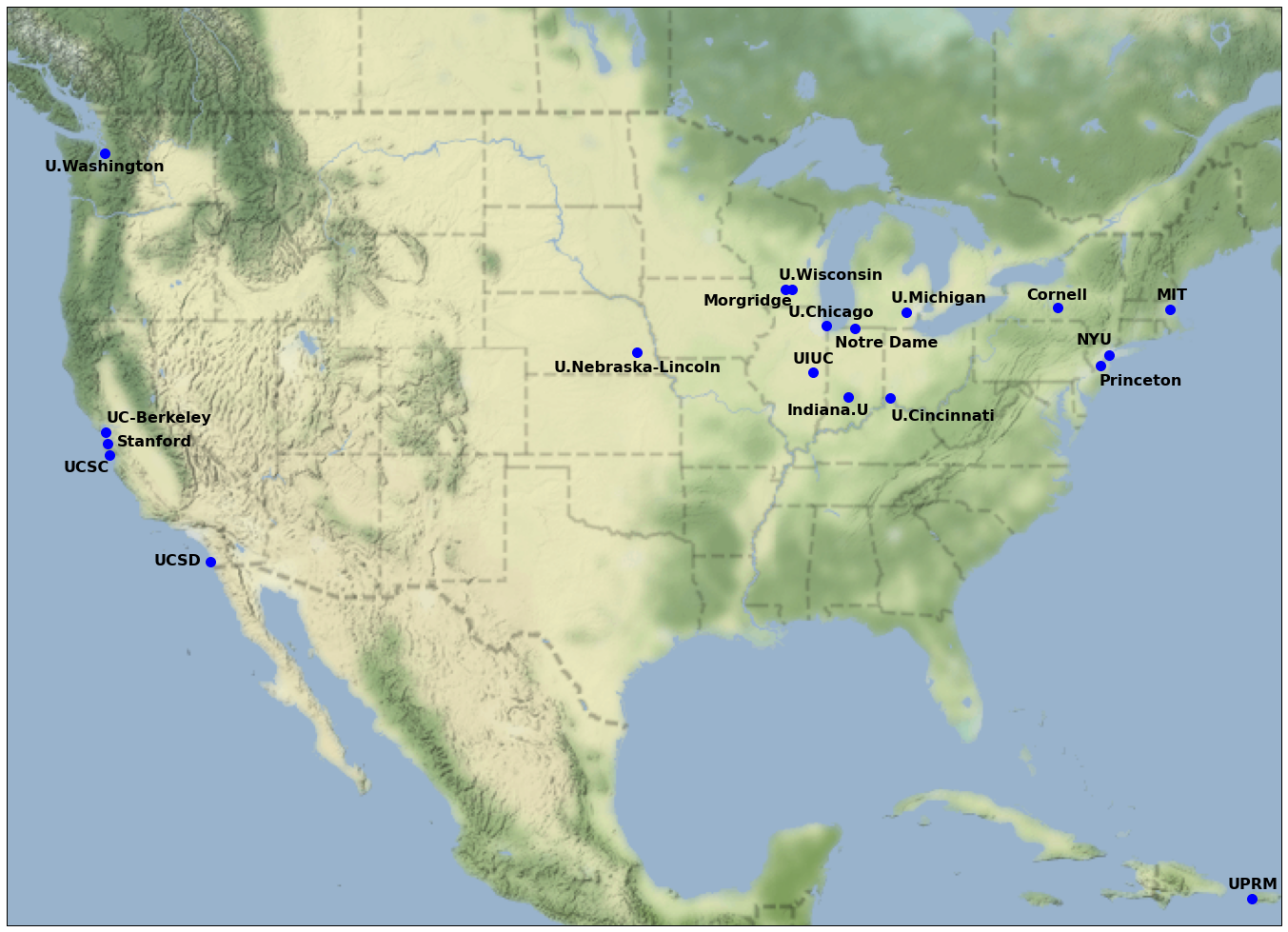}
\caption{Institutions funded during the first phase (2018-2023) 
of the IRIS-HEP software institute.}
\label{fig:irishepmap}
\end{center}
\end{figure}

In planning for the HL-LHC, it is critical that all parties agree
on the software goals and priorities, and that the efforts tend to
complement each other; for IRIS-HEP, this alignment process began
in 2016 and is ongoing to this day.  The process was started with
a HEP Software Foundation (HSF)  planning exercise in late 2016 to
prepare a Community White Paper (CWP)~\cite{CWPDOC} whose goal was
to provide a roadmap for software R\&D in preparation for the HL-LHC
era.  The community identified and prioritized the software research
and development investments required:
\begin{enumerate}
\itemsep0em
  \item 
    to enable new approaches to computing and software that 
    can  radically extend the physics reach of the detectors; and
  \item 
    to achieve improvements in software efficiency, scalability, and 
    performance, and to make use of the advances in CPU, storage, and
    and network technologies;
  \item 
    to ensure the long term sustainability of the software through 
    the lifetime of the HL-LHC.
\end{enumerate}
In parallel to this global exercise, and with funding from the
NSF, the U.S.\ community executed a conceptualization process to
produce a Strategic Plan for how a Scientific Software Innovation
Institute (\s2i2) for high-energy physics (HEP) could help meet the
HL-LHC challenges. Specifically, the \s2i2-HEP conceptualization
process~\cite{S2I2HEP} had three additional goals:
\begin{enumerate}
\itemsep0em
 \item
  to identify specific focus areas for R\&D efforts that could be part of an \s2i2 in the U.S.\ university community; 
 \item
  to build a consensus within the U.S.\ HEP software community for a 
  common effort; and
 \item
  to engage with experts from the related fields of scientific computing and
  software engineering to identify topics of mutual interest and
  build teams for collaborative work to advance the scientific
  interests of all the communities. 
\end{enumerate}
This resulted in a document, the {\it ``Strategic Plan for a Scientific Software Innovation 
Institute (\s2i2) for High Energy Physics''} \cite{S2I2HEPSP}. 
In September, 2018, NSF funded the ``Institute for Research and Innovation 
in Software for High Energy Physics (IRIS-HEP)'', a collaboration of 19 U.S. 
universities (Figure~\ref{fig:irishepmap}), to execute a program of work to 
realize that strategic plan.

IRIS-HEP also plays a primary leadership role in the international
HEP community to prepare the ``software upgrade'' which runs in
parallel to the hardware upgrades being executed for the HL-LHC.
IRIS-HEP exists within a larger context of international and national
projects.  As an intellectual hub for software R\&D, it is building
a more inclusive community process for developing, prototyping, and
deploying software.  It drives research and development in a specific
set of areas (see Section~\ref{sec:focusareas} for the plans for
the next phase of the institute) using its own resources directly,
and also leveraging them through collaborative efforts.
After a two-year startup phase, IRIS-HEP began an execution phase
and has the technical team, project management, and governance
structures in place necessary to achieve the large-scale vision
expected of an institute-class entity.  Accomplishments of this
phase include the construction of an analysis pipeline based on the
Python Data Science ecosystem; delivery of a new, vectorized Kalman
Filter algorithm into use at CMS; transitioning the LHC cyberinfrastructure
to a new, HTTP-based bulk data transfer protocol; rolling out the
use of a new authorization paradigm based on capability tokens; and
delivering a composable, Kubernetes-based Facilities R\&D platform
for the community to use.

IRIS-HEP is in its fifth year. It has executed on the strategic
plan laid out by the \s2i2 conceptualization project, but has also
started evolving beyond the initial conditions of the plan.  Thus,
the institute has led a planning process including both community
planning workshops (see Appendix~\ref{workshoplist}) and internal
exercises.  This updated strategic plan is a vision for what must
be accomplished in the next five years by the institute to enable
HL-LHC physics.

\newpage

%% file: 020-science-drivers.tex
\section{Science Drivers}
\label{science_drivers}


IRIS-HEP aims to deliver the software required for the HL-LHC science
case, in particular enabling the discovery of BSM physics and the eventual study 
its details. To understand why discovering and elucidating
BSM physics will be transformative, we need to start with the key
concepts of the SM, what they explain, what they do not, and how
the HL-LHC will address the latter.

In the past 200 years, physicists have discovered the basic
constituents of ordinary matter and they have developed a very
successful theory to describe the interactions (forces) among them.
All atoms, and the molecules from which they are built, can be
described in terms of these constituents.  The nuclei of atoms are
bound together by strong nuclear interactions.  Their decays result
from strong and weak nuclear interactions.  Electromagnetic forces
bind atoms together, and bind atoms into molecules.  The electromagnetic,
weak nuclear, and strong nuclear forces are described in terms of
quantum field theories.  The predictions of these theories are
extremely precise, generally speaking, and they have been validated
with equally precise experimental measurements.  The electromagnetic
and weak nuclear interactions are intimately related to each other,
but with a fundamental difference: the particle responsible for the
exchange of energy and momentum in electromagnetic interactions
(the photon) is massless while the corresponding particles responsible
for the exchange of energy and momentum in weak  interactions (the
$ W $ and $ Z $ bosons) are about 100 times more massive than the
proton.  A critical element of the SM is the prediction (made more
than 50 years ago) that a qualitatively new type of particle, called
the Higgs boson, would give mass to the $ W $ and $ Z $ bosons.
Its discovery at the LHC by the ATLAS and CMS Collaborations in
2012~\cite{HIGG-2012-27,Chatrchyan:2012xdj} confirmed experimentally
the last critical element of the SM.

The SM describes essentially all known physics very well, but its
mathematical structure and some important empirical evidence tell
us that it is incomplete.  These observations motivate a large
number of SM extensions, generally  using the formalism of quantum
field theory, to describe BSM physics.
For example, ``ordinary" matter accounts for only 5\% of the
mass-energy budget of the universe, while dark matter, which interacts
with ordinary matter gravitationally, accounts for 27\%.  While we
know something about dark matter at macroscopic scales, we know
nothing about its microscopic, quantum nature, {\em except} that
its particles are not found in the SM and they lack electromagnetic
and SM nuclear interactions.  BSM physics also addresses a key
feature of the observed universe: the apparent dominance of matter
over anti-matter.  The fundamental processes of leptogenesis and
baryongenesis (how electrons and protons, and their heavier cousins,
were created in the early universe) are not explained by the SM,
nor is the required level of CP violation (the asymmetry between
matter and anti-matter under charge and parity conjugation).
Constraints on BSM physics come from ``conventional" HEP experiments
plus others searching for dark matter particles either directly or
indirectly.

The LHC was designed to search for the Higgs boson and for BSM
physics -- goals in the realm of discovery science.  The ATLAS and
CMS detectors are optimized to observe and measure the direct
production and decay of massive particles.  They are now
measuring the properties of the Higgs boson more precisely to test
how well they accord with SM predictions.

Whereas ATLAS and CMS were primarily designed to study high mass particles
directly, LHCb  was designed to study heavy flavor physics where
quantum influences of very high mass particles, too massive to be directly detected at LHC, are manifest in lower
energy phenomena.  Its primary goal is to look for BSM physics in
CP violation (CPV, defined as asymmetries in the decays of particles and their
corresponding antiparticles) and rare decays of beauty and charm
hadrons.  As an example of how one can relate flavor physics to
extensions of the SM, Isidori, Nir, and Perez \cite{Isidori:2010kg}
have considered model-independent BSM constraints from measurements
of mixing and CP violation.  They assume the new fields are heavier
than SM fields and construct an effective theory.  Then, they
``analyze all realistic extensions of the SM in terms of a limited
number of parameters (the coefficients of higher dimensional
operators)." They determine bounds on an effective coupling strength
couplings 
of their results is that kaon, $ B_d $, $ B_s $, and $ D^0 $ mixing
and CPV measurements provide powerful constraints that are complementary
to each other and often constrain BSM physics more powerfully than
direct searches for high mass particles.

The Particle Physics Project Prioritization Panel (P5) issued their
{\em Strategic Plan for U.S.\ Particle Physics}~\cite{p5Final} in
May 2014.  It was very quickly endorsed by the High Energy
Physics Advisory Panel and submitted to the DOE and the NSF.  The
report says,
{\em we have identified five compelling lines of inquiry that show great
promise for discovery over the next 10 to 20 years.
These are the Science Drivers:
\begin{itemize}
 \item
  Use the Higgs boson as a new tool for discovery
 \item
  Pursue the physics associated with neutrino mass
 \item
  Identify the new physics of dark matter
 \item
  Understand cosmic acceleration: dark matter and inflation
 \item
  Explore the unknown: new particles, interactions, and 
  physical principles.
\end{itemize}
}  
The HL-LHC will address the first, third, and fifth of these
using data acquired at
twice the energy of Run 1 and with 100 times the
luminosity.
As the P5 report says,

\vskip 0.1in
\begin{quotation}
{\em
The recently discovered Higgs boson is a form of matter never
before observed, and it is mysterious. What principles determine
its effects on other particles? How does it interact with neutrinos
or with dark matter? Is there one Higgs particle or many? Is the
new particle really fundamental, or is it composed of others? The
Higgs boson offers a unique portal into the laws of nature, and
it connects several areas of particle physics. Any small deviation
in its expected properties would be a major breakthrough.

The full discovery potential of the Higgs will be unleashed by
percent-level precision studies of the Higgs properties. The
measurement of these properties is a top priority in the physics
program of high-energy colliders. The Large Hadron Collider
(LHC) will be the first laboratory to use the Higgs boson as a
tool for discovery, initially with substantial higher energy running
at 14 TeV, and then with ten times more data at the High-
Luminosity LHC (HL-LHC). The HL-LHC has a compelling and
comprehensive program that includes essential measurements
of the Higgs properties.
}  
\end{quotation}

\vskip 0.10in
\noindent
{\bf Summary of Physics Motivation:}\ \
The ATLAS and CMS collaborations published letters of intent to do
experiments at the LHC in October 1992, about 30 years ago.  At the
time, the top quark had not yet been discovered; no one knew if the
experiments would discover the Higgs boson, supersymmetry, technicolor,
or something completely different.  Looking forward, no one can say
what will be discovered in the HL-LHC era.  However, with data from
100 times the number of collisions recorded in Run 1 
the next 20 years are likely to
bring even more exciting discoveries.

\newpage

%% file: 025-computing-challenges.tex
\section{HL-LHC Software and Computing Gaps}
\label{sec:gaps}

With the start of the HL-LHC era approximately 5 years away, the
broad outlines of the remaining software capability gaps to deliver
HL-LHC science have emerged.  Bridging these ``gaps" requires the
community to maintain a robust R\&D program between now and the
start of the HL-LHC. This will make a direct impact on the quantity
and quality of the science that can be performed.

The cyberinfrastructure requirements for HL-LHC are driven by science
needs. The experiments have assessed the needs to enable the precision
and discovery potential of the HL-LHC data set based on extensive
simulation analyses~\cite{cmstp, atlastp, hl-lhc-yellowreport2019}. These
analyses probed the capabilities of the upgraded detectors, the
potential of 3000~$fb^{-1}$ datasets, and requirements driven by the increased
event complexity that comes with the much higher luminosity of
HL-LHC.  One important outcome of these studies were estimates for
the needed event rates into the online trigger and the offline
computing facilities of the experiment. While these estimates,
particularly the offline ones, are still evolving as LHC data
analysis continues, they serve as the baseline for defining the
overall cyberinfrastructure needs for HL-LHC.

\begin{iriscolorbox}{\textcolor{white}{\bf HL-LHC Software and Computing Gaps}}
The four software and computing gaps discussed at length in this section are:
\begin{enumerate}[label=G\arabic*.]
\item \textbf{Raw resource gaps}: The HL-LHC dataset will be enormous.
Event complexity and count will each go up by about an order of
magnitude.  If no improvements to algorithms or resource management
techniques are made, the HL-LHC experiments will simply be unable
to process and store the data necessary for the science program.
\item \textbf{Scalability of the distributed computing
cyberinfrastructure}: It is insufficient to buy cores and disk alone --
the cyberinfrastructure used by the experiments must also scale to support
the volume of hardware.  This challenge is especially acute when
it comes to data transfers: both the software must be ready and the
shared networking resources (e.g., ESNet in the US) must be
appropriately managed.
\item \textbf{Analysis at scale}: Analysis at the HL-LHC will be
markedly different for two reasons: (a) the scale of the datasets
involved and (b) the use of next-generation techniques (such as
the latest machine learning techniques) to increase the scientific reach of each
result.  The former will require users to heavily utilize
dedicated `analysis facilities', optimized for high data rate I/O
and the latter will require new services and data management
techniques to be developed.
\item \textbf{Sustainability}: HEP is a facilities-driven science
- the cyberinfrastructure assembled for an experiment must last or
evolve on the decadal scale.  This limits some strategies to cyberinfrastructure -
for example, it is impossible for LHC to ``do it yourself" and own
the entire software stack.  Specific sustainability strategies must
be implemented even at the R\&D phase to ensure that the cyberinfrastructure put in
place at the beginning of the experiment is one the community can
afford.
\end{enumerate}
\end{iriscolorbox}

\subsection{Raw Resource Gaps}
\label{sec:computing_challenges}

Even before Run 1 started in 2009, a dedicated scrutiny group
annually reviewed the raw resource requests from the LHC experiments.
The input to these projections come from a few physics inputs
(planned seconds of the LHC runtime, physics events recorded per
second, average luminosity) and technical inputs (average size of
event, simulation time per event, reconstruction time per event);
combined with an experiment's computing model, guidance for the
aggregate CPU, disk, and tape needs were produced for the next three
years.

Similarly, starting in about 2016, the community has provided
predictions about the resource needs for HL-LHC.  These estimates
have larger uncertainties - it is unclear what performance gains
R\&D will provide the community and the expected compute purchasing
power a decade out - but have helped the community to understand
potential resource gaps.

\paragraph{Background:} The HL-LHC computing infrastructure must
provide sufficient raw resources in order to carry out the science
programs of the HL-LHC experiments. There are two aspects where
commodity hardware is used at large scale: the online system includes
a large compute facility to select events using a software-based
trigger; and the offline system, which is responsible for carrying
out numerous tasks including data processing, simulation production
and analysis. These cyberinfrastructure needs are typically tracked
in terms of compute (CPU, GPU, etc), rapid access storage (disk),
long-term storage (tape), and wide-area network needs. Experiments
regularly estimate compute requirements for both the long term (next
5-10 years) and short term (next 1-3 years) based upon their
understanding of what resources will be needed to support the HL-LHC
scientific program. Focusing on the US university program, we discuss
primarily the compute and disk storage requirements of the offline
computing system.

The same basic approach is used by the experiments to derive both
short and long term resource projections. The approach is simply
to start from today’s computing model and operational plan from the
accelerator and fold in research and development outcomes that will
reduce (or rarely increase) the required computational resources
in the future. These estimates include models for data processing,
needs for simulation production, analysis processing and data
management. Projections have numerous sources of uncertainty,
including the LHC schedule, the changes in the scope of the physics
program as LHC results progress, as well as evolution in software
performance due to on-going R\&D. Naturally, these are substantially
larger in the long term projections.

As a means of comparison, these resource needs are compared with
projections of the distributed computing infrastructure in place
given assumptions about future budgets, or alternatively simply the
percentage by which resources will increase each year. Just as the
model projections are uncertain,  the resource projections are
uncertain given variations in technology evolution and market forces.
Combined, the compute requirements and resource projections gives
guidance on the potential raw resource gap.

\begin{table}
\begin{center}
\begin{tabular}{|l|c|c|c|}
\hline
\rowcolor{msdarkblue}
 & \textcolor{white}{\bf Run 3} & \textcolor{white}{\bf Run 4} & \textcolor{white}{\bf Run 5} \\
\rowcolor{msdarkblue}
\textcolor{white}{\bf Parameter} & \textcolor{white}{\bf (2022-2025)} & \textcolor{white}{\bf (2029-2032)} & \textcolor{white}{\bf (2035-2037)} \\ \hline
\rowcolor{verylightgray}
LHC Energy [TeV] & 13.6 & 14.0 & 14.0 \\ 
\rowcolor{mslightblue}
Average pileup & 50 & 140 & 200 \\
\rowcolor{verylightgray}
Running time [Msec/year] & 6 & 6 & 6 \\ 
\rowcolor{mslightblue}
Integrated luminosity [$fb^{-1}$/year] & 80 & 270 & 350 \\ \hline
\end{tabular}
\label{tab:lhcperf}
\caption{Summary of the expected data volumes collected by ATLAS and 
CMS during a typical year of LHC Run 3 and during the HL-LHC program.}
\end{center}
\end{table}

\begin{figure}
\includegraphics[width=0.48\linewidth]{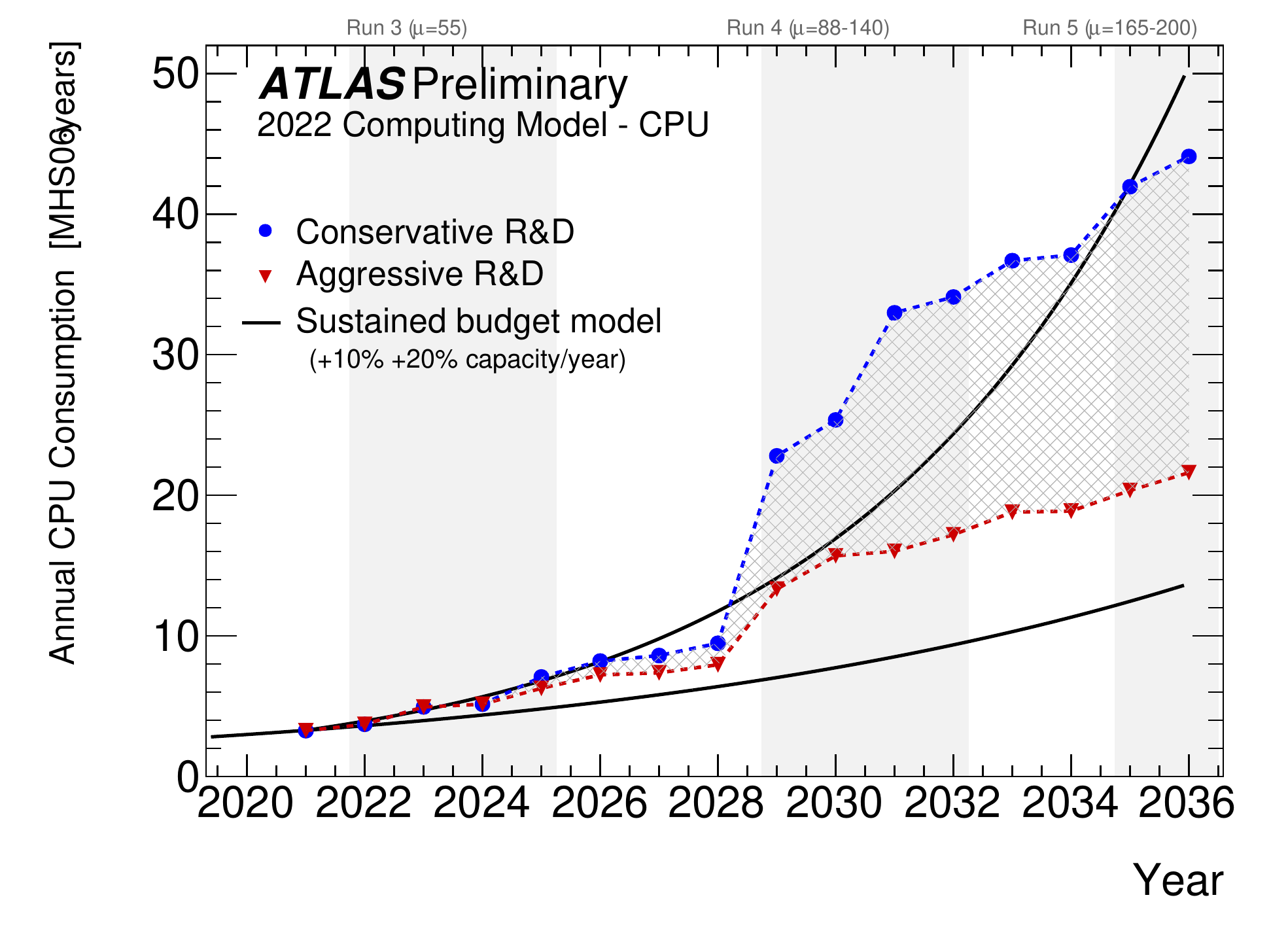}
\includegraphics[width=0.48\linewidth]{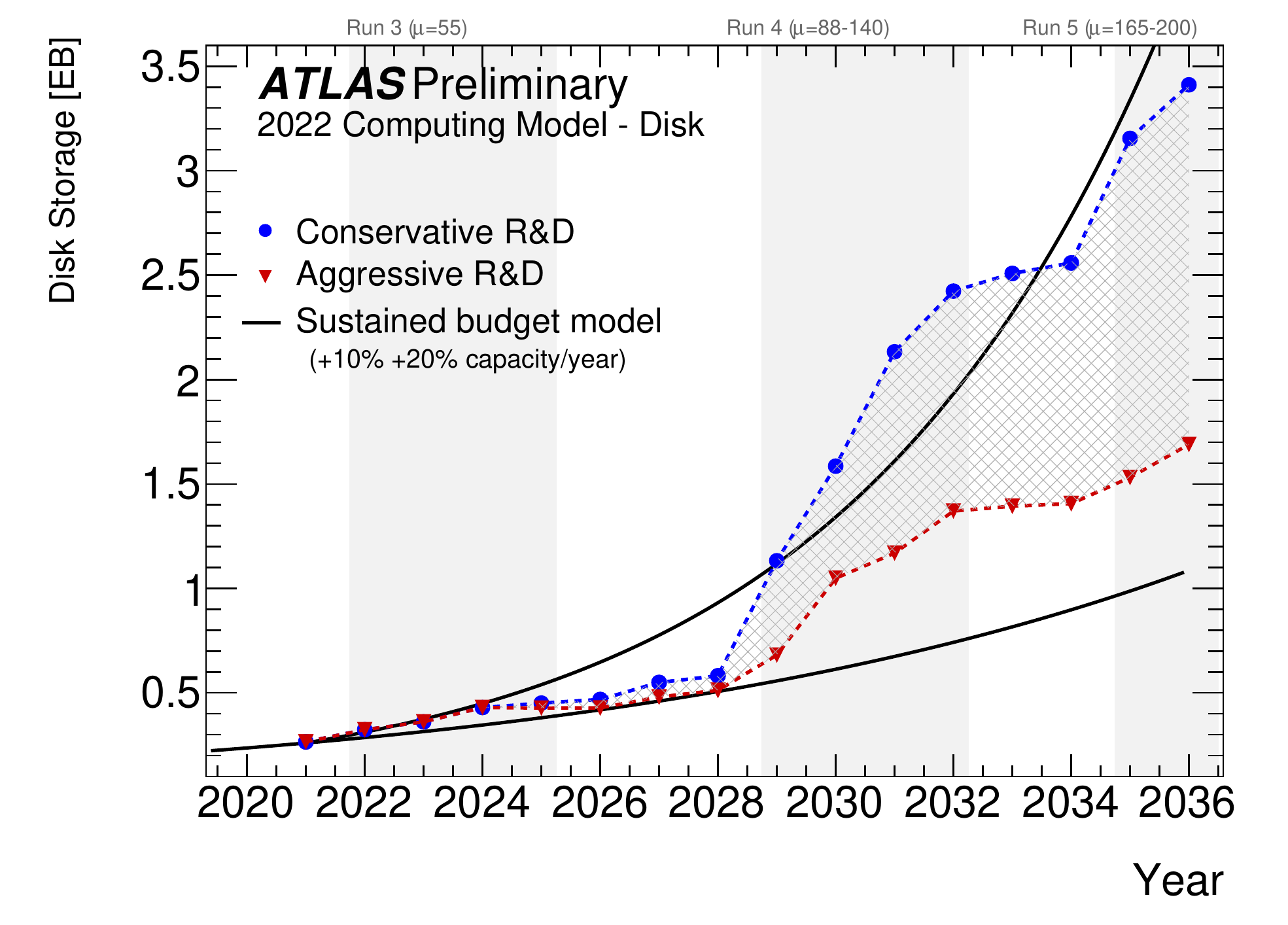}
\includegraphics[width=0.48\linewidth]{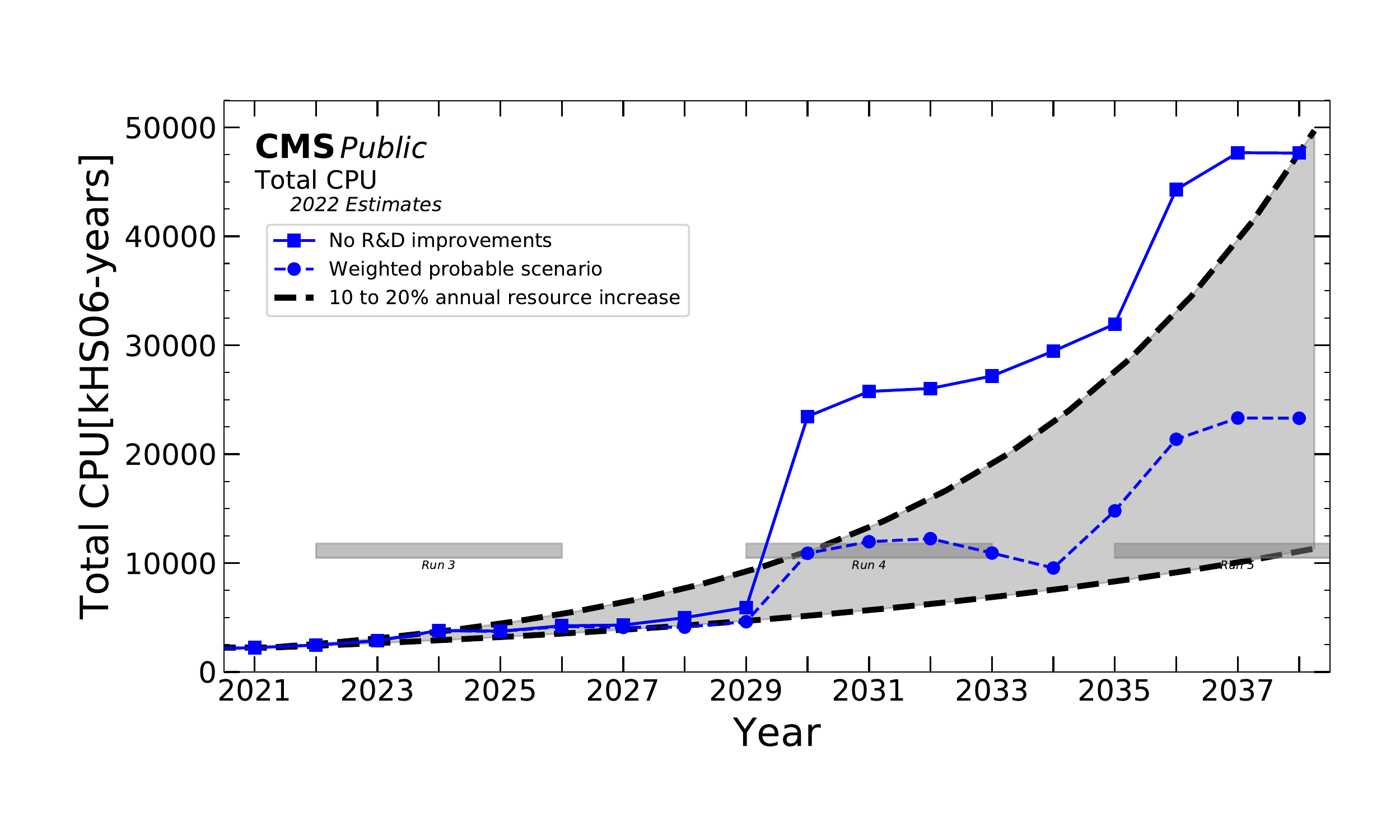}
\includegraphics[width=0.48\linewidth]{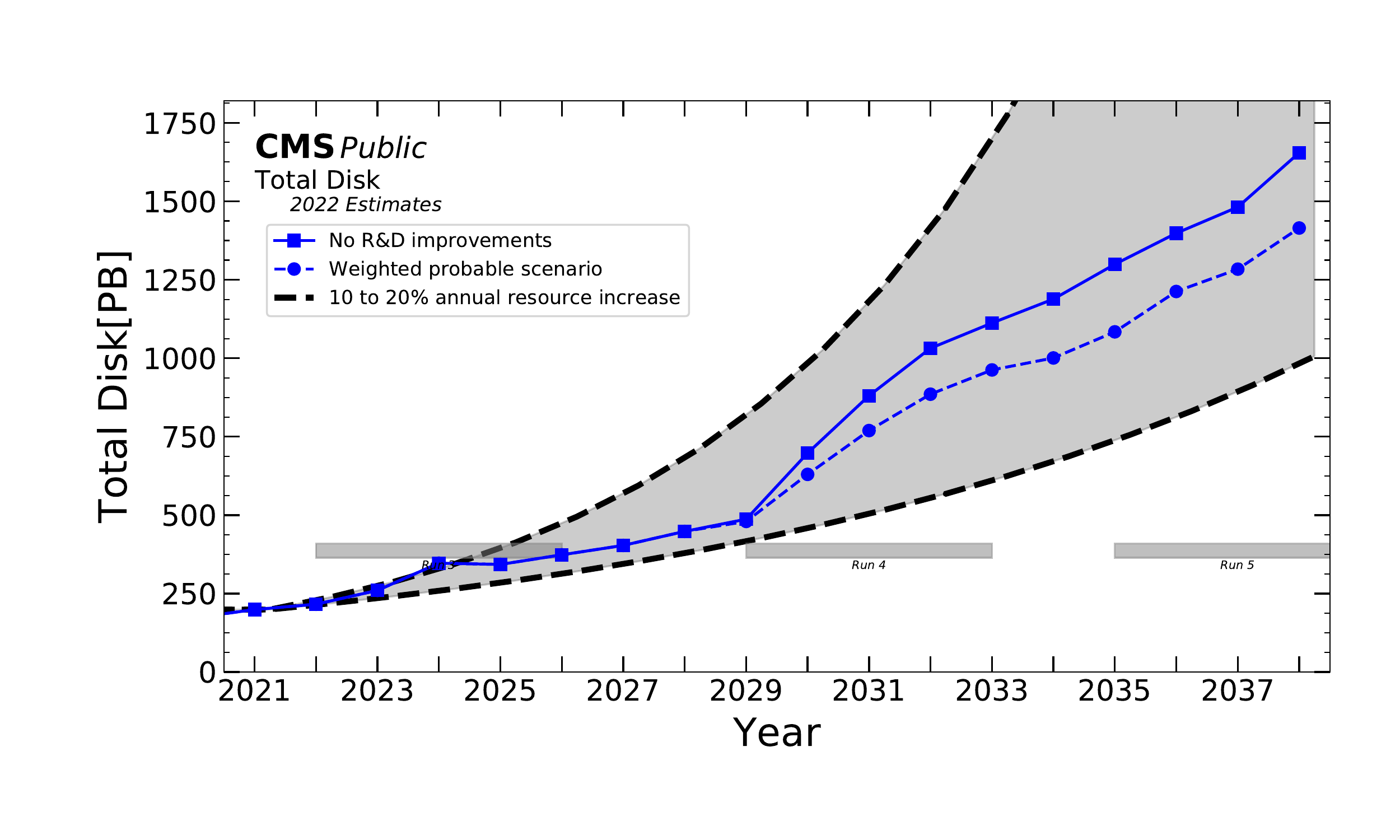}
\caption{
\label{fig:2022projections}
Current Atlas (top) and CMS (bottom) projections for CPU (left) and disk (right) requirement evolution through the first two data taking runs of HL-LHC. Each projection includes baseline estimates, and estimates including improvements that can be achieved through R\&D between now and the start of HL-LHC. Each experiment has published detailed documents~\cite{atlas2022reviewdoc,cms2022reviewdoc} on these projections for a LHCC recent review~\cite{2021lhccreview}. } 
\end{figure}

\paragraph{Resource Projections:} Figure~\ref{fig:2022projections}
shows ATLAS and CMS projections for CPU and disk needs through the
first two running periods of the HL-LHC program. As summarized in
Table~\ref{tab:lhcperf}, Run 4 (starting 2029) is expected to have
up to 140 simultaneous $pp$ interactions during each bunch crossing
(each 25 ns), while Run 5 is expected to have up to 200 such
interactions. The integrated luminosity during this period is
expected to be up to 350~$fb^{-1}$. During Run 4, the ATLAS experiment
currently expects around 10 kHz of events entering the offline
system, while CMS around 5 kHz. These can be compared to around 5
kHz of data collected by each experiment during the 2022 run. In each case, the full raw is kept for all of these events.  


Experiments have also started to incorporate the impact of future
R\&D outcomes on their resource projections. For example, the
adoption of small, columnar, data formats for analysis can substantially
reduce the disk requirements of an experiment. Similarly, a speed
up in event reconstruction application timing will reduce the compute
requirements. As R\&D outcomes are naturally uncertain, extrapolations
are done with different assumptions of what is most likely to be
realized from ongoing research. Projections are generally conservative
in this regard, as developers typically realize gains in performance
not anticipated years in advance of their research.

The other primary uncertainty component is driven by the needs of
the HL-LHC physics program. The experiments have gained substantial
knowledge during Run 2 which has lead to a broadened science program
during Run 3. In particular, both CMS and ATLAS have substantial
datasets for $b$-physics studies that were not included in the
HL-LHC projections for event rates made in the mid-2010s. 
This was made possible by continued emphasis on performance
optimization to reduce resource needs.
We anticipate that the experiments will continue to refine their needs
for HL-LHC as the LHC Run 3 program (which has just completed its
first year of operations) progresses.

\begin{iriscolorbox}{\textcolor{white}{\bf Bridging the Resource Gap}}
IRIS-HEP will continue to develop advanced reconstruction and trigger algorithms (Section~\ref{sec:recotrigger}), with a particular emphasis on charged particle tracking. It will also invest in an emerging area, Translational AI (Section~\ref{sec:ai}), with significant potential.
\end{iriscolorbox}

\subsection{Scalability of the overall distributed computing cyberinfrastructure}

While the HL-LHC experiments have not yet completed their technical
design reports, they are expected to utilize the successful distributed
computing-based approach of the LHC.  Distributed cyberinfrastructure
provides flexibility in placement of resources and allows the
community to take advantage of opportunities such as new resources
wherever they may be.  It also helps the infrastructure scale as
no single site needs to be large enough to support a single experiment.

One tradeoff of the distributed approach is the higher reliance on
middleware and wide-area network (WAN) services.  While parts of
the global LHC community have dedicated network paths solely for
the experiments, in the US most facilities heavily utilize the
shared ESNet fabric.  ESNet provides connectivity for site-to-site
transfers within the US and the transatlantic connectivity to CERN.

The \textit{scalability} challenge for the HL-LHC largely revolves
around scaling up data rates for this shared resource, demonstrating
this capability, and ensuring the network providers have sufficient
tooling to monitor and manage HL-LHC traffic.  This is an area where
purchasing hardware is not enough - the entire stack, including
file transfer software, must be tuned and prepared for the challenge.

From the raw resource requirements for the HL-LHC, one can derive
reasonable estimates of the WAN data rates that must be sustained
(e.g., if hypothetically 100PB of raw data taken at CERN is assigned
to Fermilab to process in the last three months of the year, we
know at least 110Gbps sustained for 90 days is necessary to move
the inputs).  As of 2021, the Worldwide LHC Computing Grid (WLCG)
\cite{Campana:DataChallengeOutline} estimates the worldwide set of
facilities will need to transfer at an aggregate of 10Tbps for
HL-LHC, including 900Gbps to Brookhaven National Laboratory, 1.6Tbps
to Fermi National Accelerator Laboratory, and 2,500 across the
transatlantic link.  In 2021, ESNet ran a requirements review
\cite{Zurawski:ESNetReq} that estimated a nominal Tier-2 site would
need 400Gbps of connectivity at the startup of HL-LHC.  These numbers
are approximately 20x larger than what is done today.

To meet these scalability requirements, the community has defined
the Data Grand Challenge (Section \ref{sec:domachallenge}) as a
series of escalating biennial exercises that culminate in 2025 at
100\% of the expected HL-LHC data transfer scale.  The Data Organization,
Management, and Access (DOMA) strategic area has defined a complementary
sequence of work that uses the data challenges as a periodic
integration point where progress against the goal can be measured.

The Data Grand Challenge helps prepare the cyberinfrastructure for
the scalability challenge in data movement.  Another aspect of the
scalability challenge is the computing middleware - acquiring all
the disparate resources into a virtual resource pool and executing
scientific workflows (data reconstruction, simulation, analysis
dataset derivation).  These managed compute resources will also
grow by an order of magnitude.  However, the compute scalability
challenge is more modest as some of the resource growth is canceled
out by the increase in parallelism of each compute tasks, meaning
the number of overall running tasks remains roughly the same
order of magnitude.  For compute, the experiments share a common
software layer of ``Compute Entrypoints" (CEs) used to submit ``pilot"
jobs that acquire capacity from remote batch systems.  The OSG-LHC
group helps to manage the CE and other distributed middleware running
on U.S. LHC computing facilities.

Because the scalability of the CI affects how many distributed
computing resources are available to HL-LHC experiments, meeting
this challenge is directly related to the HL-LHC science goals:
without input data moved to the processing location and the ability
to process at scale, HL-LHC science cannot be performed!

\begin{iriscolorbox}{\textcolor{white}{\bf Bridging the Cyberinfrastructure Scalability Gap}}
Many Data Organization, Management and Organization (DOMA) activities in 
IRIS-HEP (Section~\ref{sec:doma}) are designed to help scale bulk data transfer and bridge this
gap. A dedicated Data Grand Challenge (Section~\ref{sec:domachallenge}) activity will
work with the community to measure progress and deliver for HL-LHC.
\end{iriscolorbox}

\subsection{Analysis at scale}
\label{sec:analysis-at-scale}


A physics analysis in the HL-LHC era will be more than 10 times
larger than a Run 1- or 2-era physics analysis as measured by event
count alone.  New techniques increasing physics sensitivity and
controlling systematic errors will be more important in HL-LHC
physics analyses.  Technologies which empower smaller, more agile
analysis teams will be required to do the breadth of science expected
at the HL-LHC and to democratize participation across the collaborations.
The analysis model must thus change to fulfill the HL-LHC physics
program.

\begin{wrapfigure}{r}{0.4\textwidth} 
\begin{center}
\includegraphics[width=0.37\textwidth]{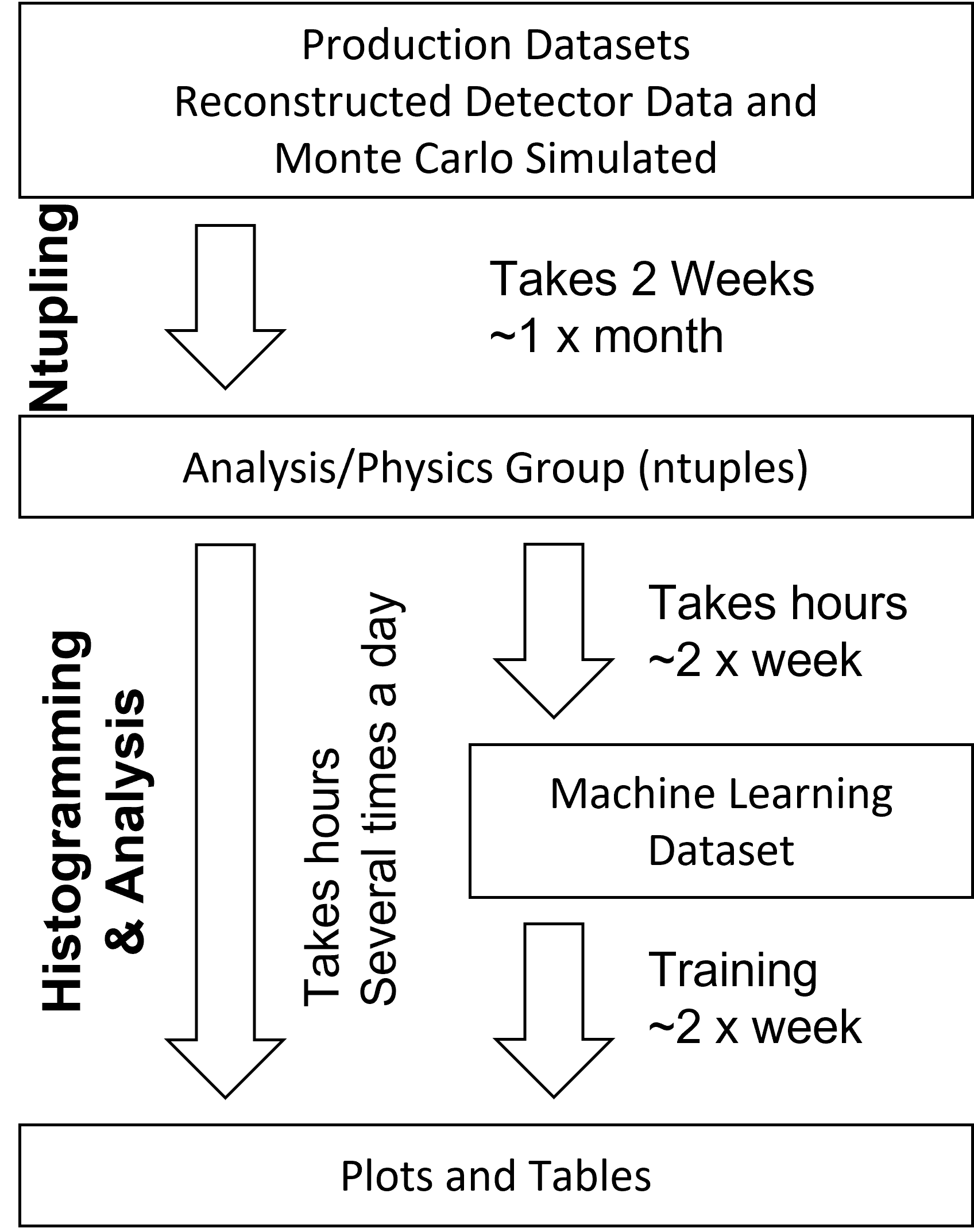}
\caption{Data and Analysis workflow for a traditional Run 2 and Run 3 physics analysis at the LHC.}
\label{fig:datareduction}
\end{center}
\end{wrapfigure}

Analysis refers to the final stage of the physics pipeline - taking
official datasets produced by an experiment and turning them into
published scientific results and discoveries.
This means spinning through the data to build histograms and other
aggregated quantities.
In Run 1 and 2, the largest analysis datasets are 100's to 10,000's
of gigabyte. In the HL-LHC they will regularly range into 100's to
1,000's of {\em terabytes} in size. Three aspects of the way our
community performs this work as shown in Figure \ref{fig:datareduction}
are not sustainable in the HL-LHC era. First, the intermediate
datasets that are produced for each analysis or group of analyses,
largely containing copies of the official data, will take up
ever-larger amounts of disk resources that will no longer be within
budget. Second, creating derivative datasets from the official data
takes 1-2 weeks with current approaches.  Scaling these costs by
the 10x additional data, a new idea that requires new data, selection
cuts, or new datasets would require months.  Innovation and
leading-edge science becomes prohibitively expensive for students.
Third, there is not enough automation to support smaller analysis
teams. Processing this data at scale, in a timely manner, will
involve novel methods of data delivery and distributed processing;
where small clusters and sometimes even laptops are enough for
HL-LHC, a larger `analysis facility' with advanced supporting
services will be necessary.

Different types of physics analyses use different tools and have
different data access patterns. The ATLAS and CMS experiments have
published papers detailing some of the important physics analyses
for the HL-LHC~\cite{hl-lhc-yellowreport2019} and attempting to
quantify how much data is needed and the expected physics reach for
the HL-LHC.  From this, we highlight three analyses that can be
used as flagships or exemplars through the next phase of IRIS-HEP
to illustrate our tools and approaches:
\begin{itemize}
\item \textbf{High-volume}: An example physics analysis that will
require extremely large datasets is the top quark mass measurement
in the $t\overline{t} \rightarrow {\rm l+jets}$ channel. The large
backgrounds in this analysis channel mean the best acceptance occurs
with the largest dataset. The backgrounds are modeled with Monte Carlo simulation,
which makes for very large simulated datasets. This is a prototypical
high-volume analysis.
\item \textbf{High-complexity}: Di-Higgs measurements probe the
Higgs self coupling, which allow us to probe the source of Electro-Weak
Symmetry Breaking - which underlies the recently discovered Higgs
mechanism. This is one of the flagship measurements for the HL-LHC.
Even with the full dataset from the HL-LHC, the sensitivity to the
Higgs self-coupling is just barely possible. This is a good example
of an analysis that will need every tool to improve its sensitivity;
Machine Learning and the use of many control regions to control
systematics are some examples.
\item \textbf{Baseline}: Finally, a $t\overline{t}$ cross section
measurement is a well understood and easily reproduced measurement.
There is also Open Data available now on which scale tests can be run.
\end{itemize}

Analysis at scale requires many individual tools to work together.
IRIS-HEP and the community created the Analysis Grand Challenge
(AGC) to integrate across the Analysis Systems, DOMA, and Facilities
R\&D areas and to demonstrate progress toward the HL-LHC goals; it
is the mechanism IRIS-HEP will use to show the analysis at scale
gap is closing. The AGC uses modern tools built by Analysis Systems,
data delivery, and prototype analysis facilities to perform all
steps of a physics analysis. This includes initial reading of the
datasets, building histograms, determining statistical and systematic
errors, performing the statistical analysis, and, finally, producing
the plots for the paper that contains the physics results. In the
current phase of IRIS-HEP, AGC uses CMS Open Data and partially
implements the \textbf{baseline} $t\overline{t}$ cross section
measurement. Over the next phase, the Institute can demonstrate it
is closing the gap by expanding the baseline and implementing a
realistic \textbf{high-volume} Higgs self-coupling measurement and
a \textbf{high-complexity} $t\overline{t}$ top mass measurement.

\begin{iriscolorbox}{\textcolor{white}{\bf Bridging the Analysis at Scale Gap}}
The Analysis Systems activities in IRIS-HEP (Section~\ref{sec:analysis-systems}) 
develop many key software elements to bridge this gap, while leveraging 
DOMA (Section~\ref{sec:doma}) projects which enable efficient data delivery
to analysis. The Analysis Grand Challenge (Section~\ref{sec:agc}) is
designed to work with the community to integrate the full system and 
demonstrate the scale required for HL-LHC.
\end{iriscolorbox}


%% file: 026-computing-challenges-sustainability.tex
\subsection{Sustainability}
\label{sec:sustainability}


The long term sustainability of the software ecosystem is critical
for HEP, given that the HL-LHC and other facilities of the 2020s will be
relevant through at least the 2030s. Key components of `scalability' include 
improved adaptability to new challenges, software longevity, and efficiency. 
We aim to ensure that the software will 
be easier to develop and maintain so that it remains available in the 
future on new platforms, meets new needs, and is as reusable as possible. 
Sustainability is also closely connected to the available workforce;
a balance is needed between the continuity of individuals -- 
as active effort or intellectually -- through the desired sustainability period 
and the need to attract and integrate new people over time.
The LHC community is {\bf in danger of a sustainability gap} in 
the HL-LHC era: care must be taken to ensure we don't have more 
software than we can sustain with the workforce we will have during 
the HL-LHC program.

Over the past few years, IRIS-HEP has
been working with the community~\cite{sustainability-blueprint-indico,sustainability-blueprint-note} 
to improve software sustainability through increased intentionality around how and what 
software is developed. The specific activities are centered both {\em (a)} on defining and 
evolving the software ecosystem and {\em (b)} workforce development.

\paragraph{Software Ecosystem:}

As a facilities-based science and with computing playing a key role within the facility, one 
can expect that some limited amount of ``operations'' effort will be 
associated to computing needs for the lifetime of the facility, but 
clear strategic choices need to be made to ensure the cyberinfrastructure is manageable within the operational effort. Given the nature of the current LHC 
and HEP software ecosystem, two paths are particularly relevant: 

\begin{itemize}
\item {\bf Identification and consolidation of redundant HEP-specific solutions:} For a number of historical and organizational reasons, many HEP software solutions are developed within the context of single experiments. In cases where the experiments actually have similar needs, this has led to multiple solutions to the same problem.
\item {\bf Adoption of solutions used by a wider scientific or open source community:} By moving to more widely used solutions the base of support for sustainability issues typically also becomes larger.
\end{itemize}

Both of these paths effectively boil down to increasing the size of the 
community using a given software element. Most software products cannot 
survive and thrive without {\em some} level of dedicated effort and 
``ownership'' by some institution or long running project. In cases
where increasing the size of the community does not significantly increase
the scope of the software, the increase effectively increases the impact
of effort invested. Concentrating available community effort
on a single solution will ultimately lead to better, more sustainable 
solutions.

For example, instead of maintaining standalone distributed high-throughput computing services within the Institute, IRIS-HEP's OSG-LHC team \ref{sec:fabric} contributes to the larger OSG Consortium.  In this way, common services needed can be shared across the NSF Science and Engineering, greatly reducing the total effort needed in this area.

While some software has been delivered for LHC Run 3, efforts to conslidate
and leverage must continue until software solutions and computing systems are 
deployed for HL-LHC. This focus on sustainability is integrated into all
of the Institute activities, and the choices we make, described 
in Sec.~\ref{sec:focusareas}.

\paragraph{Workforce Development and Evolution:}
People are also key to software sustainability. The Training 
activities of IRIS-HEP (Section~\ref{sec:training}) are central
elements for working  with the community to develop more sustainable software 
practices and skills from the ground up. The Institute's Blueprint activity aims to
build community consensus around more sustainable choices and
activities.

The Institute plays a driving role in particular for the earlier stages of 
the software lifecycle. It then partners with other organizations (the 
experiments, the US LHC operations programs, specific institutions) for 
the later elements of the lifecycle and to develop sustainability 
paths for for the long run. \\

Finally, we note that  \textbf{cross-disciplinary and broadly used software with 
a self-supporting community democratizes science, eliminating the need for 
specialized technical knowledge} and enabling researchers to gain equal access to help.

\begin{iriscolorbox}{\textcolor{white}{\bf Bridging the Sustainability Gap}}
Engineering for sustainability is built into all IRIS-HEP software activities. 
Strategically, IRIS-HEP Analysis Systems (Section~\ref{sec:analysis-systems})
activities are enabling the adoption of more broadly used data science tools. 
DOMA activities~\ref{sec:doma} work to deliver common software tools across
HEP. Most importantly, the IRIS-HEP Training activities~\ref{sec:training} 
and the Training Grand Challenge~\ref{sec:tgc} are leading the community
to prepare the workforce for the HL-LHC era.
\end{iriscolorbox}

\newpage

%% file: 050-institute-role.tex
\section{The Institute Role}
\label{sec:role}

\subsection{Institute Role within the HL-LHC Community}
\label{sec:s2i2role-hep}
\begin{figure}[htbp]
\begin{center}
\includegraphics[width=0.99\textwidth]{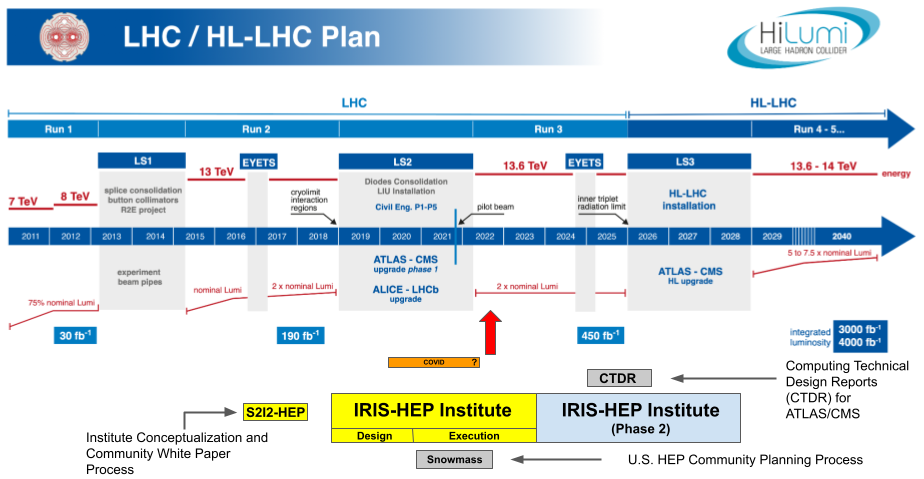}
\caption{Timeline for the LHC and HL-LHC~\cite{HLLHCTIMELINE} as of 
January 2022, indicating both data-taking periods and ``shutdown''
periods which are used for upgrades of the accelerator and detectors.
Data-taking periods are indicated by the (lower) red lines showing
the relative luminosity and (upper) red lines showing the center
of mass energy.  Shutdowns with no data-taking are indicated by
blue boxes (LS = Long Shutdown, EYETS = Extended Year End Technical
Stop). The planning and award periods for the current IRIS-HEP award
are shown in yellow.}
\label{fig:hllhctimeline}
\end{center}
\end{figure}

The mission of IRIS-HEP is to execute directly a set of R\&D
activities and to serve as an intellectual hub for the larger
community R\&D effort required to close the HL-LHC Computing Gaps.
The timeline for the LHC and HL-LHC is shown in
Figure~\ref{fig:hllhctimeline}; a 2nd phase for IRIS-HEP will operate
roughly in the 5 year period from 2023 to 2028, after the end of
funding for IRIS-HEP through the end of the R\&D window before
HL-LHC starts in 2029. This time period will coincide with two
important steps in the ramp up to the HL-LHC: the delivery of the
Computing Technical Design Reports (CTDRs) by the ATLAS and CMS
experiments in $\sim$2025 and Long Shutdown 3 in 2026-2028. The
CTDRs will describe the experiments' technical blueprints for
building software and computing to maximize the HL-LHC physics
reach, given the financial constraints defined by the funding
agencies.  For ATLAS and CMS, the increased size of the Run 3 data
sets relative to Run 2 are not expected to be a major challenge, and changes
to the detectors will be modest compared to the upgrades anticipated
for Run 4.  As a result, ATLAS and CMS have an opportunity to
continue prototyping and deploying some elements of HL-LHC computing
during Run 3, particularly during the end of year technical shutdowns,
and to perform infrastructure scaling tests alongside the production
workloads.  In contrast, LHCb made a major transition in terms of
how much data will be processed at the onset of Run 3 and HL-LHC
will be a smaller change; in the next phase of IRIS-HEP, the team
would start working on early research necessary for Run 5.

IRIS-HEP exists within a larger context of international and national
projects that are required for software and computing to successfully
enable science at the LHC, both today, and in the future. Most
importantly at the national level, this includes the U.S.\ LHC
``Operations Programs,'' jointly funded by DOE and NSF, as well as
the OSG Consortium.  At the international level, important partners
include both the HEP Software Foundation (HSF) and the Worldwide
LHC Computing Grid (WLCG). Both international organizations serve
primarily as coordination and consensus-building bodies and are
helpful for projects like IRIS-HEP that want to communicate across
the entire field. HSF focuses primarily on physics software whereas
WLCG coordinates the facilities, middleware, and operations of the
global production infrastructure.

The Institute's mission is through cooperative, community process
for developing, prototyping, and deploying software.  A strength
of an institute-scale approach is that it becomes greater than the
sum of its parts, and the larger community efforts it engenders
produce better and more sustainable software than would be possible
otherwise. Consistent with this mission, the role of IRIS-HEP is
to:
\begin{enumerate}
\item 
  Drive the software R\&D process in specific focus areas using its
  own resources directly, and also leveraging them through collaborative
  efforts (see Section~\ref{sec:focusareas}).
\item
  Work closely with the LHC experiments, their U.S.\ Operations
  Programs, the relevant national laboratories, and the larger HEP
  community to identify the highest priority software and computing
  issues and then create collaborative mechanisms to address them.
\item 
  Serve as an intellectual hub for the larger community effort in
  HEP software and computing.  IRIS-HEP brings together a critical
  mass of experts from HEP, other domain sciences, academic computer
  science, and the private sector to advise the HEP community on
  sustainable software development.  Similarly, the Institute serves
  as a center for disseminating knowledge related to the current
  software and computing landscape, emerging technologies, and
  tools.  It will continue to provide critical evaluation of new
  proposed software elements for algorithm essence (e.g. to avoid
  redundant efforts), feasibility and sustainability, and provide
  recommendations to collaborations (both experiment and theory)
  on training, workforce, and software development.
\item
  Deliver value through its (a) contributions to 
  the development of the CTDRs for ATLAS and CMS and 
  (b) research, development and deployment of software and services.
\end{enumerate}

\subsection{The Institute as an Intellectual Hub 
}
\label{sec:hub}

One highly successful aspect of the IRIS-HEP -- reinforcing the
value of the institute structure itself -- is its role as an
intellectual hub for the broader community.  Through a series of
organized activities, IRIS-HEP brings together the entire HEP
community (even beyond the Institute's science driver of HL-LHC)
to work on specific problems, disseminate knowledge and share
experience.

Through this community Intellectual Hub role, IRIS-HEP is able to
directly impact experiments such as DUNE, EIC, Belle II, FCC, SPT-3G,
XenonNT, and NoVA via its physics software, trainings, and shared
cyberinfrastructure.  For example, the analysis tools projects and
ACTS are being evaluated and utilized by EIC, Belle II, and FCC.
The services (particularly, the software stack) produced by the
OSG-LHC are incredibly popular within the HEP community; not only
are these used by DUNE, SPT-3G, XenonNT, and NoVA but by almost
every large HEP collaboration within the US.  As part of this
Intellectual Hub component, the Institute will help coalesce and
guide the broader HEP community.

Specific activities done as part of the Intellectual Hub are the
{\it Blueprint Activity} and running the series of {\em Coordinated
Ecosystem Workshops}:

\begin{itemize}
    \item \textbf{Blueprint Activity}: 
    The IRIS-HEP Blueprint activity is designed to inform development
    and evolution of the IRIS-HEP strategic vision and build (or
    strengthen) partnerships among communities driven by innovation
    in software and computing. The blueprint process includes a
    series of workshops that bring together IRIS-HEP team members,
    key stakeholders, and domain experts from disciplines of
    importance to the Institute's mission.
    
    \vspace{1mm}
    To facilitate the development of effective collaborations with
    partners across the community, the Institute will continue to
    proactively engage and bring together key personnel for small
    “blueprint” workshops on specific R\&D topics. During these
    blueprint workshops, the partners will not only inform each
    other about the status and goals of various projects, but
    actively articulate and document a common vision for how the
    various activities fit together into a coherent R\&D picture.
    The scope of each blueprint workshop should be sized in a
    pragmatic fashion to allow for convergence on the common vision,
    and key personnel involved should have the means of realigning
    efforts within the individual projects if necessary. The blueprint
    process is by design flexible and nimble to react to needs of
    the community. As a specific example, during the \textit{Learning
    from the Pandemic: the Future of Meetings in HEP and Beyond}
    workshop~\cite{Neubauer:2021xhi} experiences with HEP virtual
    events during the COVID-19 pandemic were shared and ideas for
    best practices in future meetings were developed, including the
    hybrid format for event participation which is increasingly
    common in HEP.
    
    \vspace{1mm}
    The ensemble of these blueprint workshops will be a process by
    which the Institute can establish its role within the wider
    HL-LHC R\&D effort. The blueprint process is also a mechanism
    by which the Institute and its various partners can drive the
    evolution of the R\&D topics over time.
    
    \vspace{1mm}
    \item \textbf{Coordinated Ecosystem Workshops}:  In contrast
    to the Blueprint Activity which runs topical workshops, IRIS-HEP
    (and its precursor \s2i2 conceptualization process~\cite{S2I2HEPSP}
    which paved the way for the Institute) ran three ``Coordinated
    Ecosystem Workshops", in 2017, 2019, and 2022 (the third workshop
    was delayed due to the COVID-19 pandemic) which focus on the
    entirety of the field's R\&D.  Each workshop consists of a broad
    overview of major activities across both DOE and NSF followed
    by breakouts to coordinate sub-areas.  The goal of the workshop
    is to identify potential topics of collaboration (identifying
    commonalities that may have been overlooked due to differences
    in funding agencies or experiment), help the community develop
    a prioritization of projects based on its needs and science
    opportunities, to identify gaps in the current ecosystem, and
    to avoid starting unnecessary duplicate projects. The Institute
    will continue these series of workshops and consultant with the
    community on whether their frequency should be increased to
    annually and whether they should organize discussions around
    an expanded set of science drivers.
\end{itemize}

More generally, the Institute will continue to support activities
which develop the capacity within the larger HEP and related
communities to build and support larger \textbf{research software
collaborations} whose products are relevant not only for LHC, but
more broadly. Research software is a key intellectual product of
our research, not just a critical tool.  IRIS-HEP has inspired other
such efforts and their ongoing success will be an impactful and
important legacy of the Institute.  This includes in particular
support for activities of the HEP Software Foundation, but also
building relevant collaborations with other more recently related
funded national and international community research software
projects that can build on our experience.

\subsection{Institute Role in the Software Lifecycle}

\label{sec:s2i2role-sw}

\begin{figure}[htbp]
\begin{center}
\includegraphics[width=0.75\textwidth]{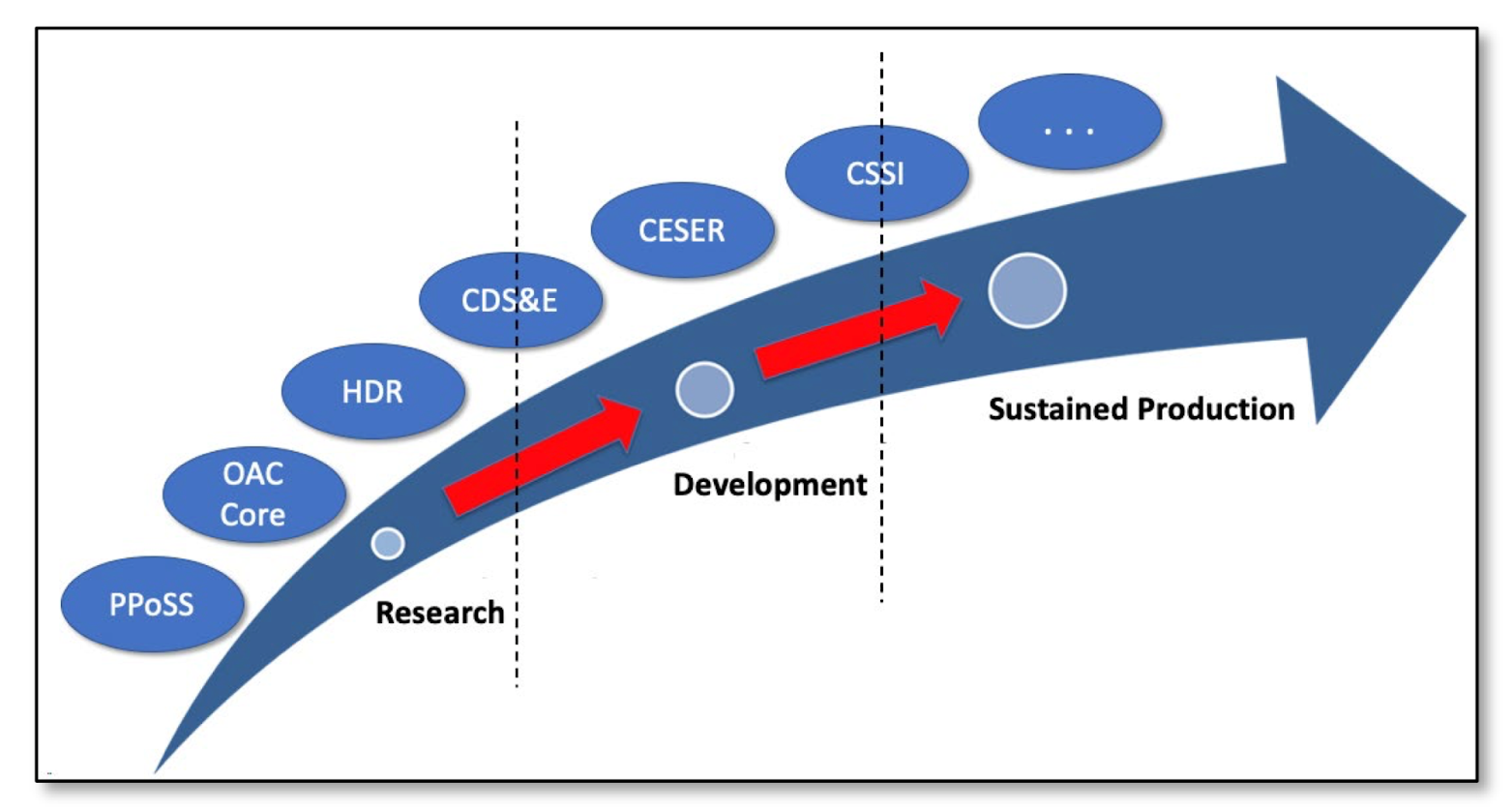}
\caption{The Software Life Cycle as seen by NSF; the Institute will bridge projects from earlier research and development phases into sustained production. Figure reproduced from ``Blueprint for a National Data and Software Cyberinfrastructure'' by NSF OAC \cite{OACBlueprint}.}
\label{fig:swprocess}
\end{center}
\end{figure}

Figure~\ref{fig:swprocess} shows elements of the software life cycle
as viewed by the NSF Office of Advanced Cyberinfrastructure (OAC),
from {\em research} in core concepts and algorithms, through {\em
development} of prototypes to deployment of software products and
long term support as part of {\em sustained production}. The community
vision for the Institute is that it will focus its resources on the
late development stage (with some exploratory work in interlinked
late-research / early-development software projects) and, in the
years closer to the start of the HL-LHC, it will partner with the
experiments, the U.S.\ LHC Operations Programs and others to
transition software into sustained production.  Compared to IRIS-HEP,
the Institute would have more of these investments in the later
stage of the lifecycle.  The experiments already provide full
integration, testing, and deployment in their internal lifecycle
processes.  The Institute will not duplicate these, but instead
will collaborate with the experiments, the Operations programs, and
the OSG Consortium on the efforts required for software integration
activities and activities associated to initial deployments of new
software products. This may also include the phasing out of older
software elements, the transition of existing systems to new modes
of working and the consolidation of existing redundant software
elements.

The next phase of the Institute will have a finite lifetime of 5 years, 
ending immediately before the start of the HL-LHC; this is still much
shorter than the planned lifetime of HL-LHC activities Thus, close
coordination with the U.S. LHC Operations programs - which will
last as long as the US participation in the HL-LHC - will be essential
to ensure the longevity and sustainability of the Institute's projects.
In its role as an intellectual hub for HEP software innovation, it
will provide advice and guidance broadly on software development
within the entire HEP ecosystem. This will be achieved through
maintaining a critical mass of experts in scientific software development
inside and outside of HEP and the cyberinfrastructure community who
partner with the Institute.

\subsection{Institute Elements}\label{institute_elements}

\begin{figure}[htbp]
\begin{center}
\includegraphics[width=0.8\textwidth]{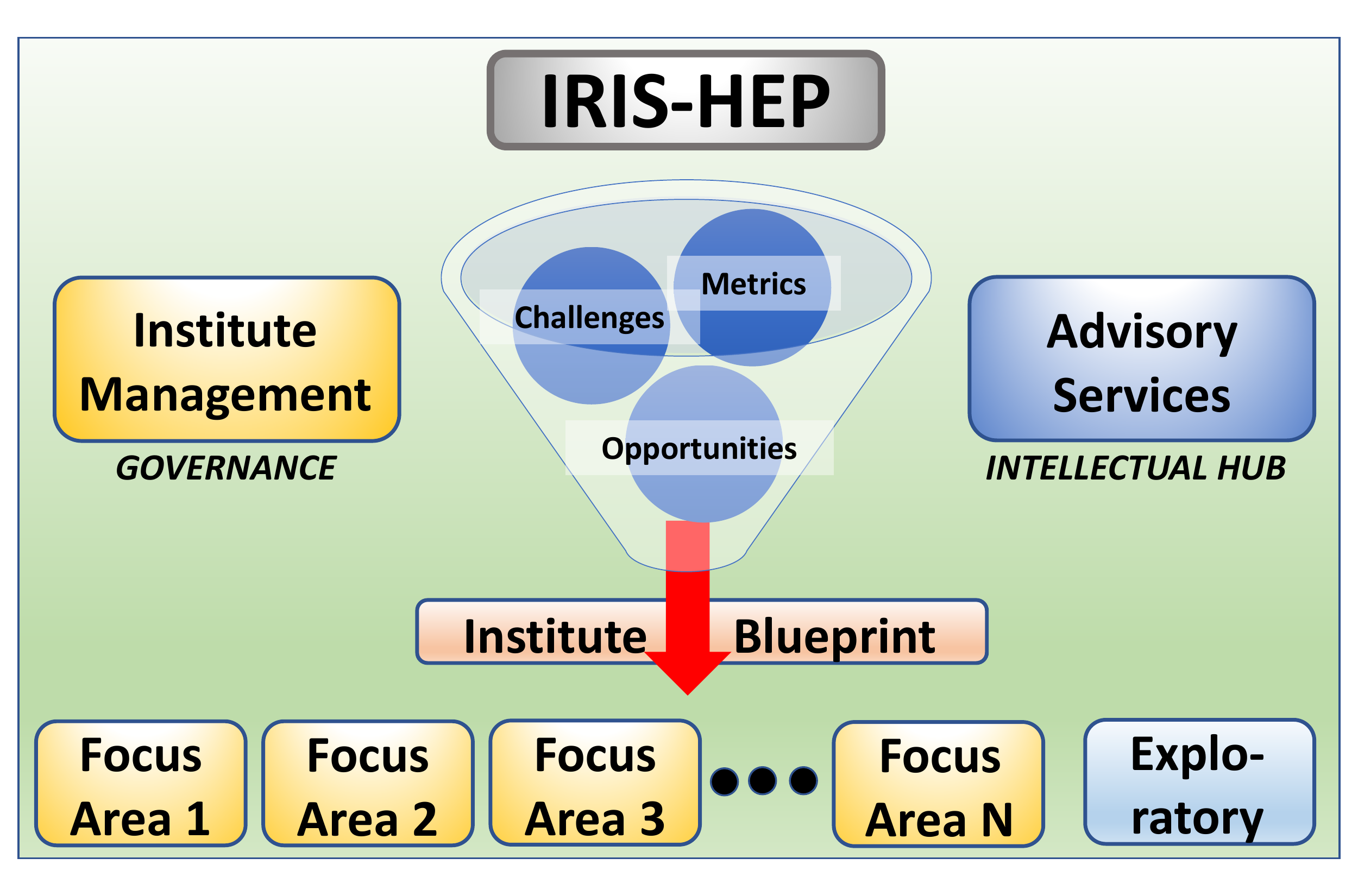}
\caption{Components of the IRIS-HEP software institute.  The focus areas envisioned for a follow-on Institute are given in Section \ref{sec:focusareas}.}
\label{fig:s2i2_elements}
\end{center}
\end{figure}

The next phase of IRIS-HEP will have a number of internal functional elements, as shown 
in Figure~\ref{fig:s2i2_elements}.

\vskip 0.1in
\noindent{\bf Institute Management:} To accomplish its mission, the institute will have a well-defined internal management structure, as well as external governance and advisory structures. Further information on this aspect is provided in Appendix~\ref{sec:orggov}.
\vskip 0.1in
\noindent{\bf Areas:} The Institute will have 7 interconnected areas of activity, which
will pursue the main goals of the Institute; these are described in Section~\ref{sec:focusareas}. The final number of areas will be contingent 
on available funding (see Section~\ref{sec:funding}). Each 
focus area will have its own specific plan of work and metrics for evaluation.
\vskip 0.1in
\noindent{\bf Blueprint \& Sustainable Software Activity:} The Institute's Blueprint Activity 
will maintain the software vision for the Institute and, 3-4 times per 
year, will bring together expertise to answer specific key questions within the 
scope of the Institute's activities or, as needed, within the wider scope 
of HEP software and computing. 
Blueprint activities will be an essential element to build a common vision with other
HEP and HL-LHC R\&D efforts, as described in Section~\ref{sec:hub}.
The blueprints will then inform the evolution of both the Institute 
activities and the overall community HL-LHC R\&D objectives in the 
medium and long term.

\vskip 0.1in
\noindent{\bf Exploratory:} Occasionally, the Institute may deploy modest resources for short term exploratory R\&D projects of relevance to inform the planning and overall mission of the Institute.
\vskip 0.1in
\noindent{\bf Advisory Services:} The Institute will play a role
in the larger research software community (in HEP and beyond) by
being available to provide technical and planning advice to other
projects and by participating in reviews. The Institute will execute
this functionality both with individuals directly employed by the
Institute and by involving others through its network of partnerships.
\vskip 0.1in

\begin{table}[h]
\begin{center}
\begin{tabular}{|l|c|}
\hline
\rowcolor{msdarkblue}
\textcolor{white}{\bf Institute WBS Area} &  \textcolor{white}{\bf Effort (FTE)} \\ \hline
\rowcolor{verylightgray}
Management/Project Office &  1.9 \\ \hline
\rowcolor{mslightblue}
Analysis Systems & 10.4 \\ \hline
\rowcolor{verylightgray}
DOMA & 2.9 \\ \hline
\rowcolor{mslightblue}
Innovative Algorithms & 8.9 \\ \hline
\rowcolor{verylightgray}
Sustainability \& Training  & 1.4 \\ \hline
\rowcolor{mslightblue}
Scalable System Laboratory & 1.3 \\ \hline
\rowcolor{verylightgray}
OSG-LHC & 5.2 \\ \hline
\rowcolor{msdarkblue}
\textcolor{white}{\bf Total} &  \textcolor{white}{\bf 32.0} \\ \hline
\end{tabular}
\end{center}
\caption{FTE distribution as of Year 5 of the current IRIS-HEP award/project, 
grouped by ``Work Breakdown Schedule (WBS)'' areas.
\label{fig:fte}}
\end{table}

\newpage

%% file: 100-strategic-areas.tex
\section{Strategic Areas of Activity for the Institute}
\label{sec:focusareas}

As an Institute focused on the software and services required to ensure
the scientific success of the HL-LHC, IRIS-HEP is part of a larger
international research, development, and deployment community. It
directly funds and leads some of the R\&D efforts, supports related
deployment efforts by the experiments, and it serves as an
intellectual hub for more diverse efforts. In a potential
second phase for IRIS-HEP from 2023-2028, the Institute's effort 
will be organized with the following focus areas:

\input{101-strategic-area-summary-include.tex}
\noindent These areas of activity were chosen primarily based 
on the {\em impact} an investment could have on the HL-LHC
Computing Gaps as outlined in Section \ref{sec:computing_challenges}
as well as continuous input from the community through workshops 
(Appendix~\ref{workshoplist}), the IRIS-HEP Steering Board and
our numerous direct collaborators.

While the strategic areas are broad - touching areas from the instant
an event is read out from the detector to the final publication of
science results - they are joined in their efforts to close the
computing gaps for HL-LHC as outlined in Section \ref{sec:gaps}, 
and, after the first phase of IRIS-HEP, intertwined and inter-reliant.
The software developed in DOMA for the production infrastructure
is largely released to the LHC facilities through OSG-LHC.  The
Facilities R\&D area provides extensive support for the testing and
scaling of Analysis Systems components.  An activity like the
Coffea-Casa prototype analysis facility crosses OSG-LHC, DOMA,
Analysis Systems, and Facility R\&D.  
Another cross-cutting activity is the support of the LHCb experiment, which
is at a different phase in its lifecycle compared to ATLAS and CMS.  Accordingly,
in Section \ref{sec:lhcb}, we aggregate together the LHCb activities that would be
distributed throughout the areas of the institute.
Even places that have little
explicit connection such as reconstruction algorithms and OSG-LHC
benefit from Institute structures like the project management and
the intellectual hub which organizes Blueprint activities.  The
institute-based approach provides the long-term vision, structure,
and alignment between distinct activities that is otherwise impossible
as a collection of small projects.

\input{110-strategic-areas-data-analysis-systems.tex}
\input{115-strategic-areas-reco-trigger.tex}

\input{120-strategic-areas-machine-learning.tex}

\input{125-strategic-areas-data-organization-management-access.tex}

\input{130-strategic-areas-facilities-rd.tex}

\input{140-strategic-areas-open-science-grid-hep.tex}

\input{145-strategic-areas-training.tex}
\input{146-strategic-areas-lhcb.tex}

\tempnewpage

%% file: 110-strategic-areas-data-analysis-systems.tex
\subsection{Analysis Systems}
\label{sec:analysis-systems}

Analysis Systems (AS) builds the tools, libraries, and pipelines that
empower a physicist to transform an experiment's production data
for physics results.
These approaches can be, and are, used for LHC analysis, though the primary goal is to unlock the ability to perform at the scale and complexity of the HL-LHC era.
While today's tools
will still function in 2029, the scale alone would make HL-LHC cost
prohibitive: some analyses today are already difficult to finish
during a PhD student's tenure.
Increasing the event count by 20x during the exploratory phase would effectively put analysis out of reach for many groups.

\begin{figure}[htbp]
\begin{center}
\includegraphics[width=0.8\textwidth]{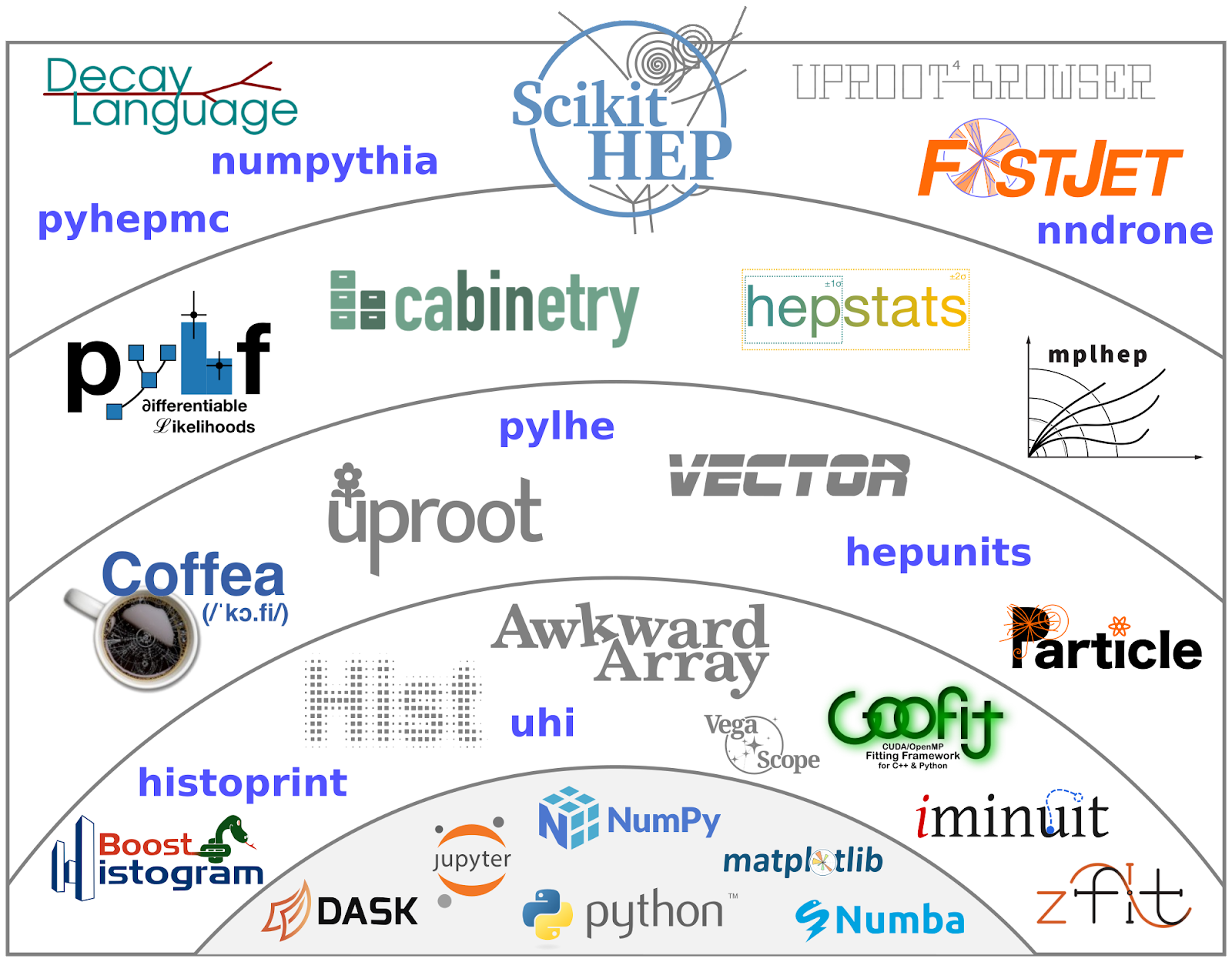}
\caption{
The components of the Python-based analysis ecosystem for HEP.
The bottom ring consists of broadly popular libraries from the Python Data Science ecosystem.
The rings surrounding the bottom contain tools
that are more specifically designed for particle physics analysis
and applications; while the bottom HEP layer is broadly used, each
layer becomes more analysis- or experiment-specific.  Many of the
HEP-specific tools are part of the Analysis System pipeline and are
either owned by or receive contributions from IRIS-HEP.  
}
\label{fig:pythonecosystem}
\end{center}
\end{figure}

This presents the LHC community with a severe computing gap (\textit{G3 (Analysis at scale)}), Section~\ref{sec:analysis-at-scale}): the current approach to analysis will not scale to the HL-LHC
dataset sizes and resource constraints.
To address this issue, IRIS-HEP has worked to build an analysis
ecosystem on the foundation of the vibrant and large open source
Python data science community, which is used abundantly in industry, and has been a driving factor in modern approaches to data analysis in astrophysics and geoscience~\cite{Harris2020}.
The ecosystem consists of individual sustainably developed analysis tools that aim to improve the analyst user experience, as seen in Figure~\ref{fig:pythonecosystem}.
The focus on consistent APIs and common
data exchange formats means this ecosystem is interoperable by
design, allowing efficient and highly scalable workflows and data
analysis pipelines --- from data ingest to final statistical analysis --- to be formed by integrating these tools.
This development focus means the tools
lend themselves to having cleaner interfaces, which is a key
ingredient for analysis reuse and reinterpretation.

An end-to-end pipeline involves much more than just software tooling.
Services are needed to deliver data and facilities are needed to
have a ``home'' where users can run analyses at much larger scale than they
do today.
Efficient delivery and analysis execution requires close integration across the DOMA, Facilities
R\&D, and Analysis Systems teams.
To facilitate such a collaboration,
IRIS-HEP has put together the Analysis Grand Challenge (AGC); see
Section~\ref{sec:agc}.  The AGC is a set of integrative, increasingly
difficult exercises which prepares the combined teams for HL-LHC
scale analysis.
The ultimate goal of these exercises is not only to enable the largest university teams to do HL-LHC analyses but to empower smaller, under-resourced teams, broadening the participation in the scientific endeavour.

\begin{figure}
    \centering
    \includegraphics[width=\linewidth]{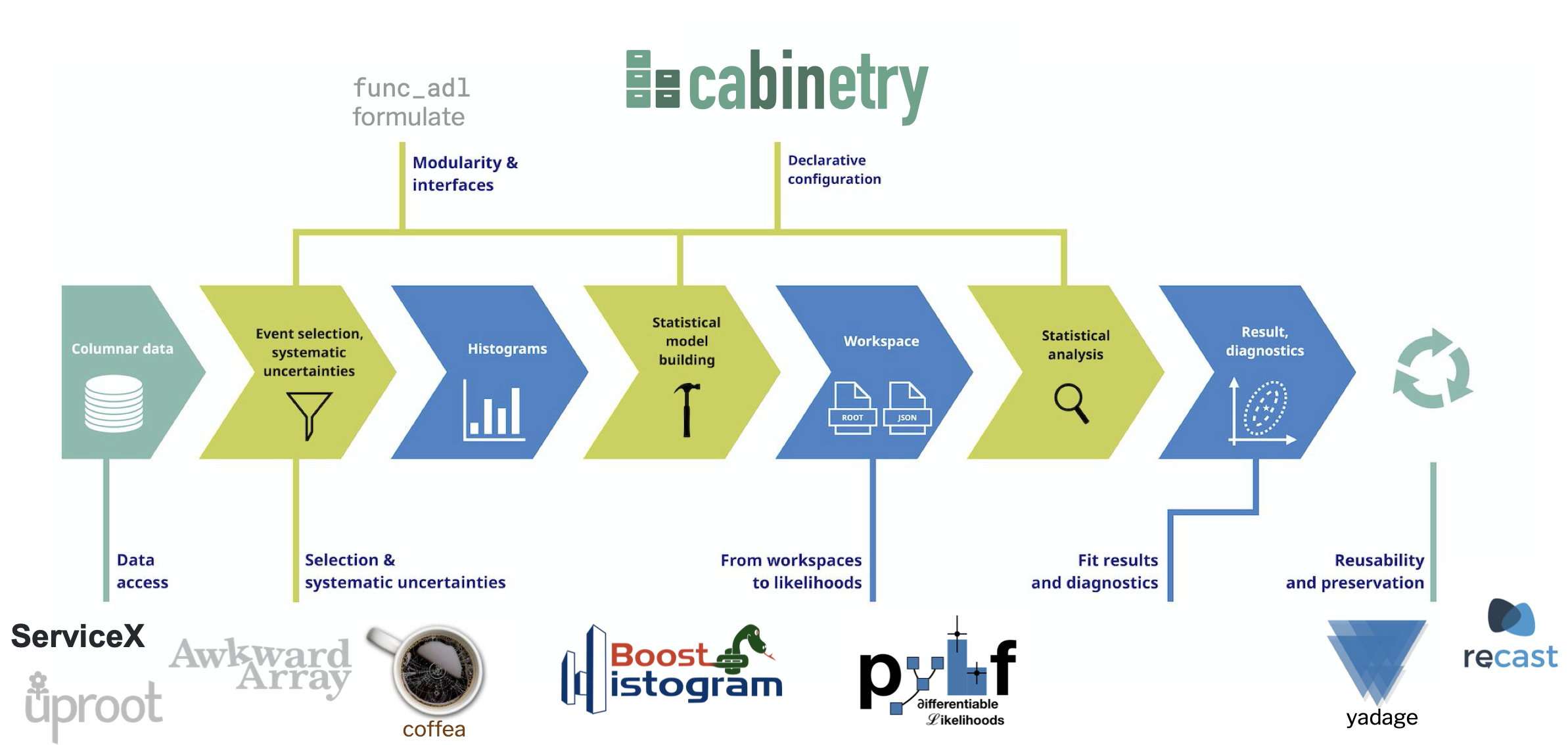}
    \caption{Representation of the Analysis Systems analysis pipeline.
    The projects shown represent the primary tools used that must integrate together to efficiently perform the analysis.}
    \label{fig:analysis-pipeline-vertical-slice}
\end{figure}

\subsubsection*{Specific Challenges and Opportunities}

In the remaining R\&D period between now and the HL-LHC, work on
improving, optimizing, and integrating the Analysis System tools with each other and others in the broad HEP analysis ecosystem is crucial.
There are also exciting new opportunities: integrating
Machine Learning (ML) applications into the analysis pipelines in a
natural way, improving our reinterpretation capabilities, and
implementation of a fully differentiable analysis pipeline.

\textbf{User experience for HL-LHC analyses}:
The AGC is an ideal
testbed for tool integration and performance tests.
The AGC is
currently using a $t\overline{t}$ cross section measurement as an analysis demonstrator. The relatively straightforward analysis has
given valuable feedback to the community showing, for example, work
needs to be done to improve how data is handed between various
stages of the analysis as well as improving the user experience for
large distributed workflows.  However, tool and integration development
work needs to be done to properly handle systematic uncertainties and make
the approach more realistic.
To add more realistic challenges in the
next phase, the AGC will adopt new flagship analyses for high-data-volume
and high-complexity cases.
Adding analyses that use even larger
datasets, similar to the $t\overline{t}$ $\ell$+jets mass measurement,
and analyses that will use ML techniques, like the Higgs self-coupling
measurement, will further exercise aspects of the integration.

\textbf{Integration of Machine Learning (ML) into the Analysis
Pipeline}: Machine Learning techniques play an increasingly important
role in analysis. By the HL-LHC almost every analysis will depend
on ML to improve physics sensitivity and reach. There are many
stand-alone tools to aid with building, training, optimization, and deployment of ML models (e.g.
MLFlow or KubeFlow), along with the more traditional approaches commonly used in the community.
There are numerous impedance mismatches between the approaches resulting in analyzers solving the same tooling problems repeatedly.
Given the centrality of ML in the ATLAS and CMS HL-LHC programs~\cite{ATLAS:2020pnm,ATLAS:2802918,CMS:2023-HL-LHC-CDR} and the institute's extensive work and expertise designing analysis workflows, folding in ML tools and services coherently into the overall ecosystem presents an opportunity for improving the user experience for analysts using ML.

\textbf{Analysis Reinterpretation}: Reinterpretation of physics
analyses is a key technique to extend the physics reach of existing
analyses.  It allows the physicist to rule out newly proposed
models using previously published analyses and extend the reach
of analyses using new observations.  Re-running a complex workflow
requires all the steps to be portable to new environments and a
workflow language specification to connect the steps.  The REANA
reproducible analysis platform~\cite{reana_github}, is used to
orchestrate the workflows across distributed resources at scale.
REANA fits well with a batch-oriented workflow and its unclear how
it and other preservation services will need to adapt for newer
services and tools.  The challenge will be balancing innovation for
new analyses while maintaining the important reinterpretation
capabilities for the field.

\textbf{Fully differentiable Analysis System pipeline}: 
ML techniques
have historically been applied in isolation in particle physics analyses.
Neural networks are often trained as a final analysis discriminant to optimize for signal and background separation, though ML techniques hold even more potential for analysis.
For example, systematic uncertainties are not included in the training which means that the network optimization may depend on variables for its separation power that are not well understood.
Expanding network training to include information from other parts of the analysis --- like information from the likelihood function used for sensitivity analysis and systematic variations of nuisance parameters --- requires the analysis pipeline to be fully differentiable so that gradient calculations can be passed between stages.
The neos project~\cite{neos_software} has demonstrated the value of differentiable analysis systems by optimizing a simplified example analysis with respect to the expected sensitivity with a process that is aware of the modelling and treatment of systematic uncertainties~\cite{Simpson:2022suz,Dorigo:2022gqm}.
The complete vision for Analysis Systems would have the full pipeline end-to-end
differentiable, from analysis pre-selection cuts to the final sensitivity determination.
Though an end-to-end differentiable analysis has never been accomplished, making it difficult to evaluate the full impact of the approach, including the systematic uncertainties in a current analysis has shown a preliminary improvement in sensitivity.

\subsubsection*{Current Approaches and Development Roadmap}

\noindent \textbf{Integrating HL-LHC analysis with the Data Science Ecosystem} \\
\textit{Funding Scenarios} \footnote{Throughout this section, we annotate each activity with associated funding scenarios (low scenario - \$4M/year, medium / baseline - \$5M/year, high - \$6M/year).  These indicate the scenarios where IRIS-HEP would work on the activity.  This particular activity would be supported in all expected scenarios; in some cases, such as the this one, in lower funding scenarios the scope and scale would be greatly reduced.}: Low (reduced scope), Medium, High\\
\textit{Description}:
The quality of the experience between a physicist and their analysis environment is important.
Historically, physicists have been trained to write analysis code and then submit it to batch computing systems and wait for their results to be returned.
In its first phase, IRIS-HEP began examining alternate approaches that arose from the data science world of task-based computing, notebook interactive computing environments, and expressive query-based languages as tools to tighten the analysis feedback loop and make data exploration more interactive.
Analysis Systems built upon existing robust data science tools, primarily from the Python ecosystem, and created a HEP-orientated ecosystem of interoperable Pythonic data analysis tools, as seen in Figure~\ref{fig:pythonecosystem}.
The tools of this ecosystem were then arranged into an analysis pipeline, as seen in Figure~\ref{fig:analysis-pipeline-vertical-slice}, based on their data analysis operations and functionality to enable a coherent flow of data between each stage of a physics analysis.
The Uproot package~\cite{Pivarski_Uproot_2017} provides the data access to the ROOT~\cite{BRUN199781} data files delivered by ServiceX~\cite{servicex} and the rest of DOMA, which are transformed into jagged Awkward Arrays~\cite{Awkward_Array_2018} where they are manipulated for efficient array-based computations.
The func\_adl query language~\cite{func_adl_github} allows for expressive event and kinematic selection criteria to be defined in analysis logic and processed in Uproot, then given to the Coffea columnar analysis framework~\cite{coffea} for additional processing, ultimately delivering histograms of the analysis selection in the form of boost-histogram~\cite{boost-histogram} objects.
These histogram objects are serialized and then ingested by cabinetry~\cite{cabinetry} to create statistical models of the physics processed in the analysis, which are then efficiently fit using cabinetry's robust selection of APIs for common analysis tasks, powered by pyhf's efficient statistical inference and optimization techniques~\cite{pyhf,pyhf_joss}, resulting in final summary statistics for the analysis.
Analysis Systems synthesizes efficient data analysis pipelines from the integration and interactions of these HEP data science tools, which coordinate closely with each other to ensure robust interoperability and efficient computations while also allowing for expressive operations and exploration as standalone tools.
Their design allows them to also extend the reach and functionality of the Analysis Systems pipeline through interactions with additional tools in the data science world (e.g. Awkward and Dask, pyhf and JAX).

As discussed more in Section~\ref{sec:fabric}, with the trend of increasing heterogeneity in the hardware market, Analysis Facilities are expected to support more ARM-based resources as well as GPUs in addition to more traditional computing architectures by the start of the HL-LHC~\cite{ATLAS:2020pnm,ATLAS:2802918,CMS:2023-HL-LHC-CDR}.
The broader data science and machine learning worlds have been following these market trends and expanding tool support from traditional x86 architectures to include multiple ARM-based platforms.
Similarly, industry machine learning frameworks are beginning to expand support across multiple commercial GPU architectures.
By integrating with and expanding on the Python data science ecosystem Analysis Systems is well poised to exploit these trends.
Many of the Analysis Systems tools already provide support for ARM-based platforms and the statistical tools already support hardware acceleration on commercially available GPUs.
It is anticipated that as ML will become a more common component of multiple stages of analysis and more analysis tools will be able to leverage hardware acceleration the demand for GPU resources will grow (discussed further in Section~\ref{sec:doma}).
Investment now in additional GPU resources at Analysis Facilities will allow for ML workflows and services to become mainstream parts of analysis workflows and reach production levels of integration before the start of the HL-LHC.
\\

\textit{Current and Potential Future Activities}:
\begin{itemize}
    \item \textit{New data formats}: A new data format, RNTuple, is being developed for the field as part of the ROOT toolkit.  Basic reading of this data has been implemented in Uproot, but the access methods are not user-friendly and RNTuple writing has not been completed.
    As the RNTuple format is still evolving, Uproot's implementation will need to evolve with it.
    Additionally, as the Apache Parquet and Apache Arrow data formats are being considered for fast columnar data access in ATLAS and CMS~\cite{ATLAS:2020pnm,ATLAS:2802918,CMS:2023-HL-LHC-CDR}, ServiceX, Awkward, and Coffea's existing compatibility with them allows for Analysis Systems pipelines to immediately explore analyses with these new formats.
    \item \textit{Supporting heterogeneous compute resources}: As heterogenous compute resources become more abundant at Analysis Facilities and across the field supporting these platforms becomes more critical.
    Analysis Systems tools already largely support multiple ARM-based platforms and are expanding coverage for operations resources.
    \item \textit{Supporting columnar data analysis}: The Analysis Systems analysis pipeline is built on a columnar analysis approach and exploits Awkward and Coffea's design choices.
    Supporting columnar data access and analysis infrastructure are planned HL-LHC milestones for ATLAS and CMS.
    As work towards these milestones advances, AS tools will need to adapt to support interoperability with the experiment solutions (e.g. implementations in ROOT's RNTuple, ATLAS's PHYSLITE, and CMS's NanoAOD).~\cite{ATLAS:2020pnm,ATLAS:2802918,CMS:2023-HL-LHC-CDR}
    \item \textit{Filling missing serialization gaps}: Interoperability between components in an ecosystem is critical for user experience.
    While most analyses could be converted to pure Python, there are key object types that cannot be cleanly serialized / deserialized from within Python (e.g. ROOT's `member-wise serialization' mode, RooFit models, boost-histogram objects).
    By closing these gaps, the next phase of IRIS-HEP can better serve HL-LHC user communities that use these techniques.
    \item \textit{Maturing tools to production}: While most of the Analysis Systems tools have become mature projects and staples of the Scikit-HEP community~\cite{Rodrigues:2020syo} with stable APIs, large user bases, and contributions from outside of the development teams, there remain edge cases in terms of features as well as user experience and knowledge that need to be addressed to reach a production state.
    Covering the full range of Analysis Grand Challenge exercises, in addition to supporting LHC analyses already using Analysis Systems tools, provides a roadmap for reaching a feature-complete production state as well as opportunities for strengthening synergies with Scikit-HEP and coordination with facilities and US operations.
    Forming community investment and knowledge for the Analysis Systems tooling ecosystem is also required to reach a production state where the HEP community can contribute to the long term maintenance of projects beyond the scope of IRIS-HEP operations (\textit{G4 (Sustainability)}), Section~\ref{sec:sustainability}).
    Analysis Systems is addressing this in multiple ways:
    \begin{itemize}
        \item Training the community: By participating in the community training and workforce development events coordinated by IRIS-HEP (described in Section~\ref{sec:training}) Analysis Systems is able to educate the HEP community on the tools and workflows being produced, while also receiving direct user feedback on the user experience.
        Analysis Systems is also closely involved with the activities of the HSF working groups, with multiple members serving as co-conveners for the HSF Python in HEP (PyHEP) working group.
        PyHEP holds an annual online workshop covering the state of Python in the field with in-depth tutorials for widely used tools.
        Analysis Systems tools have become recurring highlights of the workshop tutorials, which have hundreds of daily active participants and historically over a thousand registered participants, allowing for updates on latest techniques, workflows, and developments to be shard with the HEP user and contributor community.
        \item Engaging the broader scientific open source community: Some Analysis Systems projects, like Awkward, provide functionality that solve uses cases typical enough in scientific computing that the user base has extended beyond HEP to other scientific fields and industry.
        This has lead to a collaboration with engineers at the data science company Anaconda to create dask-awkward~\cite{dask-awkward_github} which allows Awkward arrays to be used natively in Dask computational workflows.
        To further adoption in the broader data science community and foster closer interactions with the open source tools the Analysis Systems ecosystem builds upon, multiple tools have applied to become ``affiliated projects'' with NumFOCUS~\cite{numfocus-affiliated-projects} --- a non-profit community organizing body for promoting open practices in research, data, and scientific computing which includes member projects like NumPy, SciPy, Jupyter, and Numba.
        pyhf is the first IRIS-HEP project to be accepted as an affiliated project of NumFOCUS in 2022, with other projects expected to be accepted in 2023.
        Collaborations with industry and the open source data science community allow for Analysis Systems tools to develop diverse contributor communities and maintenance support systems.
    \end{itemize}
\end{itemize}

\begin{figure}[h]
\begin{tcolorbox}
Providing high levels of interoperability between components of the Analysis Systems ecosystem with each other and with external data analysis libraries is one of the most high-impact activities of the Analysis Systems team.
\end{tcolorbox}
\label{fig:AS_integrating_HL-LHC_analysis}
\end{figure}

\noindent \textbf{Statistical Tools for HL-LHC}\\
\textit{Funding Scenarios}: Low (reduced scope), Medium, High\\
\textit{Description}: The statistical inference demands for analyses at the HL-LHC will require tools that are able to fit large and complex statistical models efficiently.
This implies tools must become easier to use ---  physicists should be able to build complex statistical models quickly using expressive APIs --- and be able to better leverage available hardware accelerators to avoid the statistical step from becoming a bottleneck.
In the first phase of IRIS-HEP, the pyhf Python library was adopted for its ability to leverage hardware acceleration and automatic differentiation for faster fits.
Similarly, the cabinetry Python library was developed to build on top of pyhf and enhance the model building experience and provide high level APIs for statistical procedures common to HEP.
The next phase of statistical tooling will focus on \emph{adding functionality} that will be necessary for HL-LHC and for a fully differentiable Analysis Systems pipeline as well as \emph{optimizing the performance} of the libraries for statistical inference on GPUs.\\
\textit{Current and Potential Future Activities}:
\begin{itemize}
    \item \textit{Fully differentiable analysis pipeline}:
    pyhf uses machine learning frameworks to exploit automatic differentiation of the constructed likelihood function to speed up statistical inference.
    The computational graph that is created for this currently remains internal to the calculation and extending this to provide passing of gradient information from pyhf to other libraries requires additional work.
    The neos project builds on pyhf to provide an example implementation of a subset of the necessary changes to be able to differentiate through pyhf-based calculations, and will serve as an external comparison for pyhf's development, as well as a future external user for validation.
    \item \textit{Hardware acceleration optimization}: The first phase of IRIS-HEP prioritized the user experience of fitting statistical models on GPUs rather than the performance optimization.
    A focus on improving the performance across its GPU-enabled computational backends is expected to deliver 2x or more speedups.
    This optimization work would provide opportunities to strengthen the relationship between the IRIS-HEP team and NVIDIA software engineers working on their RAPIDS AI product.
    \item \textit{Adoption of unifying statistical model standards}: The High Energy Physics Statistics Serialization Standard (HS3)~\cite{HS3} --- which will allow for unification of model specification across the HEP statistical landscape --- has\ recently been proposed and outlined by maintainers of tools in the HEP statistical modeling ecosystem, including the pyhf developers, with a planned publication of the new specification the end of 2023.
    Having pyhf and cabinetry support the HS3 standard would allow for increased interoperability (\textit{G4 (Sustainability)}), Section~\ref{sec:sustainability}) and remove the need for model serialization conversion between pyhf and other statistical libraries like RooFit and Bayesian Analysis Toolkit (BAT)~\cite{Schulz:2021BAT}.
    \item \textit{Efficient orchestration of template histogram production}: cabinetry provides convenient high level APIs for building and steer template fits, however at the moment the template histogram production steps are sequential and not optimized.
    This can be a bottleneck when approaching analyses with a large number of systematic uncertainties as this requires repeated sequential reads and compute operations for each systematic variation applied.
    The larger analyses of the HL-LHC will require cabinetry to have optimized read and decision operations based on variations in addition to efficient parallelization of histogram production.
\end{itemize}

\begin{figure}[h]
\begin{tcolorbox}
The performance, flexibility, and interoperability of the Analysis Systems statistical tools presents opportunities for significant impacts on HL-LHC analysis.
\end{tcolorbox}
\label{fig:AS_statistical_tools}
\end{figure}

\noindent \textbf{Analysis Preservation and Reinterpretation} \\
\textit{Funding Scenarios}: Medium (reduced scope), High\\
\textit{Description}:
The physics value of ensuring that LHC and HL-LHC analyses are preserved in a robust and complete state such that they are usable for reinterpretation of new physics signatures is significant.
In addition to reducing the analysis team time for investigation of new physics models, it reduces the total compute needed compared to an analysis being created for the first time.
Additional work will be required to find ways to map the analysis preservation and reinterpretation technology that exists for the current LHC analyses to the Analysis Systems workflows that have been designed to run on Coffea-Casa infrastructure, or, alternatively, to map the entirety of an AGC-like analysis into a computational node in a larger analysis reinterpretation workflow (i.e., implemented with REANA).\\
\textit{Current and Potential Future Activities}:
\begin{itemize}
    \item \textit{Coordinating development of analysis preservation and reinterpretation technology}: IRIS-HEP Analysis Systems team members are developers of the RECAST software~\cite{recat_atlas_github}, used for ATLAS's complete analysis preservation approach, and coordinate with the REANA development team to ensure compatibility of RECAST and REANA releases.
    This coordination work and development work will be extended to also include support for and integration of Analysis Systems workflows with these existing technologies.
    \item \textit{Development of Coffea-Casa compatible reinterpretation workflows}: To ensure that the AGC analyses preserved with modern HEP technologies do not limit the analysis computation, exploration of the implementation paradigms is required.
    One approach is to embed the entirety of the analysis as a node in a larger reinterpretation workflow (i.e. in REANA), but would present additional deployment restrictions.
    An alternative approach is to develop approaches that allow for the current preservation and reinterpretation technologies workflow graphs to be executed within Coffea-Casa infrastructure for tighter integration.
    It is unclear which is the correct implementation paradigm and requires exploration of options and then engineering of an appropriate solution.
\end{itemize}


\noindent \textbf{Enabling Differentiable Analysis for HL-LHC} \\
\textit{Funding Scenarios}: Medium (reduced scope), High\\
\textit{Description}:
A goal of Analysis Systems is to make performing more advanced analyses easier and faster through considered design of the tools and well executed implementation.
Often selection optimization with respect to a specific variable is performed in analyses at the LHC, but it is performed by variations in incremental discrete steps.
Being able to differentiate through the selection step allows for exact optimization of the variable, and similarly if the entire analysis toolchain is able to be differentiated through the entire analysis can be end-to-end optimized by optimizing the final discriminant with respect to the model parameters.
This is a large scale project that will require new development and design effort for all core Analysis Systems projects (Awkward, Uproot, func\_adl, Coffea, histogram libraries, pyhf, and cabinetry) and will be an excellent integration exercise for Analysis Systems.\\
\textit{Current and Potential Future Activities}:
\begin{itemize}
    \item \textit{Supporting automatic differentiation in AS tools}: To achieve a fully differentiable Analysis Systems pipeline requires that all components are able to handle differentiable operations as well as the propagation of gradient information.
    There are ongoing efforts to add automatic differentiation support for Awkward and expose pyhf's internal automatic differentiation to other operations, though implementing support for the remaining tools and then efficiently integrating all the tools to support end-to-end optimization requires substantial development work.
\end{itemize}

\noindent \textbf{ML Integrated with the Analysis Pipeline} \\
\textit{Funding Scenarios}: Medium (reduced scope), High \\
\textit{Description}:
As analysis scale and complexity grows at the HL-LHC the use of machine learning in all aspects of analysis will grow as well.
It is critical to be able to support ML workflows at any stage of the Analysis Systems pipeline and to have clear examples of how ML workflows and applications can be integrated into the existing pipeline.\\
\textit{Current and Potential Future Activities}:
\begin{itemize}
    \item \textit{Integration of ML operations into AGC analyses}: There are ongoing efforts to expand the AGC benchmark analyses to also include ML components.
    The first stages of this involve taking ML models that are used widely by the LHC experiments and integrating their deployment as an additional stage of the Analysis Systems pipeline.
    In addition to this initial step, adding support for ML models to be trained efficiently on GPUs on Coffea-Casa infrastructure and then used directly in the Analysis Systems pipeline will also be required.
    \item \textit{Provide MLOps infrastructure}: The process of being able to robustly retrain machine learning models, test and evaluate them, and then deploy in a versioned manner into a production analysis environment (broadly known in machine learning research as ``MLOps``) requires substantial orchestration and maintenance.
    As the Analysis Systems ecosystem and pipelines are designed to be extensible to allow analysts to quickly iterate on analyses without constraining them, providing MLOps for Analysis Systems would allow for faster integration of user ML models into analyses.
    Providing MLOps infrastructure requires development of expertise and software infrastructure in MLOps system development and deployment, but presents large added value for turnaround times on physics analyses that rely heavily on ML approaches.
    MLOps implementation will be a point of close collaboration with Facilities R\&D to ensure efficient use of GPUs and other compute resources without interruption or added burden.
    \item \textit{Support high level ML workflows}: In addition to MLOps infrastructure users will need high level workflows that allow them to iterate through the ML lifecycle --- from experimentation to deployment to the Analysis Systems pipeline --- in a reproducible manner.
    Open source MLOps orchestration tools like MLFlow and KubeFlow have been developed by industry to make it easier for researchers to manage the complexity of the end-to-end process of building and deploying ML models.
    Leveraging these industry standards to create reproducible workflows quickly will provide an improved user experience for physicists using complex ML models for analysis.
\end{itemize}

%
\noindent \textbf{Handling complex data with Awkward Array} \\
\textit{Funding Scenarios}: Low (reduced scope), Medium, High\\
\textit{Description}: One hallmark of HEP data is its complexity; it is fundamentally deeply nested, uses variable-sized arrays, records, mixed types, and missing data from records.  This has made HEP data difficult to use with standard data science tools like NumPy, which often assumes fixed-sized, rectilinear data.  Awkward Array fills this gap, providing a NumPy-like interface to this so-called ``jagged" data.
Awkward has reached a state of maturity and has an ecosystem of tools that are being built around it, including with non-HEP-specific focus like the collaboration with Anaconda.
It is foundational in the infrastructure built by IRIS-HEP and the Analysis Systems pipeline is built on top.\\
\textit{Current and Potential Future Activities}:
\begin{itemize}
    \item \textit{Supporting automatic differentiation}: A fully differentiable pipeline requires Awkward-array and to be able to perform operations that propagate gradients through calculations.
    This will require additional substantial development work.
    \item \textit{Speeding up calculations}: Awkward provides users with natural Python interface to HEP data; now the interface is established, increasing the overall speed is critical.  While its NumPy-like interfaces are \emph{much} faster than native Python, it has not achieved the same speed as native C++ code.
    A C++ JIT backend --- where operations can be combined and lowered directly to machine code --- would eliminate many of these issues.
    \item \textit{ROOT integration}: Outside Python, the defacto toolkit for data analysis is ROOT.  The AS team aims for smooth interoperability between ROOT and the rest of the ecosystem.
    For example, ROOT's RDataFrame approach to analysis needs to be extensively tested and potentially expanded when processing Awkward Arrays.
\end{itemize}

\begin{figure}[h]
\begin{tcolorbox}
Complexity of data and analyses at the HL-LHC will grow.
Planned improvements to Awkward Array will provide the data handling and manipulation capabilities necessary to approach these challenges.
\end{tcolorbox}
\label{fig:AS_complex_data}
\end{figure}

\subsubsection*{Impact and Success Criteria}


In the first stage of IRIS-HEP, Analysis Systems has provided leadership in the end-user facing HEP software community, with members of Analysis Systems holding convenerships in the HEP Software Foundation Python in HEP working group and administrator roles in the Scikit-HEP community project, and provided a nucleation site for software discussions --- generating many of the IRIS-HEP Blueprint Activity meetings and driving development discussions and best practices in the IRIS-HEP Slack.
The largest impact Analysis Systems has had is providing mature, interoperable analysis software that physicists at the LHC experiments have been able to adopt already into their workflows and analyses.
Most notably, the Coffea project used actively for analysis in CMS is not an IRIS-HEP project, yet has adopted many components of the Analysis Systems ecosystem --- including Awkward, Uproot, and hist --- as core dependencies to better integrate into the analysis design philosophy of Analysis Systems, which also affects analyses in CMS.
Additionally, as of December 2022, the pyhf library has been used by ATLAS to publish 24 full statistical models of published ATLAS results, and become a core dependency of analysis frameworks used in the ATLAS SUSY group for ATLAS Run II results.
While the adoption of collections of individual components of the Analysis Systems is a step forward, the realization of the full benefit of Analysis Systems comes from the integrated use of the ecosystem as a whole.
Investment in Analysis Systems as part of an NSF-funded software institute will bring more LHC analyses into strong positions for adoption of Analysis Systems pipelines before the HL-LHC begins and allow for the full impact of the analysis ecosystem designed and built in the first phase of IRIS-HEP on the landscape of modern physics analyses.
Having Analysis Systems as a focus area brings momentum and ecosystem continuity, with existing teams of IRIS-HEP developers ready to build on the success of the analysis ecosystem and community.

The high-impact outcomes for analysis activities in Analysis Systems will be tightly coupled with the success and participation in the AGC.
The AGC will provide multiple opportunities for demonstration of the planned Analysis Systems milestones, covered in detail in Section~\ref{sec:agc}, including implementation of the high-complexity di-Higgs search AGC analysis that will exploit the added machine learning components of the Analysis Systems pipeline and the improvements in statistical tools.
The capstone of these demonstrations will be the 2026 AGC demonstration of a fully differentiable analysis which will leverage advances across the Analysis Systems ecosystem.
These demonstration efforts will compound in value as they additionally serve as roadmaps to full adoption of Analysis Systems pipelines for the LHC experiments.\\

\noindent \textbf{Success Criteria – Milestones \& Deliverables}:

The milestones and deliverables outlined below focus on the first three years of an Institute.
These high-level items would be expanded and later years filled in as part of the execution of any project.

\begin{enumerate}[label=D4.\arabic*.]
\item All core components of the Analysis Systems pipeline fully support distributed analysis.
\textbf{May 2024}.
\item Demonstration of running a full analysis, that is suitable for addition to the AGC, that uses machine learning.
\textbf{July 2024}.
\item Demonstration of running a full analysis, that is suitable for addition to the AGC, that is able to use statistical models defined in the unified HS3 serialization format (planned for release at the end of 2023). 
\textbf{December 2024}.
\item Demonstration of the Analysis Systems pipeline being used as part of a reinterpretation workflow (e.g., a node within a larger graph executed the REANA analysis platform).
\textbf{December 2024}.
\item All core components of the Analysis Systems pipeline support integration of differentiable operations and passing of gradients.
\textbf{July 2025}.
\item Demonstration of reinterpretation of an AGC analysis through use of a reinterpretation platform. \textbf{December 2025}.
\item Demonstration of an analysis, suitable for addition to the AGC, that has been optimized end-to-end through use of automatic differentiation of the Analysis Systems pipeline.
\textbf{December 2026}.
\end{enumerate}

\noindent \textbf{Success Criteria – Metrics}:

Metrics are a useful tool to provide management with quantitative insight about progress toward overarching goals.  High-level metrics we expect to be applicable for the Analysis Systems area are below:

\begin{enumerate}[label=M4.\arabic*.]
\item The number of users and analysis groups that use Analysis Systems tools and/or pipelines in early stages of analysis.
Collecting feedback from these early adopters allows to understand better user experience related issues.
The goal is to add at least two analyses per year.
\item The number of Analysis Systems tools that support integration with automatic differentiation.
The goal is half of all relevant projects by December 2023 and all relevant projects by end of December 2024.
\item The number of analyses that use machine learning that are able to perform aspects of model training or run model inference as part of the Analysis Systems pipeline.
The goal is one by the end of July 2024, and one from each target LHC experiment by the end of July 2025.
\item The number of publications, conference notes, or public notes from each of the target LHC experiments that cite the use of the Analysis Systems tools and analysis pipeline.
The goal is at least one by the end of December 2024, and at least one from each target LHC experiment by the end of December 2025.
\item The number of Analysis Systems tools that have at least one regular contributor or maintainer from the broader scientific developer community.
The goal is at least one for each project that is undergoing active feature development --- projects that are at a mature stage of their life cycle and only providing bug fix releases are not expected to have additional non-IRIS-HEP maintainers. 
\end{enumerate}

%% file: 115-strategic-areas-reco-trigger.tex
\subsection{Reconstruction and Trigger Algorithms}
\label{sec:recotrigger}

%
%
%

Algorithms unpack raw detector data and transform it into data structures that can then be used to identify interesting events: either in real-time by the software {\em trigger} system, or later during {\em reconstruction} prior to analysis. These algorithms are run in three distinct contexts by the LHC experiments. First, in the software trigger system, where a keep or reject decision must be made for around one million events per second during standard data taking conditions in HL-LHC. This process is challenging as only around 1\% of these events (or 0.01\% of the total collision rate of LHC) can be retained. Second, these algorithms are run `offline' (outside of the detector environment at CERN) to reconstruct data providing data in experiment-specific data formats to analysts. Third, these algorithms are run to reconstruct Monte Carlo simulations to provide the same data formats to analysts. Together these use cases drive the computational processing needs of HL-LHC experiments. While already true during Run 3, the resource needs of these algorithms increases more quickly than other aspects for several reasons. First the event rates, both into the software trigger system and out of it (and into the offline) are expected to grow substantially. More importantly, the event complexity - driven by the number of simultaneous overlapping `pileup' events that occur at each $pp$ interaction will increase by a factor of three to four; the reconstruction computational cost scales approximately quadratically with the pile up. Finally, to achieve the needed science reach for HL-LHC, more advanced -- and at the same time more complex -- detectors are required. All three effects drive up the computational processing resources needed for the planned HL-LHC physics program.

Amongst the online trigger and offline reconstruction algorithms, the most resource intensive ones are the {\em tracking} algorithms: algorithms that together identify where charged particles have traversed the inner part of the detector and their kinematic properties. This is no surprise as the number of electronic channels to be digitized from our detectors for each bunch crossing is dominated by those of the inner tracking detectors. Despite being the focus of continuous research during Run 1 and Run 2 of LHC, the compute needs of traditional approaches continue to increase rapidly with increased event complexity and require significant innovation.  This re-engineering -- in some areas, a complete rethinking of the approach -- is central to ensuring the HL-LHC physics program and a critical factor affecting the scale of processing power needed both the online and offline environments.

\subsubsection*{Specific Challenges and Opportunities}

For pattern recognition algorithms, the period between now and the start of HL-LHC is a time for research consolidation and algorithm integration into the experiment's trigger and event reconstruction pipelines. Work involves making algorithms more robust and more easily maintained as well as the process of algorithm validation.

{\bf Incorporating abstraction libraries} Code written for the HL-LHC will need to last a decade or more -- three or more generations of a typical supercomputing environment.  It becomes costly to write against low-level vendor libraries; the more flexible our algorithms can become, the more sustainable they will be as hardware (particularly, accelerators) radically changes. To function efficiently throughout the HL-LHC program we need to identify the best approaches for the largest cross-section of algorithms. We plan to adopt the strategies decided on by the field, for example the results from HEP-CCE, in cases where a consensus has been reached. One possible development path forward is the use of abstraction libraries, such as Alpaka, Kokkos, etc, that can effectively compile kernels for the set of hardware architectures that they support. 

{\bf Algorithm interoperability} While developments of new tracking algorithms typically happen outside of the context of a full tracking pipeline, their ability to interoperate with this full pipeline is critical. One example is data structures that are well suited for data exchange between algorithms. Another is having the flexibility to minimize data movement across a heterogeneous infrastructure.  Interoperability aspects are critical as algorithms mature and are adopted by experiments into their full tracking pipeline.

{\bf Algorithm design and optimization} Algorithms must be designed and evolved to exploit parallelism, have optimized data structures, and use compute and memory efficiently. Tracking algorithms must expose as many parallelization opportunities as possible to the underlying environment. Exposing parallelism is mandatory to make use of accelerators but are also important for CPU-based approaches where instruction-level parallelism is required to use all the available silicon effectively. Given stagnant improvements in memory latencies and bandwidth compared to the computational power, the use of optimized data structures greatly affects algorithmic performance. This includes both data exchange between tracking algorithms and from tracking algorithms to the subsequent steps in the workload. Considerations here include efficiency for data access in highly parallel environments, optimized CPU-to-GPU data movement, and to ensure vectorized calculations can be performed.
As algorithm developments mature and turn to integration and optimization, they should undergo a tuning process to reduce compute and memory needs. This is typically best done when data exchange and algorithm interoperability aspects can be properly included in the analysis.

{\bf Machine Learning approaches to tracking} ML has the promise of delivering novel, innovative algorithms. However, a key milestone is understanding how these algorithms can provide a better or more sustainable approach to improved speed or physics performance for particular aspects of the tracking pipeline. Graph Neural Networks (GNNs) are the current best candidate approach, however the broader ML research community should be followed to identify new opportunities now that our community has a better understanding of the strengths and weaknesses of ML approaches applied to tracking problems.

{\bf Integration and validation} Integration of a new algorithm or approach into an experiment is extraordinarily challenging; impressive performance improvements count for nothing if the whole dataset must be reprocessed due to a bug in the code.  Trigger algorithms are even more conservative: an improperly rejected event is lost forever! The acceptance and adoption of low-level algorithms can be a long and time-consuming process.
Experiments have therefore set up a detailed validation process to evaluate and eventually adopt new algorithms. Given their complexity, these processes require close and intensive collaboration with developers realize the benefits and in this regard institutes like IRIS-HEP with its significant resources, long time scale and broad vision can play a unique role.

\subsubsection*{Current Approaches and Development Roadmap}

The primary focus for the next phase of the Institute will be maturing research directions and begin consolidation of the work for each experiment's tracking program for HL-LHC. Multiple algorithmic approaches will most likely make up the tracking for each experiment, experiments may take a different approach online than offline, and experiments may take different choices from each other. These are all natural choices, as tracking is a multifaceted problem (seed finding, track finding, track fitting) and has optimization points that vary depending on the experimental apparatus design and computing technical design (e.g., fast and approximate online vs best-possible precision offline). \\

\noindent {\bf ACTS development and integration}

\noindent {\it Funding Scenarios:} Low (reduced scope), Medium, High

\noindent {\it Description:} A Common Tracking Software (ACTS) project is an international, open-source project developing an experiment-independent set of track reconstruction tools. The main philosophy is to provide high-level track reconstruction modules that can be used for any tracking detector. The description of the tracking detector’s geometry is optimized for efficient navigation and quick extrapolation of tracks. Converters for several common geometry 
description languages exist. Having a highly performant, yet largely customizable implementation of track reconstruction algorithms was a primary objective for the design of this toolset. Additionally, the applicability to real-life HEP experiments plays major role in the development process. Apart from algorithmic code, this project also provides an event data model for the description of track parameters and measurements. Key features of the project include: tracking geometry description which can be constructed from standard community geometry libraries a simple and efficient event data model, performant and highly flexible algorithms for track propagation and fitting, basic seed finding algorithms.

\noindent {\it Current and Potential Future Activities:}
\begin{itemize}
\item {\em Full track reconstruction chain}: Within IRIS-HEP, the primary focus has been on the development of core track reconstruction algorithms within ACTS and its application to the ATLAS experiment. ACTS is foreseen to be the replacement for the current ATLAS tracking algorithms for the HL-LHC because it will provide speed advantages by exploiting modern C++ and built-in multithreading. At present, a number of algorithms have been developed and the integration of ACTS into Athena (the ATLAS physics framework) is ongoing. The major milestone will be to run a full track reconstruction chain on the ATLAS ITk geometry within athena. Once this has been achieved a detailed assessment of the physics and computational performance can be made and the algorithms can be tuned.

\item \textit{Machine Learning and hardware acceleration}: ACTS has R\&D lines which are pursued in parallel to the main project. 
The traccc project is developing a demonstrator of an end-to-end track reconstruction algorithm that runs on GPUs. Algorithms, such as the seed finding, which have already been ported to GPU, have demonstrated promising speed improvements, however the full end-to-end track reconstruction chain is needed to prove if GPUs are an appropriate choice for HL-LHC.

\item \textit{Tracking in the event filter trigger}: ATLAS is currently exploring different options for tracking at the event filter level. One possible option would be using ACTS as part of a fully software-based event filter trigger. IRIS-HEP has already dedicated a small amount of person-power for exploratory R\&D and this is an exciting avenue for the next phase. Both the CPU and GPU versions of the code will be explored and the performance compared between the two. The adoption of ACTS for the trigger would decreases long-term maintenance costs for the ATLAS experiment as a single code base would be used for online and offline track reconstruction.
\end{itemize}

\noindent {\bf mkFit development and integration}

\noindent {\it Funding Scenarios:} Low (reduced scope), Medium, High

\noindent {\it Description:} MkFit is a re-engineering of traditional Kalman Filter tracking to leverage the capabilities of modern CPU architectures. Most of the research to date has been on the most time-consuming tracking components, trajectory finding. MkFit has advanced sufficiently to become the baseline tracking for CMS in Run 3 and is used to find more than 90\% of CMS tracks. For these tracks, the trajectory finding is now significantly faster than the track fitting algorithms, which illustrates potential benefits gained by re-engineering conventional algorithms to match modern compute architectures.

\noindent {\it Current and Potential Future Activities:}
\begin{itemize}
\item {\em Kalman parameter fit}: Evaluating methods to apply the mkFit approach to the final Kalman parameter fit. This includes evaluation in the context of CMS data structures appropriate for vectorization-aware CPU algorithms or accelerators.
\item {\em Consolidating and achieving algorithm speed up}: The mkFit software was deployed in production as part of CMS's Run 3 software; we expect to  apply lessons learned from initial 2022 deployment and apply improvements for HL-LHC geometry.
\item {\em Improving interoperability within CMS tracking chain}: While mkFit was developed as a standalone project, it is ultimately meant to operate efficiently within the CMSSW framework. For example, in the next phase we will design efficient mechanisms to take track seeds from modernized pixel tracking (the ``Patatrack" algorithm~\cite{Bocci:2020pmi} as input. 
\end{itemize}

\noindent {\bf Line Segment Tracking development and integration}

\noindent {\it Funding Scenarios:} Low (reduced scope), Medium, High

\noindent {\it Description:} Line Segment tracking (LST) is designed to take advantage of the CMS-specific tracker detector layout which includes doublet layers. This geometry allows track segments to be initially found within each set of doublet layers, a process which is highly parallelizable. These doublets can then be linked together to create track seeds and eventually full track candidates. Initial implementations of this algorithm are inherently parallel, implemented already on NVIDIA's CUDA GPU programming language, and should be easily portable to other accelerator platforms.

\noindent {\it Current and Potential Future Activities:}
\begin{itemize}
\item {\em Mature algorithm}:  The LST approach shows potential to meet both speed and correctness requirements necessary for CMS in the HL-LHC era.  However, an open research question is precisely how much speedup remains achievable; more work is needed to increase the ``realism" of the implementation (removing simplifications and scaffolding that were used to first demonstrate the original concept) and achieve the desired impact.
\item {\em Adopt hardware abstraction layer}: The current R\&D version of the LST algorithms were written using CUDA; it cannot be run on non-NVIDIA resources such as those available on DOE's (AMD-based) Frontier, currently the fastest supercomputer in the world.  A hardware abstraction library will be required for an algorithm to survive through the HL-LHC era.
\item {\em Increasing parallelism}: The current LST algorithms are memory-hungry compared to what's available on modern GPUs (and what might be affordable during the HL-LHC era).  The amount of work assigned per GPU device - and hence the total parallelism and speedup of the algorithm - is limited by the available memory, leaving GPU cores idle.  An initial goal of the next phase is to reduce the memory footprint to fully leverage the available silicon.
\end{itemize}

\noindent {\bf Machine Learning approaches to tracking}

\noindent {\it Funding Scenarios:} High

\noindent {\it Description:} Unlike traditional tracking approaches, a GNN pipeline scales linearly, not quadratically, with data density -- making it an interesting option for the HL-LHC environment. Various research efforts have established pipelines for researching GNN architectures and training approaches applied to different tracking algorithms. While these have not caught up to the technical (speed) or physics (accuracy) performance relative to the highly-tuned existing algorithms, they are rapidly advancing and potential remains to help close the raw resource gap for HL-LHC. 

\noindent {\it Current and Potential Future Activities:}
\begin{itemize}
\item \textit{More Compact GNNs}: Evaluating novel GNN techniques and have focused partially on the inner detector (pixel-based) tracking. Small and thus quick to evaluate GNN architectures are promising but need further research to reach adequate physics performance. One important use case is the online system, where computational speed is more important than precision. 
\end{itemize}


\noindent {\bf Defining tracking pipelines for ATLAS and CMS}

\noindent {\it Funding Scenarios:} Medium, High

\noindent {\it Description:} Individual algorithms are insufficient; they must be combined into full tracking pipelines. The performance challenges for this process are in the data exchange at algorithm boundaries as well as balancing overall physics performance with computational costs. 

\noindent {\it Current and Potential Future Activities:}
\begin{itemize}
\item Experiments are in the process of defining tracking pipelines over the next five or so years. For example, a likely vision for CMS tracking in HL-LHC is one that begins with pixel seeds created by Patatrack, then unpacking / clustering of the outer tracker information and matching outer tracker hits using LST. The combination of these two could be done on an accelerator which avoids the costly moving data between GPU memory and main memory. After these steps, a more traditional Kalman filter approach via mkFit would then be used to complete the track finding and do the final track fitting. This setup would be appropriate for both online and offline environments and they could meet the differing goals of the two applications through different configurations. Piecing together such a pipeline requires active involvement of algorithm developers to properly configure and optimize it.
\end{itemize}


\subsubsection*{Impact and Success Criteria}

Investment into the research of techniques for pattern recognition algorithms -- specifically charged-particle tracking algorithms -- will enable a tracking approach that meets the goals of the HL-LHC scientific program; perform within the resource budget used for Run 3 tracking; and be flexible enough to exploit evolving computing hardware through the HL-LHC era.  Primary success metrics are related to achieving superior {\em technical performance} and delivering superior {\em science performance} relative to state-of-the-art algorithms and other R\&D approaches:
\begin{itemize}
\item {\bf Performance}: Both online and offline, the critical metric is event throughput per unit (compute) cost. Given current performance estimates, a factor of two reduction in the full tracking pipeline would translate into a 25\% reduction in compute for CMS at the HL-LHC.
Large factors have already been demonstrated in individual tracking algorithms through ongoing research. Now developments must realize these gains in practice. In the next phase of IRIS-HEP, our research aims to {\bf achieve factors of two to three in speedup of critical tracking algorithms} in both the offline and trigger environments.

\item {\bf Correctness}: Physics performance -- computing the ``correct" result for the underlying inputs -- is critical. The HL-LHC program sets a minimum threshold of algorithm's physics performance to have the desired scientific reach. This is particularly critical as tracking underlies many other reconstruction components, meaning that deficiencies will propagate throughout the physics data formats for analysis. Given the excellent physics performance of current algorithms, the primary goal for this metric is to maintain current levels while focusing on improving potential gaps. Gaps foreseen include tracking efficiency in particularly dense environments (eg, in high-$p_t$ jets), for highly displaced tracks (eg, from long-lived particles), and to further reduce the level of fake tracks (tracks ``found" by the algorithm that were not part of the actual physics).  
\end{itemize}

{\bf Tracking code written but not adopted by the experiment has minimal impact}. Unlike the analysis environment, where a variety of approaches are taken by different groups, each HL-LHC experiment can only afford to run one set of the most resource-intensive tracking algorithms online and offline. The next five years will include a long process of evaluation and decision making to define the basis for HL-LHC tracking within each experiment. This process must be tracked in order to facilitate adoption of promising research directions and to cut short those that will not be adopted; {\bf the structure and long-term view of an institute is critical for the desired impact}.

Research in algorithms for event reconstruction and trigger applications build upon numerous previous and still on-going NSF investments and serve as an excellent training ground for early-career HEP researchers. For example, three postdocs in this area who were associated to IRIS-HEP have since taken faculty positions: Philip Chang is now faculty at the University of Florida (worked on LST tracking at UCSD); Xiaocong Ai is a professor at Zhengzhou University (worked on ACTS at UC, Berkeley) and Dylan Rankin is a professor at Pennsylvania University (worked on hadron calorimetry reconstruction algorithms on accelerators at MIT). \\

\noindent \textbf{HL-LHC Computing Gap Impact}:
Research in Reconstruction and Trigger Algorithms impacts gaps {\it G1 (raw resource usage)} and {\it G4 (Sustainability)}. 
As a major consumer of resources - estimated to be a majority consumer in the case of CMS - investment in this area has the largest impact in closing the resource gap.
To be sustainable, implementations must be both flexible and adaptable across hardware platforms and generations. High-energy physics depends on being able to make effective use of both specialized high-performance (online) and commodity (offline) hardware as well as high-performance computing. Today this means having codes engineered well enough to be adapted to new hardware architectures either through an interoperability layer or by other means. As the LHC community has no control over the hardware market, and only indirect control over the hardware architectures at the distributed resources, tracking algorithms must design in mechanisms to support hardware architectures most prevalent during the HL-LHC era. HEP detector designs will continue to evolve beyond the initial HL-LHC phase. In the energy-frontier, there is R\&D for both future colliders and so-called phase-3 LHC upgrades; the software should be sufficiently adaptable  to be used for these or other future experiments. While achieving the ultimate performance from tracking algorithms in a specific detector geometry typically requires detailed tuning, algorithms which seek applicability beyond a single experiments must demonstrate how easily they can be used by another and whether there is any trade off in performance. \\

\noindent \textbf{Success Criteria - Milestones \& Deliverables}

\noindent {\bf D2.1} ATLAS demonstrators  (2024,Q2). The ATLAS experiment is currently defining  a list of demonstrator projects. Selected demonstrators need to be released by Q2 2024 after which time they will be evaluated by the experiment for physics performance, technical performance and overall readiness. \\

\noindent {\bf D2.2} CDR/TDR documents of experiments (2024). Both experiments have a timeline to prepare documents describing their software and computing model for HL-LHC and to submit documents to support the LHCC review of the WLCG.  Performance and readiness of potential tracking algorithms  is an important component of these documents. \\

\noindent {\bf D2.3} Downselect and validation by experiments (2026 or later). Transitions from research to evaluation and validation by experiments must be complete for final adoption by experiments. Timelines will be established closer to the start of HL-LHC operations. \\

\noindent \textbf{Success Criteria - Metrics:}

\noindent M2.1. Number of deployed tracking algorithm components that meet or exceed the performance requirements of the HL-LHC science program. \\

\noindent M2.2. Number of deployed tracking algorithm components that achieve factors of  two or more reduction in resource needs for HL-LHC in either the offline and trigger environments.\\

\noindent M2.3. Number of deployed tracking algorithm components that are able to run efficiency with abstraction libraries as adopted by experimental frameworks.

\tempnewpage

%% file: 120-strategic-areas-machine-learning.tex
\subsection{Translational AI}
\label{sec:ai}

Cutting-edge research in artificial intelligence and machine learning -- ``AI'' for short -- directly benefits the physics program of
the HL-LHC. In many ways, AI is transforming
the way field does physics. A dedicated effort in Translational AI will increase adoption of promising new techniques and reduce the time and effort needed to deploy these solutions in the experiments. This would significantly impact all aspects of HL-LHC physics, including trigger,
reconstruction, simulation, and analysis.

HEP has a long history of using AI and has been very active in
embracing and contributing to modern AI research. The field has developed experts in AI
and is beginning to see wider adoption of the techniques by
non-experts. 

New AI techniques are able to work with the complex, low-level data, which generally enhances performance or sensitivity compared to traditional approaches. Similarly, new AI techniques targeting simulation, multivaraite unfolding, and anomaly detection are providing qualitatively new capabilities that can change the HL-LHC physics program in more fundamental ways.  

Exploratory AI R\&D is often done outside of experiments. This work
occurs in the context of other NSF projects like IAIFI~\cite{IAIFI}
and A3D3~\cite{A3D3}, or in small research teams, or is explicitly
funded as part of experimentalist's base physics research grants.
There are a number of barriers to the adoption of this research in
the experiments. For example, the papers written by AI researchers
are often not easily translated into the HEP context without the
aid of expertise in both fields - making them accessible only to
the ML experts within the field. Even if the papers are accessible,
the implementation of the new techniques require further development
to be adapted to the experiment's context. Finally, the integration
of ML techniques into an experiment's workflow is often accomplished
with bespoke solutions, and often different groups within one
experiment will approach this differently.

\begin{figure}[h]
\begin{tcolorbox}
The faster the community can absorb new AI research and translate it
into use by experiments, the more impact it will have on the physics program. 
\end{tcolorbox}
\end{figure}

Ad-hoc collaborations and communication between AI research groups and experiments have
formed around collaborations between AI and particle physics
researchers. These links have grown a small, but active, group of
AI experts in particle physics. The communication does run in both
directions: as particle physics provides opportunities for use-inspired
research in AI, these collaborations do influence AI research
resulting in work that is well-matched to the field's problems.
This healthy communication must be maintained and strengthened as the
field must find ways to bring advanced techniques to a broader
audience. Graph
Neural Networks (GNNs) are an illustrative example. This type of AI model is
particularly well suited to our problem space. 
A multidisciplinary team including physicists and AI researchers at Deep Mind have prepared a review article and position piece outlining this development~\cite{Shlomi:2020gdn}. 
GNNs are now used
for bottom-quark identification~\cite{atlasftagGNN2022}, replacing
an RNN-based solution, and are being explored for tracking, jet
reconstruction, and tagging.  However, there are many advances
remaining before GNNs could be a solution for tracking (see Section
\ref{sec:recotrigger}).

Coordination is needed to guide the evolution from this period of
rapid R\&D, prototyping, and bespoke solutions for deployment to a
more mature and established set of practices for ML in various
contexts (e.g. trigger, reconstruction, simulation, analysis). US
funding agencies have invested resources in making foundational
progress in AI. This new strategic area will improve the research
deployment pipeline, help to further normalize the use of ML, and
provide a set of best practices that can be used by all researchers,
not just experts.

\subsubsection*{Specific Challenges and Opportunities}

The challenges and opportunities that will help transform how ML
is used at the HL-LHC can be split into three broad categories.

\paragraph{Bringing research from the AI Community to the HL-LHC:}
Most AI research happens outside the experiments. For example, a
team of Computer Science or Statistics researchers will develop a
new technique. Another common pattern is that of a small team,
including some people from the HL-LHC community and some people
from CS, will use greatly simplified detector data to publish a
paper directly addressing problems in the field. The new techniques
are not easily incorporated into the experiment. Currently, bespoke
solutions are employed for development, training, and deployment.
Worse, the target environments - trigger, reconstruction, and
analysis - all require different solutions for inference. There is
an opportunity to work with the AI research community and the
particle physics community to build tools that span the domains and
make the inclusion of new AI techniques more straightforward.

The NSF and DOE fund a large amount of foundational AI research.
Two NSF institutes, IAIFI and A3D3, are good examples of this work.
IRIS-HEP's broad reach via its intellectual hub role and work on
Analysis Facilities and training infrastructure provide opportunities
for bringing  this research into the field.

\paragraph{Improving HL-LHC's interaction with the AI Research
Community:} Our problem space is interesting to AI researchers.
Delivering realistic data to the researchers in a way that a
non-domain expert can understand is not easy, however. This is
further complicated by some experiments not wanting to release their
simulation data openly. Finally, it is desirable to use experiment-agnostic
data when collaborating with AI researchers. A toolchain exists to
generate experiment-agnostic data~\cite{DELPHES2014}, but it does
not fully represent the complexity of an HL-LHC detector, and it
writes data in a field-specific format (ROOT). Improvements to these
tools and data formats will make collaboration simpler. With this
work and a better understanding of the challenge and benchmarking
tool set, the community can participate in public challenges more
effectively and collaborate more easily with non-HEP AI research
teams.

\paragraph{Sustainable AI at the HL-LHC:} The AI landscape is
evolving rapidly. New techniques replace old ones. People move from
one task to another or leave an experiment entirely. As AI is used
more and more in reconstruction and the trigger, it becomes important
to have a process in place to maintain it. The data streaming from
the detector will evolve over the course of the HL-LHC (expected
to run more than a decade) - due to changes in the position of the
detector components or due to radiation damage, for example. The
retraining of such a mission-critical ML algorithm cannot be dependent
on a (former) student's laptop configuration. The field needs to
develop infrastructure and best practices around the long-term
sustainability of ML algorithms in reconstruction and the trigger.

\subsubsection*{Current Approaches and Development Roadmap}

Translational AI will be a new area for IRIS-HEP, and there is already a tremendous amount of work on AI in the field.
IRIS-HEP has at least two places it can offer a unique contribution that play to its already existing strengths: helping to bridge the foundational AI research community and HL-LHC physicists, and helping to build sustainable AI within the experiments. The work to accomplish this first relies on building consensus in the field and then helping to execute on that consensus. It is a mix of
blueprint meetings and concrete development projects. It would be
a new area for the next phase of IRIS-HEP, having appeared in the
initial \s2i2 strategic plan but being largely de-scoped due to
available resources. Some of the tasks are important for the long-term
sustainability of the experiments and will have to be solved by the
field; IRIS-HEP can make important cross-experiment contributions
and help set the direction for the community. The current work and roadmap is much more exploratory than in the other areas of this strategic plan as this is a new area for IRIS-HEP and a covers a lot of ground the field has not yet given careful thought to. 
\vspace{\baselineskip}

\noindent\textbf{Training and Community}\\

\noindent\textbf{Building Community Consensus} \\
\textit{Funding Scenarios}: Medium \\
\textit{Description}: Holding {\bf Blueprint meetings} (as part of
the Intellectual Hub activity in Section \ref{sec:hub}) will help
build community consensus and guide the exploratory activities in
this area. These meetings will work to improve the understanding
between experiments and groups in the experiments and also explore
best practices from outside the field of particle physics and improve
collaboration with the AI research community. The activities
listed below are all meetings planned to help bootstrap the activities
in this new area.\\
\textit{Current and Potential Future Activities}:
\begin{itemize}
    \item \textit{Streamlining ML Data Delivery}: Different
      cyberinfrastructure is used to process data during prototyping,
      training, and deployment. Frequently, the methods and code to
      extract the input data to build training/prototyping datasets
      are different from the way the same task is performed for a
      deployed solution. A common set of tools to address this will
      reduce deployment time and may allow uniform access to national
      or international resources for complex training. This work may
      also allow caching of training datasets reducing the DOMA burden.
    \item \textit{Deploying Trained Models in different environments}:
      There are experiment hardware and software experts for the
      trigger, reconstruction, and analysis. Each is a radically
      different environment in which we wish to run ML algorithms.
      The way ML is prototyped and deployed in each is quite different,
      resulting in drastically different infrastructure. This meeting
      will understand the different environments and explore how to
      make the training environment as common as possible.
    \item\textit{Using the AI Benchmarking and Challenge infrastructure
      in HEP}: With the overall goal of using more common tools in
      training, deployment, benchmarking, and challenges, this blueprint
      meeting will bring together the various communities to understand
      the current state-of-the-art and how it could evolve to help
      quickly move new ideas and solutions between communities. There
      is a lot of community expertise already existing. Challenges
      and benchmarking efforts are initiated, organized, and reported
      in workshops like ML4Jets, ML4PS, NeurIPS, Hammer \& Nails,
      MODE, Aspen, IPAM, Dagstuhl, and MIAPbP. A recent DOE award,
      ``A Fair Universe: Unbiased Data Benchmark Ecosystem for
      Physics'', is explicitly addressing the benchmark side of this
      story.
    \item\textit{Disseminating Advanced ML Techniques within the
      field}: An advanced library of ML techniques - translations of
      ML papers - is hard to build. For most physicists reading an
      ML paper is difficult due to the different problem domains and
      vocabulary (sometimes even for the same concept!). Nor is it
      possible to have one or two people translate all papers into
      some sort of common example format as the approach won't scale.
      This meeting will pull together people from similar institutes,
      such as the eScience Institute at the University of Washington
      (which actively deals with a similar problem), and experts
      within the field of particle physics to discuss scalable ways
      of pushing advanced AI techniques into the experiments.
    \item\textit{Foundational Models}: Many areas of AI are adopting
      a pattern referred to as \textit{foundational models}, where a
      large, multi-purpose model is used as a common backbone for
      many tasks. While the foundation model may be expensive to train
      over large amounts of data, those costs are amortized over
      relatively inexpensive fine tuning or task-specific modules.
      Thus far, HEP has not adopted this pattern, but it offers
      potential advantages. However, it is also not clear that this
      pattern holds well for HEP use cases. A blueprint to explore
      this pattern would be valuable.
    \item\textit{Prompt-based assistants based on language models}:
      Recently released models such as GitHub Copilot and ChatGPT,
      allow one to quickly translate instructions formulated as plain
      English text prompts into code or other actions. This blueprint
      meeting would gather experts on such models and experts from
      HEP and HEP/ML to develop a program of research that can be
      followed in the next years (in or outside of IRIS-HEP) to use
      the power of these new models.
\end{itemize}

\noindent\textbf{Training and Dissemination} \\
\textit{Funding Scenarios}: High \\
\textit{Description}: Getting advanced AI techniques into the hands of physicists requires a training program addressing multiple levels - especially when this must be scaled out to reach the full field of HL-LHC physicists. Many users of AI at the HL-LHC are still learning the basics - so their needs must be addressed in a way that they can later then use the more advanced techniques. \\
\textit{Current and Potential Future Activities}:
\begin{itemize}
    \item\textit{Basic Machine Learning Skills}: The HEP Software Foundation and IRIS-HEP and many others collaborate on building a library of training resources for basic software skills using the Software Carpentry's model. The library already contains two ML-related trainings - one for basic PyTorch training and one using GPU's to accelerate training; advanced training materials will be added to cover LHC-specific topics instead of generic basic topics.
    \item\textit{Advanced Techniques}: A library of example uses of advanced ML techniques will be built. This library will provide physicists who have absorbed the more basic training the ability to apply the techniques with best practices being integrated from the blueprint meetings.
\end{itemize}

\vspace{\baselineskip}
\noindent \textbf{Making Fundamental AI Research Sustanibly Work for the HL-LHC Experiments}\\
\\
\noindent\textbf{Sustainable AI at the HL-LHC} \\
\textit{Funding Scenarios}: High \\
\textit{Description}: Currently, AI solutions deployed in the experiments are rapidly replaced by new and improved solutions that are developed through a relatively ad-hoc process. As the AI solutions stabilize, systems will be needed to streamline the retraining or fine-tuning of these AI models in response to changing running conditions. This is akin to the routine recalibration and data reprocessing that the experiments manage, but brings in different computing challenges (access to GPUs, multiple passes over batches of training events, etc.) as the algorithmic patterns encountered in training are different than those found with recalibration. This also adds new requirements to the metadata that tracks run conditions and software versioning. There is an opportunity to design systems that can leverage computing resources that are well suited for AI, and reduce inefficiencies in the data transfer and redundancies that will be encountered without a more streamlined solution.  \\
\textit{Current and Potential Future Activities}:
\begin{itemize}
    \item \textit{Retraining Challenge}: Working with reconstruction authors within an experiment to use community tools to automate the retraining of a reconstruction component. Once implemented, track its use and updates required to keep it working as run conditions and the detector evolve.
\end{itemize}

\noindent \textbf{Benchmarks and AI Challenges} \\
\textit{Funding Scenarios}: High \\
\textit{Description}: Coordination is needed to guide the evolution from this period of rapid R\&D, prototyping, and bespoke solutions for deployment to a more mature or established set of practices for ML in various contexts (trigger, reconstruction, simulation, analysis). There is an opportunity to establish the cyberinfrastructure components (both human and technical) that can serve as a bridge between the vibrant AI R\&D activities taking place outside of the context of the individual experiments and the experiment-specific software and computing frameworks.  \\
\textit{Current and Potential Future Activities}:
\begin{itemize}
    \item \textit{Community Challenge and Benchmark Infrastructure Efforts}: Use the roadmap from the blueprint meeting ``Using the AI Benchmarking and Challenge infrastructure in HEP'' to form a program of work to better integrate challenge and benchmark tools in to the HL-LHC's ML infrastructure.
    \item \textit{Using Common Infrastructure in the Experiments}: Close the gap between the backend infrastructure used for benchmarks and challenges and what is used in the experiments. The experiments have multiple environments that would require separate studies (trigger, reconstruction, analysis). Especially important will be supporting prototyping efforts as new techniques are explored inside the experiments.
\end{itemize}

\noindent\textbf{Hardware Accelerators} \\
\textit{Funding Scenarios}: High \\
\textit{Description}: R\&D on AI with accelerators (e.g. FPGAs) and in highly constrained computing environments (e.g. trigger) requires both specialized hardware and specialized expertise. This poses a high barrier to entry and hinders participation from a more distributed and diverse group of teams. There is an opportunity to facilitate research and lower the barrier to participation by investing in a centralized resource with dedicated expertise and hardware to support distributed research on AI in these highly constrained computing environments.\\
\textit{Current and Potential Future Activities}:
\begin{itemize}
    \item\textit{Making ML on FPGA's Accessible}: Work with the A3D3 institute~\cite{A3D3} to best understand how to move their R\&D work into a place where others in the experiment can make use of it. In particular, the software pipeline and testbeds that can be hosted at a community resource (e.g. an Analysis Facility). The goal of this work is to make the barrier of entry to developing ML trigger solutions lower.
\end{itemize}

\vspace{\baselineskip}
\noindent \textbf{Improving AI Research on HEP Data}\\

\noindent\textbf{Expanding connections between AI and HEP Communities} \\
\textit{Funding Scenarios}: High \\
\textit{Description}: Data from HL-LHC detectors is ideal for many types of AI research. The problems that need to be solved (e.g. jet tagging, track reconstruction, signal identification) require learning complex processes, physics concepts, and even geometric relationships. Further, the problems solved can have large impacts on the physics ability of the HL-LHC and resource requirements. The common tools in the field to generate data suitable for ML are more tuned for internal use. \\
\textit{Current and Potential Future Activities}:
\begin{itemize}
    \item \textit{Simulating a Generic HL-LHC Detector}: Tools like DELPHES implement a simplified simulation of an LHC detector. This tool set needs to be updated to reflect some of the complexities of an HL-LHC detector: adding more modern tracking architectures and an active calorimeter much like CMS's next-generation forward calorimeter. This work would also include building a more automated pipeline for producing datasets that could be used in challenges.
    \item\textit{Data Formats For Research}: The data formats used by the HL-LHC community are mostly ROOT based, which is not easy for AI researchers and others outside the field to use. Tools like NumPy and Pandas DataFrames or Arrow Tables are more commonly used. This task will work on building these common formats as first-class citizens.
    \item \textit{AI Challenges}: As the work above comes together this will enable the field to build and run AI challenges. Running a successful challenge is far from simple, however. Working with others within the field that have run challenges, a set of best practices will be built by example.
\end{itemize}

\subsubsection*{Impact and Success Criteria}

\noindent\textbf{Success Criteria – Milestones \& Deliverables}:

The milestones and deliverables outlined below focus on the first three years of an Institute. These high-level items would be expanded and later years filled in as part of the execution of any project.

\begin{enumerate}[label=D4.\arabic*.]
\item Each Blueprint meeting will have one of its products a report. A milestone will be associated with the circulation of the report to the community. The expected dates will be tied to the schedule (about 4 months after the occurrence of the blueprint meeting). {\bf Various}
\item Retraining Challenge completed: a ML algorithm, used in the reconstruction or trigger, has its retraining automated. {\bf December 2025}. 
\item Launching an FPGA testbed (like SSL). {\bf December 2026}.
\item Run a Challenge with entrants from the AI Research Community. {\bf December 2027}
\item Access to a FPGA for ML training is possible on a Analysis Facility. {\bf December 2027}
\item A training is run locally and non-locally from the training and testing infrastructure. Normalizing training means that researchers can take advantage of GPU's for training around the world, or on the infrastructure hosting the training service, and have the same user experience.
\end{enumerate}

\noindent \textbf{Success Criteria – Metrics}:

Metrics are a useful tool to provide management with quantitative insight into progress toward overarching goals.  High-level metrics we expect to be applicable to the Translational AI area are below:

\begin{enumerate}[label=M3.\arabic*.]
\item The number of foundational trainings added to the HSF training library to help physicists incorporate ML into their workflows.
\item The number of advanced ML techniques added to a library as exemplars for the field.
\item Number of  AI / ML models or use cases that are integrated into experimental infrastructure using the processes developed by the institute.
\item Number of times the FGPA service at an Analysis Facilities are being used per quarter.
\end{enumerate}

\tempnewpage

%% file: 125-strategic-areas-data-organization-management-access.tex
\subsection{Data Organization, Management and Access (DOMA)} 
\label{sec:doma}

%
%
%

Given the centrality of ‘data’ to all things the HL-LHC will do, it is not surprising the investment area of data organization, management, and access is a key part of all four HL-LHC Computing Gaps.  The immense volume of data is the easiest to conceptualize – starting with the predicted trigger rate, days of data taking per year, and event size, the HL-LHC experiments can make leading-order estimates of the raw, processed, and simulated data sets necessary for the desired scientific scope.  Feed these numbers into the current resource requirement modeling and already the disk requirements needed outpace the fixed-budget scenario.  The problems compound, however, once one takes into account this data must be moved about the shared national cyberinfrastructure, analyzed at much higher data rates than today at facilities, and done using common infrastructure to reduce sustainability costs.

\begin{wrapfigure}{r}{0.4\textwidth} 
\begin{center}
\includegraphics[width=0.37\textwidth]{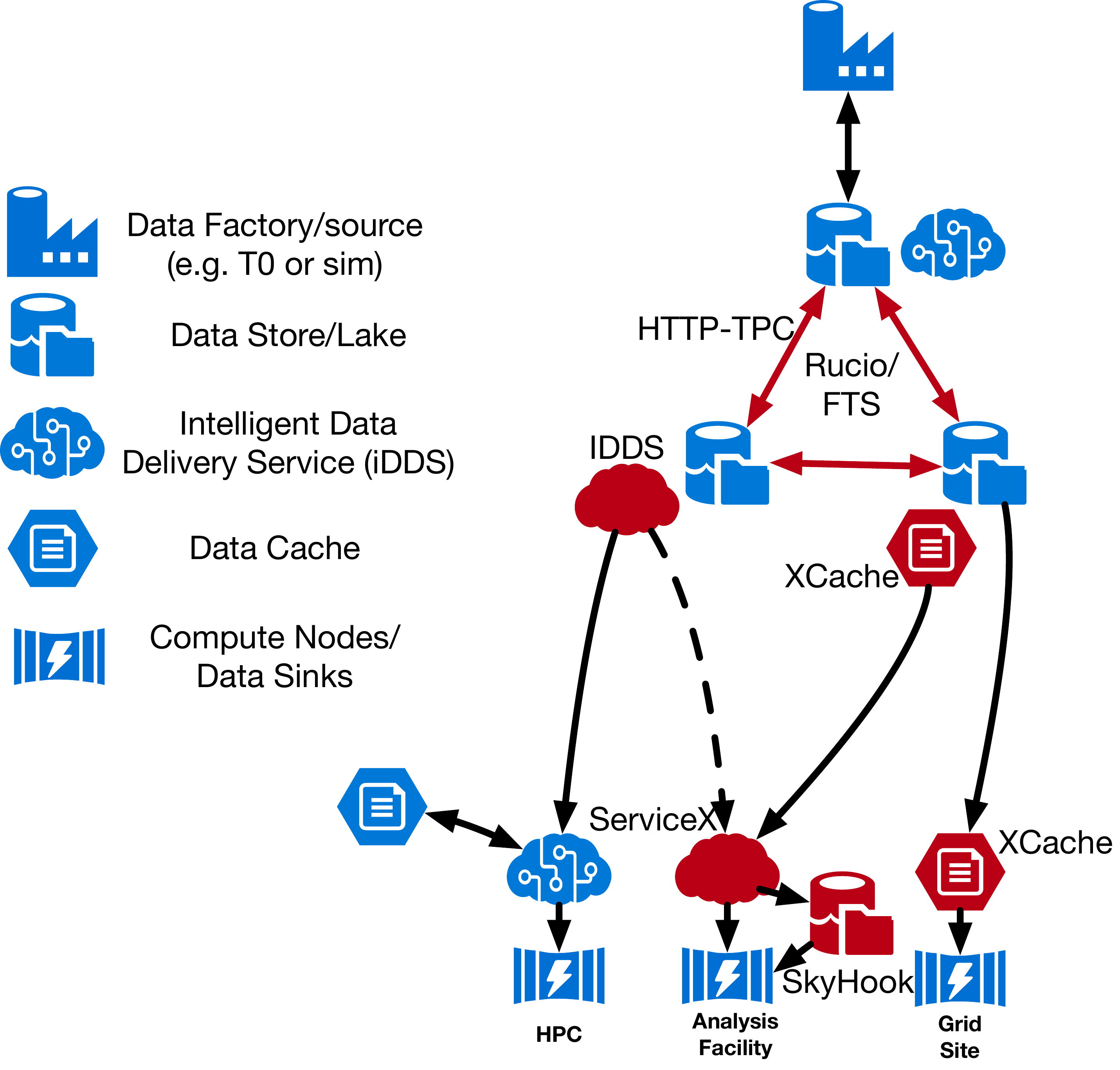}
\caption{Schematic showing the DOMA-related components in the production infrastructure.  The components that DOMA started or contributed to in IRIS-HEP are colored in red.  The dashed arrow from IDDS to ServiceX is one of the planned projects for the next phase.}
\label{fig:doma}
\end{center}
\end{wrapfigure}

Together, these challenges in the DOMA area illustrate the fact that HL-LHC science is not achievable by simply purchasing more servers, disks, and network switches - but otherwise leaving the approaches untouched.  A robust R\&D program is needed across both the production infrastructure and analysis facilities sub-areas of DOMA building upon the successes in the last 5 years.  For bulk data movement, the entire LHC production infrastructure has migrated from the niche GridFTP protocol to the industry-standard HTTP, demonstrating the capability to both deliver from R\&D to production and also execute system-wide changes.  The community has planned and executed the first biennial data challenge, DC21, as part of the Data Grand Challenge (Section \ref{sec:domachallenge}, showing the ability to coordinate on the global scale and provide integrative milestones for technology projects.  Finally, during IRIS-HEP, the Coffea-Casa analysis facility platform was developed; Coffea-Casa provides a “first look” location for new analysis system services to intersect with users.  The areas where the DOMA team has been active during IRIS-HEP are shown in Figure \ref{fig:doma}.

\subsubsection*{Specific Challenges and Opportunities}

In the remaining R\&D period between now and the HL-LHC, we see opportunities in the DOMA area continuing work in preparing the coordinated cyberinfrastructure or HL-LHC data volumes, transitioning to newer technologies used by industry and elsewhere in scientific computing infrastructures, and delivering data to analysis facilities. \\

\textbf{Scaling up data volumes to HL-LHC}: There is a need for innovation, development, and integration in nearly every service in the bulk data transfer infrastructure used by the LHC community to prepare it for the HL-LHC data volumes.  Opportunities include benchmarking the existing software on available R\&D testbeds such as the National Research Platform (NRP) \cite{NRP}, utilizing next-generation networks such as FABRIC \cite{FABRIC}, and leveraging engineered network paths as in the ESNet SENSE project \cite{SENSE}.  The community has the opportunity to utilize these testbeds as a proving ground for techniques between the proposed biennial global data challenges (Section \ref{sec:domachallenge}) and as part of any regional / US-specific exercises. \\

\textbf{Technology transitions for bulk data movement}: Several of the technologies historically used for the bulk data movement stack are at end-of-life, causing difficulty in the sustainability of the infrastructure.  For example, the WLCG-wide tape archival protocol, Storage Resource Management (SRM), is based on a custom transport layer (not Transport Layer Security, TLS, which is ubiquitous across the Internet) and uses a remote procedure call protocol, SOAP, which is long out of favor.  This has created a new opportunity to reconsider parts of the ecosystem - for example, the community has already replaced the use of the niche GridFTP protocol with the industry-standard HTTP - resulting in potential leaps in sustainability through leveraging industry standards.  During the next 5 years, we see the following potential transitions necessary for the community:
\begin{itemize}
\item Moving the authorization scheme from identity-mapping based (mapping a credential to a local identity) to capability based (a credential is tied to a specific action).  Capability-based schemes provide for more fine-grained authorization - only the actions intended for a job must be sent along with it.  This transition requires coordinated changes across the software stack and integration in the ecosystem.
\item Migrating bulk data movement to IPv6.  IPv4 is considered a legacy protocol by the IETF meaning future innovations in the networking layer will be exclusively done for IPv6; the transition to IPv6 for all bulk traffic should be completed.  For example, packet extensions that enable monitoring of network flows on ESNet are only defined for IPv6.
\end{itemize}

\textbf{Delivering data to analysis facilities}:  The hallmark of HL-LHC analysis will be the size of the dataset (event count and volume), the need to deliver initial results at lower latency, and the use of a broader range of “data science” and machine learning tools from outside physics.  The data volume and complex environment is expected to cause users to move from laptop-scale machines to analysis facilities that specialize in high-data-rate I/O, ML-support services (such as inference services), user interfaces (such as pseudo-interactive task-based computing or notebooks), and advanced data management services.  These facilities, either from new approaches as in Section \ref{sec:facilities} or built on top of existing, batch-oriented analysis facilities, will need advanced data delivery and management services.

For example, an important trend in HEP analysis is the movement to Python tools and columnar analysis. In addition to giving analyzers experience with popular data science tooling, columnar analysis has the potential to unlock vectorization and performance benefits by defining operations on columns as opposed to individual events.  To successfully leverage these analysis techniques, services are needed to filter, transform, and deliver columns to workloads running at analysis facilities.

HEP analysis has historically been a strongly filesystem-based activity: a user will take a  dataset – often just a set of files in a directory – and apply simple transforms and filtering, resulting in a reduced set of output files in a new dataset.  Users often make only simple changes, such as adding a few derived values to the event, meaning each of these derived datasets have a high level of overlap.  Further, if a column was missing or incorrect in the original derivation, the user has to repeat the process from scratch.  The system is wasteful of both disk space and – more importantly – physicist time.  There are modern data management systems that provide the ability to do common database techniques such as virtual views and joins without needing to convert the ecosystem to a RDBMS; there is opportunity to both tap into a rich vein of industry investment (helping with the {\em G4 (Sustainability)} gap) in addition to making physicists more productive.  As the management evolves, data delivery will similarly need to change to effectively schedule and pull the data from the wider distributed cyberinfrastructure into the facility and transform the data into the desired format for data management.  This will require both services in the analysis facility and at the network layer (leveraging the NRP and SENSE work) to ensure proper prioritization and effective caching. \\

\textbf{Integration point}: Finally, the institute structure itself provides a unique opportunity.  The DOMA ecosystem for the LHC and HL-LHC is large, complex, and slow-moving, making it difficult for independent smaller-scale research projects (such as those in the NSF CSSI Elements program) to deliver their work to production.  The Institute’s DOMA strategic area would be large enough to serve as an intellectual hub, connecting between innovative ideas coming into the NSF ecosystem, R\&D testbeds at the national scale, and the production cyberinfrastructure for the HL-LHC.

\subsubsection*{Current Approaches and Development Roadmap}

The DOMA area can be partitioned into investments into the production infrastructure, focused primarily on the HL-LHC Computing Gaps for sustainability (G4), scalability (G2), and raw resource requirements (G1), and the analysis systems investments, focused on sustainability (G4) and analysis at scale (G2).  Below we outline existing approaches in each area.\\

\noindent \textbf{Production Infrastructure Projects}\\
\\
\textbf{Scaling the CI to HL-LHC data rates} \\
\textit{Funding Scenarios}: Medium (reduced scope), High \\
\textit{Description}: From the beginning of Run 3 to the first full year of
HL-LHC, the expected data volume moved per year by the LHC
experiments will rise by 20 fold.  To prepare for these data rates,
the community has defined a set of biennial data challenges as
described in Section \ref{sec:domachallenge}.  The data challenges are meant as
capstones \& milestones; the majority of the effort is in the
preparation in the years leading up to the challenge.  This project
would work on the engineering and integration of community-wide
research projects to demonstrate the readiness of the
cyberinfrastructure for the HL-LHC scale. \\
\textit{Current and Potential Future Activities}:
\begin{itemize}
    \item \textit{Demonstrating scaling of reference platforms}: The various software reference platforms (for IRIS-HEP, this is XRootD on top of a POSIX filesystem) used for bulk data transfer must be shown to scale well before trying load tests on production systems.  NSF has funded a number of experimental storage (such as NRP, providing filesystems) and network (such as FABRIC, a 1Tbps platform between San Diego and New York) resources.  By leveraging external R\&D testbeds, the community is able to attempt tests at HL-LHC scales without needing to “buy ahead” at production facilities.
    \item \textit{Leveraging engineered network paths}: By segmenting HL-LHC bulk data traffic onto specially-engineered network paths, the network connectivity providers (for US LHC, this is ESNet) have the ability to isolate user specified data flows, enhancing their accountability and at the same time allowing for prioritization, segmentation, and specialized quality-of-service techniques.  ESNet’s SENSE project \cite{SENSE} provides a mechanism for engineering the paths end-to-end; however, this functionality needs to be integrated into the LHC’s data management software (Rucio) and shown to work across multiple sites.  
    \item \textit{Organizing US regional exercises}: The WLCG-wide data challenges provide a biennial global synchronization point for facilities and technologies.  These require significant effort and coordination making it difficult to incorporate less mature technologies or regional concerns.  We foresee a need to have US-specific regional exercises focused on technologies the US LHC is investing in (such as the Rucio / SENSE integration) and challenges such as integration with US HPC sites.  An institute-scale DOMA strategic area, cross-cutting the HL-LHC experiments, would be uniquely situated to coordinate such exercises.

\end{itemize}

\begin{figure}[h]
\begin{tcolorbox}
Scaling the cyberinfrastructure to close the 20x gap between today's and HL-LHC's expected transfer rates is one of the most high-impact activities of the DOMA team.
\end{tcolorbox}
\end{figure}

\noindent \textbf{Authorization technology overhaul for the distributed infrastructure} \\
\textit{Funding Scenarios}: Medium, High \\
\textit{Description}: Capability-based authorization provides a powerful new paradigm for asserting authorizations on a distributed infrastructure.  Each experiment now has a token issuer service that asserts a specific action the bearer (hence the term “bearer token”) can perform in the experiment’s distributed resources; IRIS-HEP has worked to ensure the storage systems used by the LHC community interpret these assertions in an interoperable manner.

However, significant work remains to ensure all the software used by the community has been adopted to acquire, manage, and utilize these new credentials.  These technologies remain a headline feature planned for the next data challenge to allow sufficient time for a complete transition by the start of the HL-LHC. \\
\textit{Current and Potential Future Activities}:

\begin{itemize}
\item \textit{Coordinating evolution of token usage \& profiles}: While the intellectual approach to capability-based tokens was done in 2017 by the SciTokens project, significant coordination effort was required to expand the idea to cover all LHC use cases and to gain traction in the community.  Common interpretations of authorization schemes is a strict requirement for establishing trust when moving data between sites; the DOMA strategic area has the opportunity to continue working with the WLCG authorization working group as the profiles evolve and are updated with technologies.

Now the foundational work is done on the capability language, the community needs to better define the token acquisition and exchange workflows, for example, designing how they move from data management system (Rucio) to file transfer system (FTS) to storage endpoint.
\item \textit{Engineering a reference implementation}: The IRIS-HEP DOMA area provided a reference implementation for token-based authorization embedded in the XRootD server software.  This is used by several LHC facilities and provides an endpoint for storage and middleware developers to use for testing their own independent implementations.
\item \textit{Provide leadership and support}: New technologies always pose a challenge for system administrators, providing expertise and a direction helps to ease the burden and make changes better welcomed by the community.

\end{itemize}

\noindent \textbf{Organized delivery of data to running production workflows} \\
\textit{Funding Scenarios}: High \\
\textit{Description}: In the first four years of IRIS-HEP, the DOMA area invested in the Intelligent Data Delivery Service (IDDS) which provided the ability to do data-centric workflows within the PanDA workflow management system.  IDDS was used to help implement “disk carousels” (minimizing the use of disk buffers by the ATLAS experiment) and manage large-scale hyperparameter optimization runs; in impacts beyond HEP, IDDS was adopted by LSST to manage large-scale data processing.  Given the significant raw cost of disk storage – by some estimates, HL-LHC dedicated disk storage is more expensive than CPU – tight management of disk buffers has the opportunity to reduce their required size, allowing the community to rely more on less-expensive archival tiers. \\
\textit{Current and Potential Future Activities}:
\begin{itemize}
\item Integration between IDDS and ServiceX: Currently the column delivery service (ServiceX) for analysis facilities is entirely dependent on the input datasets being on disk and being transferred then processed.  By having ServiceX call out to IDDS to do the data processing, there’s the opportunity to use this technology to stream datasets from tape into an analysis facility, extract columnar data needed for an analysis without needing to persist the entire dataset on disk.
\end{itemize}

\noindent \textbf{Analysis Projects}\\
\\
\textbf{Delivering columnar data with ServiceX} \\
\textit{Funding Scenarios}: Medium, High \\
\textit{Description}: Columnar analysis is a relatively new and growing approach to physics analysis. ServiceX is an service designed to run inside analysis facilities to create and cache columns to meet analysts’ data delivery needs.  ServiceX was designed from the beginning to read experiment specific data formats and write selected events and properties to common columnar data formats used in industry (in addition to the HEP-native “ROOT” format). \\
\textit{Current and Potential Future Activities}:
\begin{itemize}
\item Broaden the set of usable data formats: Analysis approaches are ever-evolving as the compute models for HL-LHC are refined.  ServiceX has focused on the compact, analysis-oriented data format; however, new ML techniques often include the use of low-level objects that might be in more raw data (potentially on tape, needing IDDS integration).  By building as broad a set of use cases as possible, ServiceX can help increase the impact of future analysis facilities.
\item Integration with data management tools:  The service can currently either deliver data to a S3-like object store; both provide quite simplistic tools for the management of data.  When combined with industry techniques – for example, delivering to SkyHook (below) – there users will be able to better utilize those tools.
\item Integrate with analysis preservation systems: One reason for the durability of “a dataset as a collection of files” is in its simplicity: the only data service needed to help reproduce a workflow is a filesystem.  This is not necessarily true when a series of complex services is used to in the processing chain, including the case of delivering a column and integrating into into an existing dataset (which potentially implies the delivery service needs to be captured by the analysis preservation).  To help sustain the ability to preserve analysis, commonly-used tools such as RECAST will need to understand/presserve analyses that were run with ServiceX.
\end{itemize}

\noindent
\textbf{Leveraging Industry Data Management for HEP} \\
\textit{Funding Scenarios}: Medium (reduced scope), High \\
\textit{Description}: Analysis datasets will grow significantly \textit{and} users will demand higher event rates to improve the time-to-insight. The Skyhook project is an investment into emerging data processing architectures where additional computational resources and hardware accelerators are available in the networking and storage layer. The service is leveraging commonly used data access libraries to “push down” structured queries into these layers to reduce data movement and to create views of datasets without creating copies. A significant part of the Skyhook project is the management of metadata that enables the establishment of views as well as versioning and branching of views (similar to versioning and branching in git-based source code management). Skyhook-enabled storage systems are a natural place for maintaining this metadata and servicing it to other tools in convenient formats. The project is leveraging data and query specifications of successful open-source projects such as Apache Arrow, Parquet, Substrait, and Apache Iceberg. The Skyhook project was able to upstream a Ceph extension to the Apache Arrow project, allowing Apache Arrow dataset interface queries to be pushed down and distributed into Ceph storage servers storing Parquet files. Other parts of the Skyhook project involve coordinated processing of Apache Arrow streams of particle physics data across multiple Bluefield SmartNICs using Substrait as query specification and processing status and offloading comparisons of genomes to hard drives.\\
\textit{Current and Potential Future Activities}:
\begin{itemize}
\item \textit{Integrate with file-based data management systems}: File-based data management systems like Apache Iceberg provide functionality typically associated with databases without losing the filesystem-based approach that LHC is heavily invested in.  SkyHook is interested in utilizing the using ADL Benchmark library \cite{Proffitt:ADL} as a way to compare Iceberg’s performance with the to results presented by Graur et al. at VLDB’22 \cite{Graur:VLDB} and using py.iceberg as a mechanism to implement versioned views.
\item \textit{Alternate data modeling languages}: With such complex data as in LHC events, modeling and querying is not trivial and maps poorly to common languages like SQL; we aim to evaluate the new query and data modeling languages (such as Malloy) and compare it to results presented in \cite{Graur:VLDB}. 
\item \textit{Substrait integration}: Substrait is quickly becoming the established way to specify query plans independent of the original query language.  It appears to be a promising way to create a composable data management stack that can unify multiple query approaches.  For example, it would be able to help specify views and name cached queries and connect the LHC analysis stack to future data processing, management, and storage systems being developed in industry and wider CS R\&D.
\end{itemize}

\noindent \textbf{Joint Area Projects} \\

\noindent \textbf{Coffea-Casa, an Analysis Facility Platform for the HL-LHC} (Joint with Analysis Systems, Facilities R\&D) \\
\textit{Funding Scenarios}: Medium, High \\
\textit{Description}: One lesson learned from the first four years of IRIS-HEP is the difficulty in keeping multiple projects at different levels of maturity aligned toward a goal.  Coffea-Casa is a prototype analysis facility which aims to be a common integration point for a broad set of technologies and provide them with a first exposure to a user base of analysts.

Coffea-Casa starts with a base of the “Coffea” processing framework for low latency columnar analysis and has a modular approach for adding other services such as ServiceX, SkyHook, or scale-out of tasks into a traditional batch system. The facility provides an interactive experience for physicists that’s closer to working on a laptop as opposed to a traditional batch system-based facility.  The facility adopts an approach that allows transforming existing computing facilities into composable systems using Kubernetes as the enabling technology.  Kubernetes not only enables rapid deployment of services by developers via the DevOps methodology but also serves as a common language for service orchestration (see Section \ref{sec:facilities}), enabling the services to be easily duplicated across multiple Coffea-Casa instances.

Within IRIS-HEP, Coffee-Casa is used to execute the exercises of the IRIS-HEP Analysis Grand Challenge (Section \ref{sec:agc}) and has deployments for the U.S. CMS \& U.S. ATLAS physicist community.\\
\textit{Current and Potential Future Activities}:
\begin{itemize}
\item \textit{Maturing Coffea-Casa to production}: As the analysis facility concepts and services mature, we are working to expand the user base to help gain the experience necessary for at-scale analysis for the HL-LHC.  We will continue to utilize Kubernetes as a technology fabric for additional services and enabling portability.  Coffea-Casa will grow and scale along with the increasingly complex set of Analysis Grand Challenge exercises.  The integrations with real CMS \& ATLAS analyses will help us tune and benchmark for HL-LHC data rates.   Further, as the prototype facilities scales from today's dozens of users to potentially hundreds, we expect challenges requiring improved resource management within these facilities.
\item \textit{Strengthening the integration across DOMA R\&D}: As the integration point for analysis-oriented work in the IRIS-HEP Analysis Systems and DOMA areas, there is a continued need for Coffea-Casa to deploy new versions and data delivery services.  Integrations with existing projects need to be strengthened (such as having ServiceX authentication credentials be auto-generated on login or scaling data caching services) and starting integrations with new technologies such as bearer token support for authorizing global data access.
\item \textit{Develop approach for enabling ML-based analysis}: Machine learning techniques have a long history of use within HEP.  However, they have rarely grown to the size where their needs need explicit planning in the site infrastructure; as ML rapidly grows across science and industry, the interest in the LHC community has grown as well.  We will investigate approaches offering an interactive training environment with GPU access (more difficult than CPU scheduling as there are few GPU devices in the facilities) and pursue opportunities to integrate with national-scale GPU / ML training resources.  Additionally, industry has built several new tools for deployment on Kubernetes – such as MLflow – to manage hyperparameter optimization tracking and advanced ML pipelines.  Finally, multiple analyses have already begun to include ML inference into their workload; these would benefit from facility-local inference services to accelerate throughput.
\item \textit{Analysis preservation}: The addition of new service types (data management, ML inference services, task-based computing) creates new paradigms that need to be reflected in analysis preservation systems.  Preservation strategies need to be devised and integrated with other community projects.
\item \textit{Coordinating the growing network of LHC analysis facilities}: IRIS-HEP is not the sole entity investigating analysis approaches for HL-LHC; there are several similar efforts globally.  There’s a role for a future software institute to serve as an intellectual hub, ensure coordination and knowledge sharing between international projects, NSF-funded efforts, and the U.S. LHC facilities.
\end{itemize}

\begin{figure}[h]
\begin{tcolorbox}
The prototype Coffea-Casa facility has proven to be a valuable ``meeting point" for IRIS-HEP's analysis R\&D vision and the team serves as an intellectual hub inside and outside the institute.
\end{tcolorbox}
\end{figure}

\noindent \textbf{Core data streaming with XRootD and XCache} (Joint with OSG-LHC) \\
\textit{Funding Scenarios}: Medium, High \\
\textit{Description}: The XRootD software framework is foundational for efforts across the LHC.  It is used as a reference platform in IRIS-HEP for data streaming and bulk data transfer, as the basis for data federations in CMS, and, in its ``XCache'' configuration, as a data caching service for both production and analysis.  Sustainability of this platform is essential. \\
\textit{Current and Potential Future Activities}:
\begin{itemize}
\item \textit{Evolving bulk data transfer}: As a multi-protocol server, XRootD implements both the proprietary “xrootd” protocol and the standard HTTPS.  It is used by over half the U.S. LHC facilities for HTTP-based data transfer between sites.  As the technology stack for bulk data transfer evolves – refining HTTP-based data movement, scaling the data rates, and implementing technology – this platform will continue to lead the community.
\item \textit{Core development, integration, and delivery}: Beyond bulk data transfer, XRootD is critical software for the LHC community.  It is a collaboration with contributions from DOE labs, universities, and CERN; there is a role for the DOMA and OSG-LHC strategic areas to make contributions (core development, testing, integration, software delivery) to ensure the investments in this software are sustained.  Having an institute-class investment from NSF like IRIS-HEP participate in these activities would help the collaboration provide a broader stakeholder setup for the NSF cyberinfrastructure community.
\item \textit{Exploring event delivery}: XRootD’s multi-protocol framework includes an RPC and streaming layer which provides for a way to scalable, load-balanced data delivery.  The SkyHook project has shown the value of Apache Arrow as a high-performance, interoperable data format; however, it is currently limited to being used as part of Ceph (reasonably common, but nowhere near universal at LHC facilities).  There is potential in using XRootD as a way to deliver high-level, structured data in Apache Arrow format while leveraging the existing authentication and streaming features of the software.
\end{itemize}

\subsubsection*{Impact and Success Criteria}

DOMA has constituted a historical strength of the US university community.  Particularly, in the last 5 years, the existing IRIS-HEP team has held leadership positions within the WLCG in this area and has designed and executed the transition from GridFTP to HTTP-TPC and began the transition to tokens.  Investing in DOMA as part of an NSF-funded software institute would keep the HL-LHC community well-aligned with the activities and priorities of the broader NSF Office of Advanced Cyberinfrastructure (OAC); DOMA is uniquely situated to leverage existing distributed testbed resources in its program of work.  Existing teams are in place for services like ServiceX and Skyhook and able to build on the success of the IRIS-HEP project.  Having DOMA as a strategic area balances continuity -- completing projects that will be in progress at the end of IRIS-HEP -- with leveraging an existing productive team to start new projects.  Further, the team will be able to help sustain critical projects such as XRootD widely used across the community.

The high-impact outcomes for the production activities in DOMA will revolve around the successful participation in, and execution of, the biennial data challenges coordinated in part with WLCG.  The capstone of this effort will be the 2027 data challenge (approximately Year 4 of the next phase of IRIS-HEP) where we will aim to demonstrate bulk data transfers at 100\% of the expected HL-LHC data rates.  Another significant milestone will occur for DC23, when the majority of transfers will be run using token-based authorization.  For DC25, we expect the remainder of the infrastructure will be token-based and the community will be leveraging from network services to improve accountability, monitoring, and management of data transfers.

For analysis-related activities, the expected outcomes of the DOMA strategic area will be around the usage of new paradigms of analysis facilities and the associated data delivery and management services.  We expect these to be the most common analysis environment at the start of HL-LHC and for the Institute to demonstrate data rates to analyses at HL-LHC scale, as defined by the Analysis Grand challenge, by 2026.  ServiceX is expected to be the primary way for users to ingest official experiment datasets into their analysis environments, Skyhook or similar services will provide a way to manage data (virtual views, joins of columns) from within the environment, and for analyses to have access to production-quality, user-facing services for ML inference. \\

\noindent \textbf{HL-LHC Computing Gap Impact}:  The DOMA area is expected to make an impact on all four defined HL-LHC computing gaps.  Most of the resources will be dedicated toward {\em G2 (Scalability)} as part of the benchmarking of different software components and the participation in regional data exercises and global data challenges.  This work will provide a better understanding of our network flows and help increase the use of network services to increases the manageability of our data transfers translates on a more efficient use of our resources.  The {\em G1 (Raw resource usage)} gap will be tackled through the investment in data delivery via IDDS which helps manage the disk buffers used for annual data (re-)processing and reduce the total amount of online disk needed for HL-LHC.  For {\em  G3 (Analysis at scale)}, DOMA will put new services and techniques in production a interactive facilities, combining “local laptop-like” responsiveness with datasets at the HL-LHC scale.

Finally, for {\em G4 (Sustainability)}, the DOMA area leverages industry protocols (HTTP, JWT-formatted bearer tokens), methodologies (Kubernetes-based service orchestration), and technologies (Apache Arrow) wherever possible.  By helping the LHC community use broader ecosystems instead of developing technologies in-house, we minimize the LHC-specific pieces to those areas where the community is truly unique. \\

\noindent \textbf{Success Criteria – Milestones \& Deliverables}:

The milestones and deliverables outlined below focus on the first three years of an Institute.  These high-level items would be expanded and later years filled in as part of the execution of any project.

\begin{enumerate}[label=D4.\arabic*.]
\item Coffea-Casa used as part of one production analysis facilities in both ATLAS and CMS.  {\bf December 2023.}
\item All the U.S. LHC T2s in the US support bearer tokens on their Storage Elements for Third Party Copy transfers. {\bf December 2023}
\item Rucio/SENSE integration is included as part of the DC23. {\bf March 2024}
\item Successful execution of DC23, meeting its data transfer and technology goals.  {\bf March 2024}
\item ServiceX used for physics analyses as part of Coffea-Casa.  {\bf March 2024.}
\item Demonstrate analyses running at 200Gbps as part of the Analysis Grand Challenge. {\bf December 2024}
\item Demonstrate the capability of XRootD to scale beyond 400Gbps. {\bf June 2025}
\end{enumerate}

\noindent \textbf{Success Criteria – Metrics}:

Metrics are a useful tool to provide management with quantitative insight about progress toward overarching goals.  High-level metrics we expect to be applicable for the DOMA area are below:

\begin{enumerate}[label=M4.\arabic*.]
\item Demonstrate data rates as a percentage of the expected HL-LHC rates.  Goals for this metric are defined by the DGC in the timelines set out in Section \ref{sec:domachallenge}.
\item Number of sites participating in the Rucio/SENSE testbed.  Goal is 4 by DC23 and 8 by March 2025.
\item Percentage of U.S. LHC facilities supporting bearer tokens for data transfer.  Goal is $>50\%$  as part of DC23 and for all sites by DC25.
\item Maximum transfer rate demonstrated by the XRootD reference platform as a percentage of the desired HL-LHC rate for Tier-2s (400Gbps sustained).  Goal is 200Gbps as part of DC23 and 400Gbps by DC25.
\end{enumerate}


%% file: 130-strategic-areas-facilities-rd.tex
\subsection{Facilities R\&D and Integration}
\label{sec:facilities}
%
%
%

{\emph{Facilities R\&D}} broadly refers to activities related to the exploration and innovation of systems, services and physical infrastructure that provide platforms suitable for HL-LHC service environments and runtime ecosystems. These can be purely local facilities (platforms deployed within a local area network) or distributed, in the sense of interoperating services over wide area networks (the ``Grid" or nowadays, distributed platforms).  Two IRIS-HEP research areas (Analysis Systems and DOMA) have opted for software containers and cloud-native application management methods (Helm charts and GitOps) to standardize deployments. Driven primarily by DOMA software development and Analysis Grand Challenge scalability requirements, flexible strategies employing  cloud-native technologies were tried out in the Scalable Systems Laboratory (SSL). 

\begin{figure}[htbp]
\begin{center}
\includegraphics[width=0.99\textwidth]{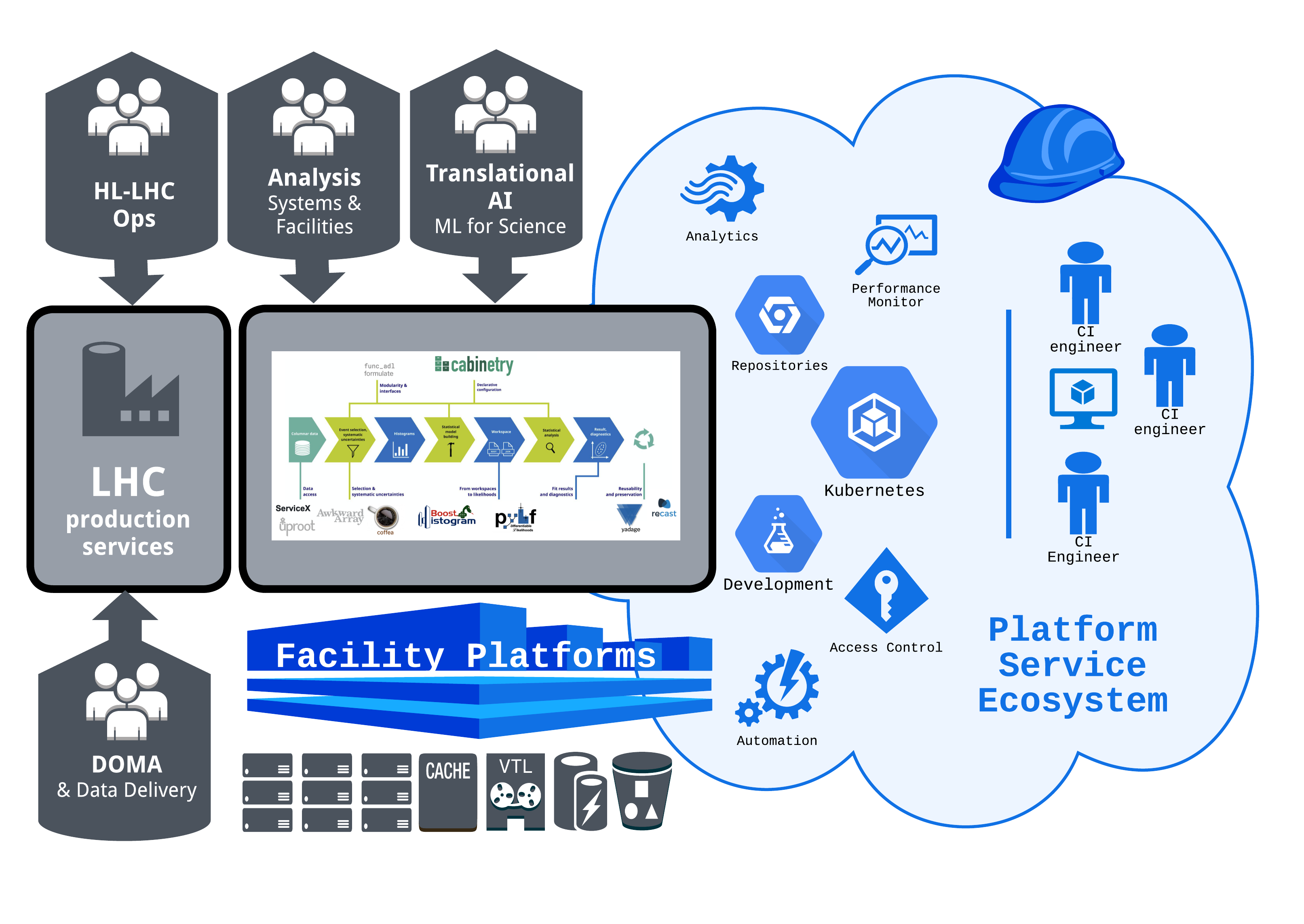}
\caption{Schematic of illustrating role of Facilities R\&D in the production and analysis ecosystem. CI engineers prepare and operate infrastructure and services in support of advanced data delivery, analysis, and other services.}
\label{fig:facilityRD}
\end{center}
\end{figure}

Indeed the concept of facility ``substrates" using Kubernetes, a strategy pioneered in IRIS-HEP blueprint meetings, ``k8s-hep meetups", and in WLCG Kubernetes workshops, are gaining popularity in production LHC analysis facilities as they offer reproducible application deployment and improved reliability of operation. The scope of facility R\&D needed for the capabilities and scale of the HL-LHC, however, extends well beyond the substrate and includes several technological areas involving a diversity of data storage systems, innovative networks, and services for continuous integration and operation.  Figure~\ref{fig:facilityRD} illustrates the relationships between infrastructure, platforms and higher level services supporting production workflows.

{\emph{Facilities Integration}} refers to activities relating to the installation, configuration, and operation of {\bf  development and pre-production} IRIS-HEP software components. Within IRIS-HEP these include  Coffea-Casa, ServiceX, Skyhook, and other services.  Other research cyberinfrastructure services finding application in the ecosystem (such as declarative processing frameworks, caching services, container image registries) likewise need to be integrated into production systems. 
Where possible development services are deployed in realistic contexts, alongside the experiment's existing systems and production infrastructure. 
This is the most immediate path to understanding which components in the overall cyberinfrastructure present the most significant scalability  challenges. Collaboration between IRIS-HEP software developers and experienced U.S. LHC Tier-2 and Tier-3 systems administrators at universities is essential to achieve sustainable systems through the lifetime of the HL-LHC.  This extends, as well, to distributed software developers in the US LHC Operations program as well as laboratory staff at the Tier-1 centers. 

\begin{figure}[h]
\begin{tcolorbox}
The IRIS-HEP Scalable Systems Laboratory is a Kubernetes DevOps platform for DOMA, Analysis Systems and  Innovative Algorithms. 
The SSL hosts infrastructure for CoDaS-HEP training events and Analysis Grand Challenge pipeline testing and tutorials.  
\end{tcolorbox}
\end{figure}

\subsubsection*{Specific Challenges and Opportunities}

Over the past ten years, WLCG production computing infrastructure has remained largely static, managed with decades-old deployment and management techniques. Tier-1 and Tier-2 facility infrastructure, continuously in operation, has been able to cope with the event processing rates and storage requirements necessary to conduct the successful physics programs of the LHC experiments to date. This won't be true in the HL-LHC era unless those facilities change. For example, data access and delivery from distributed data sources to analysis platforms will require leveraging new capabilities in the network, in computational storage systems, in facility resource managers, and service orchestration frameworks.  As we develop new capabilities to confront the data rates and processing scales of the HL-LHC, methods for infrastructure management and scaling become critical. The facility infrastructure itself must evolve. 

This was recognized during the first phase of IRIS-HEP with the SSL providing two vital roles: a testbed infrastructure to incubate DOMA and analysis system prototypes to be evaluated by end-user physicists at LHC Run2 scales, and a challenge-environment for systems administrators and CI engineers to flexibly and adaptively create the needed infrastructure platforms.  Already during LHC Run3, methods employed building the SSL have provided analysis facilities with a blueprint for infrastructure creation and operation.  It is vital that SSL inspired facility R\&D, together with pre-production service integration, continue during the next phase of IRIS-HEP. Considerable experience within IRIS-HEP and from integration with existing production services helps identify the near and longer-term challenges and opportunities for this area in the Institute.

\textbf{Evolving Infrastructure for Next Generation Tier-2 and Analysis Facilities}: Over the course of the next five years, the community will transform the historic
roles of Tier-2 and end-user analysis facilities in the WLCG computing hierarchy to adapt to different sets of needs.  For example, as opportunistic resources from leadership 
and national scale HPC centers are increasingly harnessed by the LHC experiments to fill the resource gaps (even during Run3 but more significantly for 
Run4), the Tier-2 centers, which currently are workhorses for centrally organized simulation and data derivation tasks, may increasingly become 
vital actors of the analysis and data delivery ecosystem.  At present, they are not readily adaptable to take on those challenges. 
Similarly, end-stage physics analysis will shift from the laptop (or small-scale institutional clusters) to shared analysis  facilities, hosting analysis 
and data access services capable of handling HL-LHC scale processing and data rates.  Practically speaking, we may see these facilities blending roles and 
sharing responsibilities as frameworks for orchestrating services provide new agility for resources in the ecosystem. 

To create, optimize, and reliably operate diverse services (such as analysis processing frameworks and data reformatting services) with sustained capabilities,  and to 
ensure toolset adoption by the current Run3 physics groups, a sustained research and integration effort is essential.  
This will involve working at a nexus of disparate communities: LHC physicists, LHC (\& HL-LHC) software developers, 
research cyberinfrastructure innovators, national and leadership HPC technical staff, the Research End User Group \cite{RUGS} in the Cloud Native Computing Foundation (CNCF) \cite{CNCF}.

\textbf{Distributed Infrastructure Management}: DOMA systems intrinsically rely on distributed services to implement, for example, data delivery to analysis and production facilities. XCache networks deployed at multiple Tier-1 and Tier-2 sites for ATLAS and CMS are a prime example, essentially providing the equivalent of a commercial content delivery network (CDN) but with one important distinction: In the WLCG context, these services are mostly managed across administrative domains by multiple teams, leading to delays in software updates and in some cases inefficiency of operation. Lacking a seamless DevOps pipeline for the entire distributed platform, DOMA development teams are limited in their ability to advance new capabilities or respond quickly to bugs and security issues across the platform.  Additionally, as analysis facilities scale to HL-LHC rates, we anticipate ``scaling off-site" to handle bursty workloads during high periods of analysis activity.  The off-site sources of processing cycles may come from Tier-2 production facilities, cloud resources, and potentially HPC centers (which are GPU-rich). A few groups have addressed this gap in capability through central management of services with techniques such as employed by NSF's National Research Platform (\url{https://nationalresearchplatform.org/}) which uses a single administrative domain (with full privileges) to combine resources from over 50 sites.  
In a similar manner, during the first five years of IRIS-HEP the SSL team partnered with the NSF SLATE project (\url{https://slateci.io/}) to develop a federated operations model (``FedOps") which requires only user privileges at Kubernetes-equipped clusters to deploy and operate edge services.  With this approach an XCache network spanning all US ATLAS Tier-2  facilities, and extending to half a dozen more in Europe, are managed by one person allowing rapid updates as the underlying xrootd software evolves. During Run3 and LS3, the lead up period to HL-LHC, an opportunity exists to expand the number of services managed in this fashion, evolve the delivery tools with sustainable frameworks engineered with the broader CNCF community, and strengthen the operational policies and needed security models (work to be done with OSG-LHC). 

\textbf{Integration with Production Services during Run3}: Software developed in the Institute eventually makes contact with the production ecosystem of the experiments and on production infrastructure. The sooner this happens the more quickly challenges are exposed and opportunities identified.  To that end we've made certain that software  graduated from the SSL that is ready for production deployment can have its entire lifecycle neatly packed up and reproduced elsewhere, such as at Tier-2s, national labs (Tier-1s), institutional clusters, HPC centers providing Kubernetes hosting platforms, and even within suitably equipped network testbeds (such as FABRIC). For example, ServiceX has been integrated with the production distributed data management service for ATLAS and CMS, essentially optimizing the data lookups, and exercised with production storage endpoints and production caches. Additionally, we have co-located the Coffea-Casa analysis facility software with shared Run3 analysis centers in production (both at Tier-2 and analysis facilities) to understand resource sharing, software access, identity management, data delivery, and user file system integration. The added bonus is side-by-side proximity to hundreds of physicists conducting analysis on data from LHC Run2 and Run3 thus simplifying the introduction of innovative capabilities from IRIS-HEP to new communities of users, gaining early feedback, increasing adoption.

\subsubsection*{Current Approaches and Development Roadmap}

\noindent\textbf{Facility R\&D Across the Ecosystem}  \\
\textit{Funding Scenarios}: High \\
\textit{Description}: The Scalable Systems Laboratory is a unique resource in our field as it provides a flexible and diverse DevOps platform shared by IRIS-HEP  software teams, developers from ATLAS and CMS, CI engineers from related NSF cyberinfrastructure projects, and LHC systems administrators.  Part of its appeal is the context in which it sits, being (administratively) co-located with existing production LHC Tier-3 and Tier-2 facilities and infrastructure supporting a number of OSG-LHC related services (such as hosted compute entry points). 
It has proved to be an invaluable resource for not only IRIS-HEP software development and service deployment testing, but also a number of Kubernetes-deployable research cyberinfrastructure services including a network analytics service for WLCG PerfSONAR meshes, a network visualization service, an analytics service for ATLAS software and detector conditions data distribution, an XCache analytics service for CMS, scalability testing for FuncX (Functions-as-a-Service) by its core development team, a REANA test deployment, 
tests of containerized Rucio services, a notebook service for the annual CODAS-HEP training event, and a persistent notebook portal to GPU resources serving the ATLAS machine learning community. \\
\textit{Current and Potential Future Activities}:
\begin{itemize}

\item \textit{Continued operation of the reference SSL}: Providing a testbed   infrastructure remains vital as IRIS-HEP software components continue to mature and operate with other services in the ecosystem.  During the first five years 
of IRIS-HEP we established development and ``production" SSL cluster testbeds to fit the cadence and stability requirements of the DOMA and Analysis Systems development teams. 
As these services come into production, this work will continue apace as 
developer sprints respond to user and operational feedback, new use cases and alternative strategies.

\item \textit{Constructing shared Tier-3 analysis facilities}: participate with the broader WLCG community in identifying best practices, infrastructure management patterns, and recipes forged on the SSL leading to reliable and scalable service deployments capable of supporting hundreds of users at Run3 scales in the near term, but with proven capabilities to scale up to Run4 during LS3.  

\item \textit{Evolving Tier-2 infrastructure management}: retrofitting Tier-2 facilities with substrates and higher level infrastructure management tools and services, utilizing industry standard solutions where possible. The major focus will be placed on sustainability and key capabilities needed for stable, continuous Run3 operation while introducing new services targeted for HL-LHC demonstrators and challenge problems. 
Resources from across the Tier-2 complex will be needed in coordinated fashion for 
grand challenge scalability and other proof-of-concept demonstrators. 

\begin{figure}[h]
\begin{tcolorbox}
The IRIS-HEP SSL informs the design of next generation WLCG Tier-2 centers and shared Tier-3 analysis facility infrastructure.  
\end{tcolorbox}
\end{figure}

\item \textit{Capturing facility patterns and blueprints}: Creation and curation of GitOps charts and repository actions to provide reliability and reproducibility such that both infrastructure and applications can be 
easily replicated across facilities.   

\item \textit{Identifying infrastructure (and service) bottlenecks}: a critical activity during data and analysis grand challenge exercises, and related demonstrator proof-of-concept exercises planned by the experiments, are load tests of representative software workloads that accurately mimic planned infrastructure and HL-LHC use cases.  This will require instrumenting services and infrastructure with the needed monitoring 
hooks, streaming the resulting metrics to analytics dashboards, identifying and characterizing any bottlenecks that might exist, and measuring overall performance.

\item \textit{Exploring potential cost savings with ARM processors}: valuation and quantitative assessment of ARM processor technologies for HL-LHC will be a community-wide effort during the next few years. The US ATLAS Tier-1 will provide the ATLAS PanDA workload management system with an ARM testbed capable of testing a diverse set of simulation, reconstruction and analysis tasks.  IRIS-HEP may leverage these resources to test performance and applicability for key services, for example ServiceX transformer performance on ARM. A related task will be providing infrastructure for the OSG-LHC software team to build images for ARM processors and test needed variants of the OSG-LHC middleware (coordinated work with OSG-LHC).

\item \textit{Improve resource sharing across tasks on facilities}:  It is well known that default resource allocation mechanisms offered by Kubernetes do not match the full spectrum of fair share scheduling capabilities for a diverse platform that may be simultaneously providing CPUs for ServiceX transformers, Dask workers for Coffea-Casa, HTCondor job slots for traditional batch processing. This must be dealt with during Run3 for shared Tier-3 analysis facilities which are providing new and old analysis environments.

\item \textit{Strategies for evolving storage system organization and access}: As described in the DOMA strategy above, a promising approach to accelerate access 
to data objects is to ``push" a portion of the query evaluation down into the storage  infrastructure, taking advantage of processors and potentially 
accelerators at that layer.  Skyhook provides this capability for Ceph-based object storage systems. Measurements at scale scale and assessment of performance impacts for ServiceX transformers with Skyhook will need to be conducted for representative analysis queries and processing pipelines.  

\item \textit {Future storage system R\&D activities}: As described in the preceding DOMA strategy, closing the storage gap may indeed present the greatest computing challenge for the HL-LHC era.  Over the next few years, efforts within IRIS-HEP and from across the LHC computing community will inform priorities for  facility R\&D activities in the Institute relating to promising storage technologies and approaches.  For example, creation of heterogeneous storage systems for efficient analysis data access using distributed, asynchronous object stores (DAOS) and RNTuple formats will be explored in proof-of-concept exercises. Performance comparisons of processing rates with POSIX file systems over NVMe drives, distributed dCache pools with spinning disk (and accessed via xrootd protocol), and off-the-shelf object stores for both RNTuple and columnar data formats will provide valuable guidance to the community. The Institute will need to engage broadly and participate directly in specific demonstrators driven
by the experiments. 

\item \textit{Transparent storage quality of service and ``tiered storage" models}: In many cases cost savings can be found if the expected workloads are better matched to the performance characteristics of the underlying storage systems.  
A transparent data quality-of-service that feeds this information to DOMA clients and to job workload management systems may achieve this.  For example, a service which sends tiny log files to object stores, detector simulation input data to filesystems backed by spinning disk (very low I/O is required for those tasks), and analysis outputs to fast NVMe-backed file systems for speedy iterative access. 
Information collected would permit higher level services to move data to cold storage based on access patterns: very infrequently accessed data (``ice cold data") could migrate to machines programmatically switched off, something akin to Amazon Glacier, reducing overall facility costs. Facility managers would use analytics gathered in a given year
to plan purchases for the following year, organizing procurement of storage types based on the experiment's projected needs by access category.

\end{itemize}

\noindent\textbf{Managing Infrastructure over Wide Area Networks to Achieve Scale} \\
\textit{Funding Scenarios}: Medium, High \\
\textit{Description}: Over the past several years significant advances have been made in remotely managing coordinated sets of services to more easily innovate systems while reducing operational cost.  This has evolved
our view of distributed computing resources from a cooperating grid of computing sites towards scalable research platforms offering programmatic capabilities for service orchestration, with dramatic potential impacts given the 
increased versatility of access.  In the next phase of IRIS-HEP we anticipate making facility infrastructure advances for both data delivery and analysis processing incorporating these ideas. \\
\textit{Current and Potential Future Activities}:
\begin{itemize}

\item \textit{Reducing site storage requirements through caching}: Currently, both ATLAS and CMS employ data delivery methods involving differing access and caching architectures using XRootD technology which have direct implications on facility management and storage costs.  
Central tools for management of the global system can be used to introduce new capabilities and updates which strengthen the infrastructure. 
Current storage capacity requirements at WLCG processing sites can
potentially be reduced using caches and knowledge of data placement and access history.  Adding storage-less sites to the environment  
obviously would expand the pool of available CPUs and facilitate access to opportunistic resources. Additionally, existing WLCG sites with 
sufficient WAN bandwidth capacity could potentially retire their storage endpoints altogether and offer increased CPU capacity, trading storage costs (both operational labor and equipment) for additional CPU.  
As part of this activity we will stress-test caching networks with diverse analysis payloads at scale and provide performance and operational
feedback to the DOMA development team.

\item \textit{Distributed workers to scale up analysis capacity}: Coffea-Casa analysis facilities to date have been deployed
only on single-site, Tier-3 or Tier-2 clusters.  For potential scale up to HL-LHC volumes, and to reduce waiting times on existing facilities in production, one could package and deploy lightweight service endpoints providing potentially an unlimited supply of Dask workers to a central home base (the ``Casa").  Work is needed to understand the interplay and  tradeoffs of providing centralized login servers, home servers, and local user storage services vs cloned stand-alone systems. The potential gains could be significant to reduce costs of analysis facilities (reducing their number) while vastly improving performance. Managing a coordinated network of lightweight worker providers, or ``Coffea-Pots", across administrative domains is possible with FedOps techniques we've employed in the SSL.

\item \textit{Embedding advanced filtering services in the network}: We have begun exploring deployments of ServiceX inside the FABRIC testbed which opens up new strategies for data delivery and cost assessments in distributed environments.  For example, by placing ServiceX on the server-side, at the data source, one can reduce bandwidth consumption between analysis clients and storage systems by combining filtering and reformatting functions at the source rather than copying complete files over the network.  Significant impact could be made with implications for cost reductions of both network capacity and storage, both being expensive components of facility infrastructure. In Q2 2023, as part of the NSF FAB project (FABRIC Across Borders), a FABRIC ``node" will be put into production at CERN, co-located with datasets hosted by the Tier-0. 
We will explore benefits of server-side ServiceX in operation at the Tier-0 while providing fast delivery of Run3 samples to production analysis facilities in the U.S. The exercise will provide an estimate of potential bandwidth reductions on the costly transatlantic link.

\item \textit{Upgrading the WLCG software distribution and detector conditions data caching network with modern software}: We have begun 
investigations of alternate approaches for managing cache-friendly data at WLCG sites, such as using the   
Varnish (\url{https://www.varnish-software.com/}) software.  Alternate data distribution methods would aim to decrease the total cost to operate WLCG sites, helping close the overall resource gap.

\end{itemize}

\noindent\textbf{Integration with Experiment Environments on Production Infrastructure} \\
\textit{Funding Scenarios}: High \\
\textit{Description}: Throughout the development process, from prototypes to  pre-production software and systems, there are points of integration with the 
larger software ecosystems of the experiments that must be designed, developed and iteratively tested. This  requires significant and frequent contact among the development teams and with systems staff from the facilities.  In addition, to perform meaningful tests with live environments and at meaningful scales for grand challenge demonstrators, production-scale resources must be marshalled and fair-share scheduled with on-going operations. \\
\textit{Current and Potential Future Activities}:

\begin{itemize}

\item \textit{Facility integration of pre-production services}: In the first phase of IRIS-HEP, providing efficient interfaces between ServiceX and Rucio, the LHC data management service, was a significant challenge and required several optimizations to reduce latencies in queries to the central production service. As Rucio evolves with its own development roadmap this integration work will need to continue.  As discussed in the DOMA strategic plan, integration of ServiceX with iDDS provides an opportunity to scale-up transformer processor capacity, and conversely, provide PanDA brokered analysis tasks with a columnar reformatting service.  There are potential performance benefits from improved integration of ServiceX with XCache that can only be explored in 
production settings. Infrastructure and deployments for ancillary services like MinIO, the storage service for ServiceX formatted data outputs, including utilities to manage these outputs are needed. Discovering and removing experiment-specific dependencies (overlooked from early development prototypes) making them suitable for use by multiple experiments is often done in this phase. Supporting operation and user access to pre-production instances of DOMA services and analysis facilities for early adopting communities, including first draft user guides, requires expert knowledge from both domains.

\item \textit{Preparing demonstrator resources at scale}: Our approaches to date have been driven by IRIS-HEP developer priorities, analysis grand challenge goals, facility R\&D opportunities, and a desire to find a sustainable path forward for creating declarative, reproducible cyberinfrastructure capable of scaling out to wherever resources may be found. In addition to IRIS-HEP data and analysis grand challenge exercises, the LHC experiments are planning additional demonstrators as input to computing  technical design reports (TDRs).  These TDRs will include a prioritization of R\&D projects to complete during Long Shutdown 3 (S3) of the LHC. To inform those decisions, R\&D projects for TDRs  must provide a program of work towards the Run4 software and systems releases. 
They must include risks and effort estimates and proof-of-concept demonstrators which will lead to estimates on impacts for resource utilization (e.g. CPU time, GPU time, RAM, local disk and class of service, tape, and networks). IRIS-HEP will work with the computing and software management teams to align goals, configurations, data sets, and platforms to be used for these impact studies. 

\item \textit{Advanced integration with HPC facilities}: Even as HPC resources are viewed as a prime opportunity to close the processing gap during the course of HL-LHC, coupling the experiment's data and workload management services to those facilities will remain a significant challenge as high energy physics is only one of a broad 
collection of science-drivers served by those facilities.  A number of challenges have been identified in a recent workshop (\url{https://indico.cern.ch/event/1183995/}), among them data delivery into and from these facilities.  HPCs are likely to support 
Kubernetes platforms within their infrastructure, even beyond edge platforms for hosting user services, potentially providing additional options to  incorporate those resources into the LHC computing environment. Working with OSG-LHC, the SSL group will contribute
to strategies to efficiently interface experiment systems with the
resource ``APIs" of HPC facilities.

\item \textit{Infrastructure for training and on-boarding}: To ensure community adoption and maximize user feedback to developers, it is essential to continue to provide
training infrastructure for IRIS-HEP software tutorial events, experiment on-boarding events to analysis facilities, and advanced training activities sponsored by the Insitute, such as the annual Computational and Data Science Training for High Energy Physics school (CoDaS-HEP). 
Where experiment dedicated resources cannot be used (e.g. production
Tier-3 analysis centers), the SSL can provide the necessary notebooks, CPU and GPU resources for diverse groups of young researchers during
these events.

\end{itemize}

\subsubsection*{Impact and Success Criteria}

Equipping IRIS-HEP with the Scalable Systems Laboratory not only 
accelerated the pace of development of DOMA and Analysis Systems,
and their integration with experiment production services and facility systems, but due to its versatility provided a unique resource in the field of high energy physics computing.  The SSL has hosted instances of the ATLAS Distributed Computing Analytics Platform (\url{https://analytics.mwt2.org}), OSG Compute Entry points (providing access to cluster resources from over 20 universities), and provided a DevOps environment for a diverse community of CI engineers. Clearly the SSL facility and team assembled to operate it has been an important contributor to the overall NSF cyberinfrastructure ecosystem. The SSL implemented approaches for creating and managing flexible and reproducible infrastructure that others in the WLCG community are adopting. Maintaining an agile, open facility environment to incubate ideas, try out service prototypes in context, explore and cultivate new patterns for infrastructure delivery, and to confront the integration and scalability challenges over the next five years will be vital to the success of the Institute. 

It is important to note that while specific resources (dedicated clusters) and infrastructure have been assembled and labeled as ``the SSL" in the first five years of IRIS-HEP, facility R\&D is of course 
broadly scoped and distributed across the community, receiving  contributions from software developers, CI engineers, systems administrators and researchers from many universities, national laboratories and CERN.  In the Institute we have the opportunity to bring convergence to approaches from many sources that will close the gaps to meeting HL-LHC computing resource needs and capabilities.

\noindent \textbf{Success Criteria – Milestones \& Deliverables}:

As for other areas in an Institute, the milestones and deliverables outlined below focus on the first three years.  These high-level items would be expanded in later years as part of the execution of any project. Note however that at the time of writing, many proof-of-concept demonstrators and facility R\&D topics are being planned by the wider community during this time frame, potentially impacting deliverables described below. 

\begin{enumerate}[label=D5.\arabic*.]

\item Establish infrastructure and orchestration services necessary for a sustained reference SSL. {\bf December 2023}

\item With DOMA, ServiceX deployed inside FABRIC at CERN. {\bf December  2023}

\item Releases of Analysis Systems pipelines supporting distributed analysis deployed on SSL. \textbf{Continuous}

\item Provide a deployment package for upgraded software distribution and conditions data caching servers. {\bf May 2024}

\item With AS, evaluate distributed workers for Coffea-Casa. {\bf May 2024}

\item Curate and publish production Tier-3 deployment patterns including integration of traditional systems in use during Run3 and forward-looking analysis systems. {\bf December 2024}

\item With AS, support demonstration of running a full analysis suitable  that uses machine learning. \textbf{December 2024}

\item Evaluate technologies, services and infrastructure management patterns for next generation LHC Tier-2 facilities and publish for 
the community. {\bf December 2025}

\end{enumerate}

\noindent \textbf{Success Criteria – Metrics}:

Metrics are a useful tool to provide management with quantitative insight about progress toward overarching goals.  High-level metrics we expect to be applicable for the Facility R\&D and Integration area are below:

\begin{enumerate}[label=M5.\arabic*.]

\item Number of deployed analysis facilities in production operation. Goal is 6.   

\item Number of sites providing distributed workers to an analysis facility in production operation. Goal is 3. 

\item Number of storage-less sites in sustained production. Goal is 2. 

\item Number of sites with upgraded software distribution and detector conditions data caching servers in production. Goal is 7. 

\end{enumerate}

%% file: 140-strategic-areas-open-science-grid-hep.tex
\subsection{Fabric of distributed high-throughput computing services}
\label{sec:fabric}

The LHC depends on a global, distributed cyberinfrastructure to
process, move, and store data and that infrastructure will need to
evolve to meet the needs of the HL-LHC. The OSG Consortium provides
the OSG Fabric of Services, a tightly-integrated set of services
which provides for the cyberinfrastructure of the existing LHC
experiments within the US. \\

\begin{figure}[h]
\begin{tcolorbox}
Established in 2005, the OSG Consortium operates a fabric of distributed High Throughput Computing (dHTC) services in support of the National Science \& Engineering community. The research collaborations, campuses, national laboratories, and software providers that form the consortium are unified in their commitment to advance open science via these services.
\end{tcolorbox}
\end{figure}

The OSG Consortium is not a legal entity but a coordinated set of
stakeholders working toward a common goal.  A strategic area – the
``OSG-LHC” - is needed to make contributions to the OSG Consortium
to cover the existing cyberinfrastructure needs of the LHC, ensure
the U.S. LHC facilities remain operational, and integrate the U.S.
LHC facilities with the worldwide infrastructure.

Within the national distributed High Throughput Computing infrastructure,
the LHC community leads in the scale and complexity of the distributed
system.  Through its participating in the broader OSG Consortium,
work performed and knowledge gained for the LHC community has a
broader impact than LHC and HEP.  During IRIS-HEP, the OSG-LHC team
has remained closely aligned with the Partnership to Advance
Throughput Computing (PATh).  The latter has a scope of all of NSF’s
Science and Engineering community, ensuring innovations done by
OSG-LHC have a broad national footprint.  The OSG-LHC helps the
HL-LHC with both the scalability of its cyberinfrastructure and its
sustainability by sharing it with multiple domains.

Over the past 15 years, the OSG Consortium has provided a stable
foundation of software and common services to meet the HEP community’s
needs.  An important aspect that is managing change in a rapidly
changing world of software. For example, the OSG-LHC has led the
transition of the entire national distributed high-throughput
computing (dHTC) infrastructure from the niche “Grid Security
Infrastructure” (GSI) identity-based authentication model to
industry-standard bearer tokens implementing capabilities.  When
the Globus organization announced it would stop supporting the
Globus Toolkit, the OSG-LHC team helped establish the Grid Community
Forum and forked the toolkit to be the Grid Community Toolkit (GCT).
During the first four years of IRIS-HEP, a major accomplishment was
retiring the use of the GCT and replacing it with updated services.

OSG-LHC also integrates with rest of the institute. For example,
through interaction with the Facilities R\&D area, OSG-LHC began
to add containers as a ``first-class citizen" for software distribution;
both DOMA and Analysis Systems groups make use of the OSG-LHC
container registry; and DOMA's work to implement bearer tokens or
HTTP-TPC within the XRootD software is distributed in turn by
OSG-LHC.  Thus, software not only flows from the developers out to
the communities but the ideas from within the institute also affect
OSG-LHC's approaches.  Managing the full software lifecycle is a
critical component of any distributed CI and an important role the
OSG-LHC will continue to play in the future.

\subsubsection*{Specific Challenges and Opportunities}

\paragraph{Rise of the container-native facilities} Container-focused
(“Cloud Native”) compute facilities are a generational change in
how sites are operated and deliver capacity to the LHC experiments;
see Section \ref{sec:facilities} for an overview of the rapid change
in the cyberinfrastructure.  OSG-LHC has significant expertise in
delivering services as well-integrated software packages shipped
in the operating system’s default packaging format.  This historically
was the desired mechanism for most sites that deploy the distributed
services.  Container-focused sites (largely deployed on top of
Kubernetes) prefer services be delivered in containers, agnostic
to the packaging or even the container OS, and have a higher-level
service orchestration language for how services should be deployed.
OSG-LHC has started to deliver services using these formats but
there are still opportunities; for example, containers allow services
to be delivered without needing superuser privileges on the host
and container registries provide the ability to do daily scans of
OSG-LHC containers for known security vulnerabilities.

Not all sites have an even adoption of newer technologies.  There
is a need amongst the cyberinfrastructure – not specific to any
experiment – around Kubernetes community and knowledge building,
including training and workforce development. \\

\paragraph{Increasing diversity of hardware resources} The
LHC community has long benefited from a period of relatively
homogeneous resources – for well over a decade, all its pledged
resources were x86 cores with 2GB of RAM.  With the advent of GPUs,
this has slowly started to change (although at a rate limited by
the paucity of large-scale GPU workflows).  The change to heterogeneous
resources has a potential to accelerate as more ARM-based servers
are released on the market an become competitive with capacity based
on x86. 

\paragraph{Integration with HPC, ML, and non-WLCG resources} A
common thread throughout the LHC lifetime is the fact its minimum
hardware needs were met through its dedicated resources.  During
Run 2 and 3, additional HPC resources from outside WLCG were heavily
utilized but these resources were never critical-path for the science
program.  For HL-LHC, to meet projected flat budgets, these non-WLCG
resources will become essential.  As part of its experiment-agnostic
fabric, the OSG-LHC has the opportunity to help integrate NSF
leadership class and ACCESS-allocated resources into the experiments’
fabrics.

A new type of resource and service expected for the HL-LHC will be
machine learning inference services.  Users have long used inference
as part of their analysis but the complexity of the models are
expected to increase rapidly through the years.  Instead of trying
to host sufficient resources within a LHC-specific facility,
integration with a national-scale investment for inferencing would
deliver value to the community. \\

\paragraph{Operating Production Services} Unlike other strategic
areas which have a mixture of projects in various stages of the
software development lifecycle, OSG-LHC’s purpose is to operate,
maintain, and evolve production services.  OSG-LHC must ensure the
OSG Consortium’s fabric of services remains operational and delivers
value to the U.S. LHC Operations programs.  This includes all parts
of the lifecycle, including deprecation and obsolescence – for
example, OSG-LHC will retire the OSG 3.6 software release series
along with the June 2024 end-of-life of Enterprise Linux 7.  OSG-LHC
will help the community to bridge from today’s cyberinfrastructure
to HL-LHC’s.

\subsubsection*{Current Approaches and Development Roadmap}

\noindent \textbf{Broadening Service Delivery} \\
\textit{Funding Scenarios}: Medium, High \\
\textit{Description}: Services today are delivered as packages for
the standard platform on the WLCG (Enterprise Linux 7, 8, or later
on the x86 architecture).  As the community progresses to the HL-LHC,
the needs will become more heterogeneous (including ARM servers and
the HPC platforms available as part of the NSF coordinated
cyberinfrastructure in the next five years) and expect alternate
delivery mechanisms (including container images and the service
packaging such as Helm charts). \\
\textit{Current and Potential Future Activities}:

\begin{itemize}
\item \textit{Support for Heterogeneous Architectures}: Evolve the
existing software compilations and repositories to also support
non-x86 architectures including ARM.
\item \textit{Container-Native Service Delivery}: Beyond individual
packages, create and maintain repositories of service descriptions
for newer technology platforms like Kubernetes.  Work with the
community to develop and implement best practices such as reducing
privileges inside the running containers.
\item \textit{HPC Integration}: The OSG-LHC has developed a variety
mechanisms for integrating resources into the OSG Fabric of Services.
This includes both the “Hosted CE”, which integrates compute resources
over SSH connections, and supporting XRootD endpoints on using
unprivileged containers, giving access to storage resources at HPC
centers.
\item \textit{Hosting Institute container images}: other areas of
the Institute produce software outputs in the form of container
images. Working closely with the Securing an Open and Trustworthy
Ecosystem for Research Infrastructure and Applications (SOTERIA)
project, OSG-LHC provides a national cyberinfrastructre-focused
home for IRIS-HEP images.
\end{itemize}

\noindent \textbf{Maintaining the OSG Fabric of Services} \\
\textit{Funding Scenarios}: Low, Medium, High \\
\textit{Description}: Through the OSG-LHC, the OSG Consortium
maintains services essential to the operations of the U.S. LHC
facilities and their integration into the global cyberinfrastructure.
Shared services include resource usage accounting, endpoint
registration, and monitoring of service status. \\
\textit{Current and Potential Future Activities}:
\begin{itemize}
\item \textit{Packaging}: The OSG Software team, as part of the
OSG-LHC contribution to the Consortium, maintains a curated software
stack for facilities to use for services like compute Entrypoints
(CEs), storage endpoints, and the worker node runtime client.
\item \textit{Resource and Service Registration}:  OSG-LHC maintains
an authoritative list of services part of the OSG Consortium and
their association to administrators (necessary for communications
and security responses).  We aim to integrate with the CILogon-based
Single Sign On (SSO) to ensure administrators can login to the
infrastructure with their local credentials.  The same mechanisms
can be extended to help simplify workflows for new Analysis Facilities
and other services such as container registries for ATLAS analysis
users.
\item \textit{International Cyberinfrastructure Integration}:
Facilities have a requirement to report usage to the WLCG and report
monitoring to ensure the U.S. LHC organizations are delivering on
their commitments.  OSG-LHC manages accounting and monitoring
services and ensures they are adopted to changes coming from the
WLCG.
\end{itemize}

\noindent \textbf{Operational Cybersecurity} \\
\textit{Funding Scenarios}: Low, Medium, High \\
\textit{Description}: Like any cyberinfrastructure, the LHC facilities,
OSG Fabric of Services, and the user population are resources
continuously targeted by hackers.  The OSG Security area provides
effort for operational security, responsible for security exercises
and policy, responding to threats, and scanning services for
vulnerabilities. \\
\textit{Current and Potential Future Activities}:
\begin{itemize}
\item Coordination with global cybersecurity teams: Like data
transfer, having functional cybersecurity requires interoperating
with other infrastructures.  The OSG-LHC cybersecurity team is the
bridge between U.S. LHC efforts and EGI, CERN, and WLCG security
teams.
\item Improved monitoring of software artifacts and container-based
services: Many of the security controls were designed when services
consisted of packages deployed on physical hosts.  The majority of
services, however, run inside containers and are transient – meaning
logs are likely lost unless they are retained inside a central
logging service.  Security controls need to be adopted and policies
need to be updated to provide guidance to facilities on similar
approaches.
\end{itemize}

\noindent \textbf{Network Monitoring} \\
\textit{Funding Scenarios}: Medium, High \\
\textit{Description}: Whether for bulk data movement or for streaming
events to consumers, network performance is critical for the LHC
community.  The OSG-LHC Network Monitoring Area helps develop and
operate technologies for monitoring performance and understanding
network usage patterns at the application level. \\
\textit{Current and Potential Future Activities}:
\begin{itemize}
\item \textit{Packet monitoring}: Historically, network
operators have had little insight into the usage of their networks
by the LHC community which makes long-term planning more difficult.
Typically, traffic between LHC sites are all labelled as “LHC”
without regards to whether they are actually LHC traffic (versus
other experiments), user versus production, or high priority versus
low priority.  The network monitoring team is working to have
transfer services annotate packet header so network providers have
richer metadata.
\item \textit{Leadership of the WLCG Network Monitoring task force}: Networking
is another point where keeping the international community aligned
is key.  Currently, OSG-LHC co-leads the WLCG Network Monitoring
task force and is well-positioned to guide network transitions
between now and HL-LHC.
\end{itemize}

\subsubsection*{Impact and Success Criteria}

OSG-LHC is unique among the Institute strategic areas in that its
primary goal is to ensure smooth operations of the cyberinfrastructure
for the LHC, with services meeting defined SLAs; ``success" during
this period is providing a operational roadmap to the HL-LHC era.
However, there are expected outcomes and milestones beyond operations.
For example, the area would finish the migration of services to
Kubernetes and start supporting Kubernetes-only (or, at least,
Kubernetes-centric) sites.  The current OSG-LHC provides leadership
in running production services on Kubernetes and we expect these
new and converted facilities will rely heavily on its expertise.
OSG-LHC will also be essential to the deployment of next-generation
network monitoring, enabling network providers to introspect HL-LHC
flows.

In its current incarnation, OSG-LHC consists of a team who have led
the evolution of the OSG over the past 15 years, meaning there is
a wealth of accumulated expertise in running a fabric of distributed
high-throughput computing services and deep technical knowledge of
how the current system operates.  It is an essential element of any
future Institute. Without it, the U.S. LHC operations programs would
need to pick up the activities and it is unlikely that separate
implementations by experiments would be as cost-effective as a
common solution. Having the OSG-LHC be part of the future
cyberinfrastructure for the HL-LHC will ensure key continuity of
operations from now until the HL-LHC era. \\

\paragraph{HL-LHC Computing Challenge Impact}: OSG-LHC will primarily
have impact in the {\em G4 (Sustainability)} and {\em G2 (Scalability)}
computing challenges. The approach of the OSG-LHC strategic area
–- contributing to the wider OSG Consortium -– improves the
sustainability of the cyberinfrastructure by having a shared common
layer across NSF Science and Engineering domains.  We see this as
being mutually beneficial: OSG-LHC will use the commons that have
contributions from multiple projects and other projects will benefit
from the unique scalability experience of the LHC/HL-LHC community.
OSG-LHC will provide the foundational infrastructure for the HL-LHC
and is thus responsible for ensuring the cyberinfrastructure can
meet HL-LHC’s scalability goals. \\

\noindent \textbf{Success Criteria - Milestones:}

\begin{enumerate}[label=D.\arabic*.]
    \item Use of network flow monitoring as part of the next data challenge exercise; see Section \ref{sec:domachallenge}. {\bf March 2024}
    \item Successful use of tokens for data transfers during the next data challenge exercise (joint with DOMA). {\bf March 2024}
    \item Retirement of the OSG 3.6 along with the end-of-life of RHEL7. {\bf July 2024}
    \item Remove last requirements and usage of GSI / X.509 security from the OSG Fabric of Services at the end of Run 3. {\bf January 2026}
\end{enumerate}

\noindent \textbf{Success Criteria - Metrics:}

\begin{enumerate}[label=M.\arabic*.]
    \item Percent of critical OSG-LHC services meeting their agreed-upon SLAs.  Goal is 100\% of SLAs met, measured quarterly.
    \item Number of IRIS-HEP and LHC container images hosted in the OSG-LHC container registry. Goal steady increase, measured quarterly.
\end{enumerate}

\newpage

%% file: 145-strategic-areas-training.tex
\subsection{Training and Workforce Development}
\label{sec:training}

People are the key to successful software. The community is currently
building hardware upgrades and planning for an HL-LHC era which
will {\em start} collecting data  at the end of this decade, and
then acquire data for at least another decade.  People, working
together, across disciplines and experiments, over several generations,
will be the critical foundation underlying  sustainable software.

\subsubsection*{Specific Challenges and Opportunities}
Training support for software-related activities in HEP has
historically been uneven. Although most universities do provide
some relevant computer science and software engineering courses,
and many now provide introductory ``data science'' courses, many
HEP graduate students and postdocs are not required to take these
classes as part of the curriculum. As students enter the research
phase of graduate student training, many recognize the value
of such classes, but are no longer in a position to take the
classes easily. No ``standard'' recommendations exist for incoming students,
either for HEP experiments or the HEP field as a whole. Some
universities are developing curricula for STEM training in general
or ``certificate'' programs for basic data science or
software training, but these are by no means yet universal.  The
result has been that the graduate student and postdoc population
historically had a diverse spectrum of relevant skills.

\begin{figure}[htbp]
\begin{center}
\includegraphics[width=0.99\textwidth]{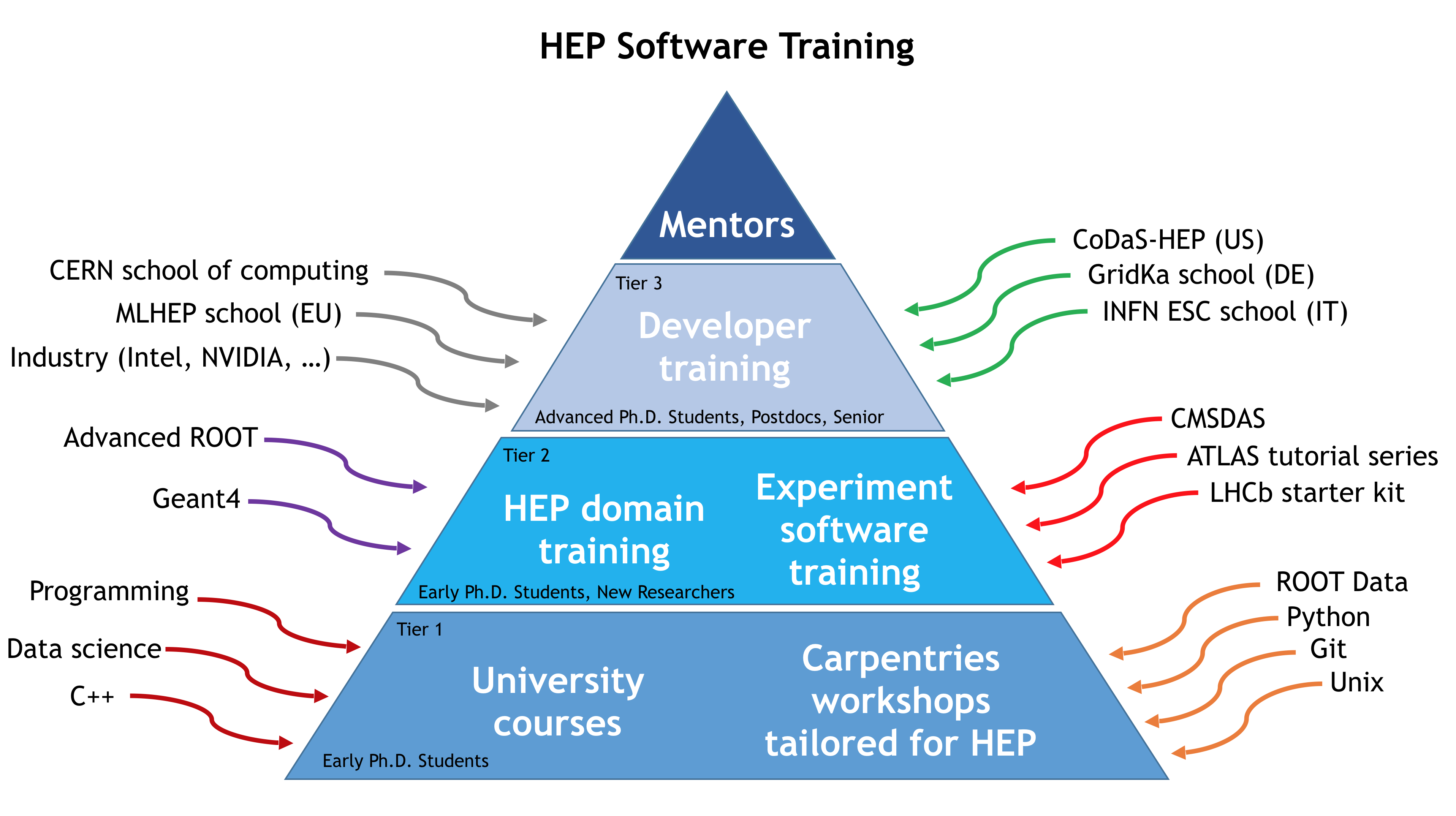}
\caption{The HSF vision for software training in particle physics, from generally required 
basic software skills through advanced developer training, including eventual 
mentoring to contribute to research software projects.}
\label{fig:training-pyramid}
\end{center}
\end{figure}

To address this, IRIS-HEP worked with HSF to develop a community
vision five years ago (Figure~\ref{fig:training-pyramid}) for a
progression of training activities which, if implemented by the
community, would create a new generation of software-enabled
scientists.  The ultimate goal was to invest today in the young
students and postdocs who would become faculty leaders driving the
research agenda in the HL-LHC era.  IRIS-HEP has been working with
the community to implement this vision.  A sustainable framework
for software training will be an important community legacy from
IRIS-HEP and position the community to succeed maximally in the
HL-LHC era.

\subsubsection*{Current Approaches and Development Roadmap}
\paragraph{Basic Curriculum:} Working with the HSF training group over the 
past few years, a complete basic curriculum (the lowest tier in
Figure~\ref{fig:training-pyramid}) has been implemented~\cite{hsftraining}
at a level appropriate for HEP students. 
It includes material developed by The Carpentries and additional modules tailored to HEP beginners. 
Using this curriculum, the IRIS-HEP software training
group and collaborators have organized more than 30 training events and trained over 1,500 students and postdocs.
The material has also been adopted by the DUNE experiment for software
training activities as well as the USCMS PURSUE summer student
program~\cite{uscms-pursue}, which focuses on broadening participation.
Driven by a community of more than 50 motivated educators, the training modules
are open-source and collaborative in nature. They span the spectrum
from robust basic software skills to advanced topics like
machine learning on GPUs.  Sustainability has been the centerpiece
of the approach, given that there are hundreds of new entrants in
the community every year. Different tools and platforms, like GitHub,
have enabled technical continuity, collaboration, and nurtured the
sense to develop reproducible and reusable software.  A significant
effort has been devoted to ensuring that the basic software curriculum 
is taught frequently enough that most students can be trained very
early in their research careers.

\begin{figure}[htbp]
\begin{center}
\includegraphics[width=0.8\textwidth]{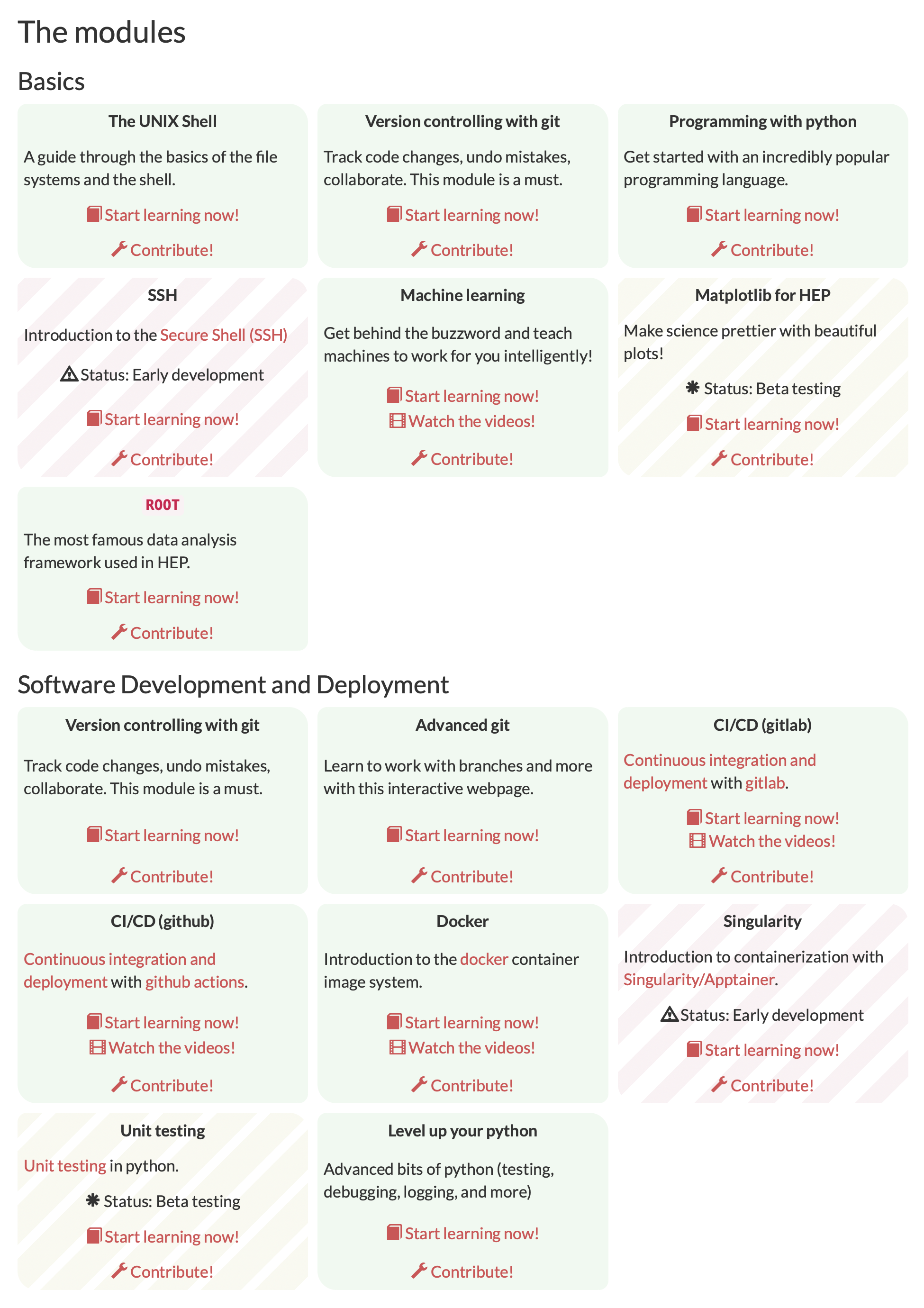}
\caption{Some elements of the HSF training curriculum. The full set can be found on
the HSF Training Center website.~\cite{hsf-curriculum}}
\label{fig:hsfmodules}
\end{center}
\end{figure}

Following the approach of The Carpentries, each training module corresponds 
to a training webpage, including verbose descriptions, key-point summaries, 
and exercises to make training interactive. Recordings for all of our 
workshops are available, and several of our lessons are accompanied 
by dedicated videos targeting individual learners. As of 2022, the 
basic curriculum is considered complete and in production use by
the community. A first set of reusable training modules on more 
advanced topics is also becoming available. (See Figure~\ref{fig:hsfmodules}
for examples, and the HSF Training Center~\cite{hsf-curriculum}
for the full set.)

\begin{figure}[htbp]
\begin{center}
\includegraphics[width=0.8\textwidth]{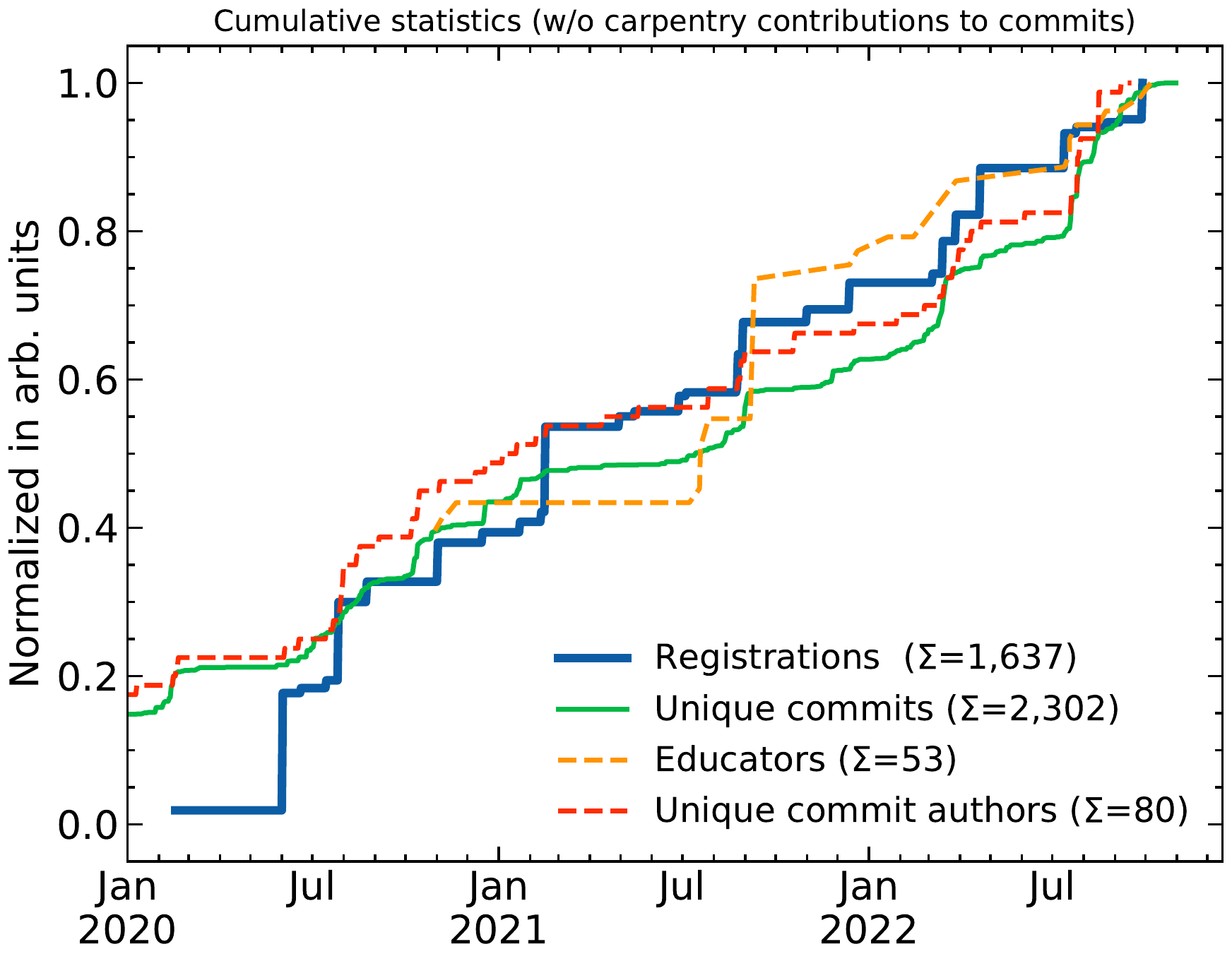}
\caption{Growing the training group. Four different cumulative metrics are overlaid: the number of registrations in our events, the number of unique commits in our repositories (excluding commits to the framework developed by The Carpentries), the number of educators registered in the HSF Training community, and the number of unique commit authors (again excluding commits attributed to The Carpentries). The absolute values of each metric are shown in the legend.}
\label{fig:training_stats}
\end{center}
\end{figure}

%

\begin{figure}[htbp]
\begin{center}
\includegraphics[width=0.99\textwidth]{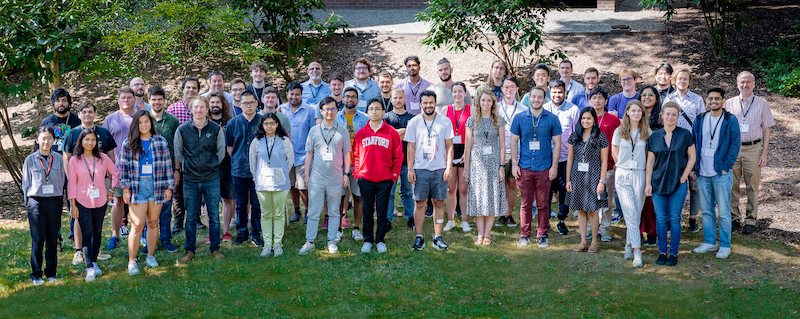}
\caption{Participants in the CoDaS-HEP 2022 summer school.}
\label{fig:codashep2022}
\end{center}
\end{figure}

\paragraph{Expert-level workshops and mentoring:}
While the basic training is typically sufficient for physicists
involved primarily in small-group data analysis, many community members 
become Cyberinfrastructure professionals and
will be involved in larger-scale software development activities for
their experiments.
Because of the long time scale of the experiments, imparting
sustainable software development best practices is crucial.
In addition, more advanced software and data science skills are 
often a strong asset when career evolution takes people to industry
or other academic research domains.
Offering orientation about different career paths for Cyberinfrastructure
professionals will be an important aspect of sustaining the demand
for software experts.

Some elements of more advanced training have been put in place.
For example, the intensive week-long Computational and Data Science for 
High Energy Physics (CoDaS-HEP) summer school was begun in 2017 with 
the aim of teaching more advanced skills such as parallel programming, 
data science tools  and techniques, machine learning technology and 
methods, code performance evaluation and collaborative use of git and GitHub.
The school consists of both lectures and hands-on sessions
and brings together young scientists from various HEP experiments 
(not only LHC) with experts from HEP and beyond.
A ``cohort'' experience is also created by bringing the participants 
(that have similar interests) together in a single location
for a week.
The school was originally developed as part of an older NSF-funded project 
(PHY-1521042) and participant support has also been provided by an
NSF CyberTraining award (OAC-1829729). Many instructors have,
however, typically come from among the personnel funded by IRIS-HEP.
The school was not run in 2020 and 2021 due to COVID, but began
again in 2022 and is expected to run annually going forward.

Finally, it is important to note that many advanced concepts, such
as architectural design decisions, cannot be easily imparted by
teaching alone but must be practiced deliberately under the guidance
of mentors that have experience developing research software used
by others. For this purpose, the IRIS-HEP Fellows program was
created. Fellowships provide an intense and personalized form
of developer training. During a period of several months, Fellows
work on a project of their choosing, closely mentored by an expert
from the field. By contributing to a HEP software project,
Fellows obtain first-hand experience in collaborative software development
and put their software knowledge into practice. Since 2019, more
than 100 fellowships have been awarded. The success of the IRIS-HEP 
Fellows program funded by NSF has also attracted additional funding from 
private foundations and industry to extend the program.
For example, twenty students from Ukraine were supported in 2022
and the Simon institute has supported around 5-8 international fellows every year since 2020.
We see already that some of the CoDaS-HEP attendees and Fellows remain engaged
with HEP research software development as they progress in their 
careers.
\paragraph{Plans for the next phase of IRIS-HEP:}
The current IRIS-HEP project has been
instrumental in advancing the vision of a training pipeline shown
in Figure~\ref{fig:training-pyramid}. It is, however, not yet 
complete: in the HL-LHC era, we want every student entering our experiments to be fully conversant in basic software
skills, with many possessing the more advanced software skills needed
to create and evolve research software in HEP. Moreover, the
available training progression will be widely recognized, and
sustainability mechanisms will be established. For example,
previous ``students'' of training activities may return as
instructors for the next cohorts. The Fellows program will
be widely recognized as an ``on-ramp'' for students to
not only build skills, but also get connected to experiment 
and community research software activities. Finally, the 
students we train in the coming years include the faculty, staff
scientists and research software engineers of the HL-LHC
era. 

%% file: 146-strategic-areas-lhcb.tex
\subsection{LHCb \& LHC Run 5}
\label{sec:lhcb}

Lastly, we describe the additional strategic interest for ongoing 
collaborative efforts within IRIS-HEP regarding LHCb, another LHC 
experiment supported from the US
primarily by the NSF. The upgrade plan for LHCb is out-of-phase with
respect to ATLAS and CMS: major upgrades happened for LHC Run 3 and are
planned for LHC Run 5.
For example, between LHC Runs 2 and 3
LHCb underwent a major upgrade:
its hardware trigger was removed and the first level trigger (Hlt1) now processes
30 MHz of beam crossings, about 5 TB/s of data.  Extending several
aspects of the software infrastructure will be important for LHCb
in Run 4 and critical for Run 5, and may serve as models or starting
points for software that is valuable to other experiments during
the HL-LHC era. 

As of Run 3, Hlt1  executes on GPUs with one instance of {\tt
Allen}\cite{Aaij:2019zbu} running on each GPU.  Currently, every
GPU receives complete events from an event building unit (an x86
CPU server) and handles several thousand events at once.  Raw detector
data is copied to the GPU, the full Hlt1 sequence is processed on
the GPU;
only selection decisions and objects used for the
selections, such as tracks and primary vertices, are copied back
to the CPU. The data rate between the two x86 server farms (the 
Hlt1 farm and the Hlt2 farm) is,
therefore, reduced by a factor 30–60.  {\tt Allen} runs inside Gaudi
in production.  The native monitoring software inside {\tt Allen}
has an interface similar to that of Gaudi so the aggregated histograms
are propagated back to the host, eventually giving them to Gaudi.
In the future, LHCb plans to use Allen for the second level trigger
as well.  This requires additional work.  An x86 compilation of
Allen can run on the WLCG.  In that case, the event loop is steered
by Gaudi, and {\tt Allen} is called one event at a time.  While
{\tt Allen} is not the only approach to using GPUs for reconstruction
and event selection entirely within GPUs, it demonstrates that this
is possible.  Other experiments might consider adopting it, or
integrating some of its features into their GPU software.  As GPUs
become part of the WLCG, it will be important to build infrastructure
to allow {\tt Allen} to use these resources effectively.  The
environment will be very different from that of the online system,
so the new infrastructure for using heterogeneous resources will
take dedicated work.  
The same infrastructure could be designed to be common
to all experiments using WLCG resources.

LHCb developed {\tt PyConf}, to make Gaudi application configuration
safer, cleaner, and simpler to debug.  It was originally developed
as a general pythonic functional framework for configuring complex
CPU workflows, but it works for  GPU workflows as well.  At the
moment, this package is specific to the LHCb version of Gaudi, but
it should be possible to generalize for other users of Gaudi (such
as ATLAS).  This could be an appropriate effort for IRIS-HEP.

Recently, LHCb demonstrated that an {\tt end-to-end} deep neural
network (DNN) can provide better efficiency {\bf and} a lower false
positive rate for finding primary vertices (PVs) using simluated
LHCb Run 3 data.  It starts with track parameters and produces
``target histograms" that a simple heuristic algorithm scans to
extract PV positions and estimated longitudinal resolutions.  Two
separate DNNs are trained, then joined with extra channels added
in the intermediate layer to allow the overall algorithm to learn
additional additional details.  The first consists of fully connected
layers.  It mimics the calculation of a heuristic kernel density
estimator (KDE) that is a one-dimensional representation of where
tracks intersect each other.  The second is a convolutional neural
network.  It starts with KDEs and produces target histograms.
Several CNN architectures produce statistically indistinguishable
results.  The preferred architecture is resembles the U-Net
architecture~\cite{https://doi.org/10.48550/arxiv.1505.04597}
developed for image segmentation in medical applications.  The {\tt
KDE-to-histograms} algorithm has been adapted to find PVs from
simulated ATLAS data, and initial results are promising.  Extending
the LHCb algorithm to find secondary vertices and understanding how
to best adapt the algorithm for use by ATLAS and CMS is a promising
avenue for IRIS-HEP.

LHCb has deployed an older version of its {\tt KDE-to-histograms}
algorithm in its CPU software stack.  Work is in progress now to
deploy the full {\tt end-to-end} algorithm inside {\tt Allen}.
NVIDIA's newest GPUs feature {\tt tensor cores} in addition to {\tt
CUDA cores}.  The former are optimized for the matrix multiplications
that characterize DNN inference engines.  On an RTX-A6000, they
nominally provide 310 TFLOPS performance compared to the  {\tt CUDA
cores}' 39 TFLOPS single precision performance.  The {\tt tensor
cores} are not yet used by {\tt Allen}.  If the {\tt end-to-end}
inference engine can be run using the {\tt tensor cores}, this will
provide much better use of the silicon.  A future direction for
research will be studying how different DNN architectures most
effectively use {\tt tensor cores}.  As many architectures can
provide the same physics performance, identifying those that provide
the best bang per buck will be important as we (HEP computing) think
about how to replace other types of heuristic algorithms with DNNs.

A natural follow-on to developing DNNs to replace heuristic PV-finding
algorithms is developing DNNs to replace aspects of Kalman filters.
In an oversimplified picture, Kalman filters have three functions:
(i) they describe trajectories, (ii) they describe the covariance
matrices along the trajectories, and (iii) they provide quality
metrics describing how well the hits used to construct the trajectories
match the projected trajectories.  In the same way that the algorithm
for finding PVs was broken into two steps using domain expertise,
the Kalman filter problem will need to be broken into steps and
each step addressed separately before the pieces can be put back
together again.  DNNs do not need to provide better physics performance
than heuristic algorithms -- they need to execute more quickly by
taking better advantage of the silicon available in GPUs.  As the
number of tracks per event increases during HL-LHC, this will become
more and more important.  While the details would change from one
experiment to another, this is another area where a cross-experiment
effort could produce a significant advance.

The work described in this section overlaps strongly with the work
described in Sections~\ref{sec:recotrigger} and~\ref{sec:ai}.  The
{\tt PyConf} work discussed in this section may help address some
of the issues related to the increasing diversity of hardware
resources in the WLCG discussed in Section~\ref{sec:facilities}.

\newpage

%% file: 300-grand-challenges.tex
\section{Grand Challenges for the HL-LHC}
\label{sec:grand-challenges}

Starting in 2020, IRIS-HEP has introduced the concept of a ``Grand Challenge" within the institute to focus one or more areas on a long-term, large-scale goal part of the institute's overall vision.  A Grand Challenge differs from a more traditional milestone or deliverable by its scale (often requiring cross-cutting teams working together), a multi-year timeframe, and the fact the entirety of the approach may not be known upfront.  The challenges are executed through a series of increasingly difficult exercises coordinated throughout the community.  The currently defined grand challenges are:

\begin{itemize}
\item {\bf Analysis Grand Challenge (AGC)}: The AGC aims to execute 
realistic analyses at the scale and complexity envisioned by the HL-LHC 
using a set of tools, facilities, and services developed within IRIS-HEP 
as an exemplar. The AGC team coordinates an annual workshop to 
demonstrate current progress against the goals and to update the vision 
and approach as necessary.
\item {\bf Data Grand Challenge (DGC)}: The DGC for IRIS-HEP is realized 
as a set of global data challenges coordinated with the WLCG.  These 
challenges, occurring biennially, bring the entire global community 
together to demonstrate aggregate transfer data rates and compare what 
is currently achievable with the HL-LHC roadmap. These challenges 
also provide an opportunity for integrating new technologies being 
worked on by DOMA into the production infrastructure.
\item {\bf Training Grand Challenge (TGC)}: To tackle the challenges of 
the HL-LHC, we need a workforce with broad software knowledge, spanning 
from basic programming skills to highly specialized training. The TGC 
defines a roadmap to efficiently scale up training activities and 
provide adequate training to create the software-skilled workforce
that will realize HL-LHC science.
\end{itemize}

Not every activity within the Institute fits into the Grand Challenge 
format. For example, the data reconstruction and tracking activities of 
each experiment are coordinated independently and, based on the differences 
in experiment timelines and production code, is not conducive to 
joint exercises. Instead, progress is typically tracked through 
improved relative performance and delivery to the experiments' 
production frameworks.

\input{301-analysis-grand-challenge.tex}
\input{302-data-grand-challenge.tex}
\input{303-reco-grand-challenge.tex}
\input{305-training-grand-challenge.tex}

\tempnewpage

%% file: 301-analysis-grand-challenge.tex
\subsection{Analysis Grand Challenge} 
\label{sec:agc}



\begin{figure}[h]
\begin{tcolorbox}
Physics analysis pipelines make use of a large number of tools and services — the Analysis Grand Challenge provides the mandatory end-to-end tests capturing the full complexity of workflows to ensure readiness of the stack for the HL-LHC.
\end{tcolorbox}
\end{figure}

The HL-LHC will be a challenging analysis environment compared to the LHC experiments of today. The data volumes will go up by a factor ${\sim}100$ and, to reach the desired physics reach of results, analysts will need new techniques and approaches as discussed in Section \ref{sec:analysis-at-scale}. What might be done on a laptop today with events stored on the local file system will need to be done on a dedicated facility in the HL-LHC era, leveraging large-scale computing hardware, advanced data delivery services, and ML training and inference environments. As described in Section~\ref{sec:analysis-systems}, new tools and services are being actively developed to address these challenges.

\begin{figure}
\centering
\includegraphics[width=0.98\linewidth]{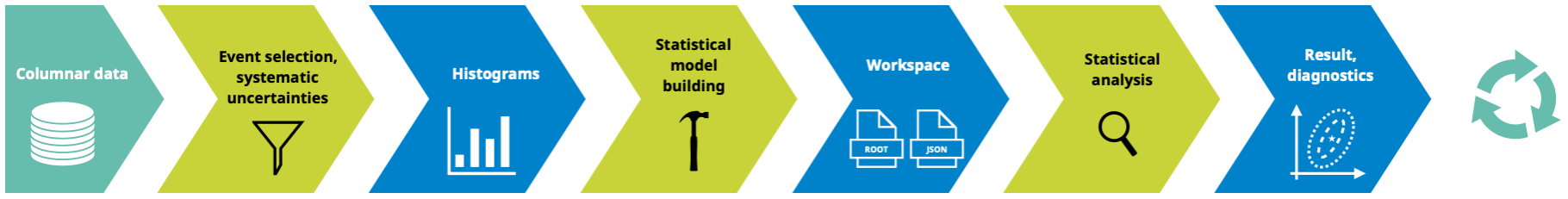}
\caption{Data analysis pipeline for the AGC. This is a simplified view of the Analysis Systems pipeline in figure~\ref{fig:analysis-pipeline-vertical-slice}.} 
\label{fig:agc-pipeline}
\end{figure}

Even if all the tools and services needed existed independently, their full potential can only be realized through integration into a coherent stack tuned to run on an Analysis Facility. The Analysis Grand Challenge (AGC) is designed to function as a series of escalating integration exercises with the ultimate goal of showcasing HL-LHC analyses at full scale and complexity. AGC brings together tools from the AS, DOMA, and Facilities R\&D areas (and elsewhere) and defines physics analysis tasks, representative of the requirements of the HL-LHC, which have to be addressed with an end-to-end analysis pipeline that requires all the ingredients to work together effectively as a system. Figure~\ref{fig:agc-pipeline} depicts the steps in analysis and by extension the components in the software stack.

The AGC connects not only to areas inside the Institute but also connects to the broader surrounding ecosystem; for example, the Coffea analysis framework used by the AGC has originally been developed by Fermilab scientists. The tools being built within and outside IRIS-HEP are designed as small pluggable libraries and steps in a toolchain rather than a framework. The AGC takes advantage of this by pulling in other tools where they are already developed.

A goal of IRIS-HEP is to help democratize science: anyone, associated with an experiment or not, can re-implement analysis pipelines with the tools and workflows of their choice. This allows the AGC to serve as a central gathering point for a larger community focused on end-user analysis in HEP. All the AGC demos and workbooks are openly accessible, and we strive to have as many components as possible based on publicly available Open Data so that anyone can try them out. At the same time, realism requires some demos are developed that run with the experiment's data. 


The AGC will be executed as a series of annual 3-day workshops; each workshop will have a set of technical capability and scale goals set by the AGC leads. The exercises will be cumulative, building on the scale, complexity, and realism that was shown the prior year. The typical workshop will have one day of demonstrations from the IRIS-HEP team, one day of tutorials (splitting into ATLAS / CMS sessions if needed to talk about proprietary analysis techniques), and ending with a day of planning where the goals for the next workshops are set and updated.

Following the success of the AGC during the first phase of IRIS-HEP, there are opportunities for extending these exercises through the start of the HL-LHC. We expect that the AGC will continue to provide a platform for coordinated efforts and to organize information exchanges between different areas, programs, and other entities (e.g., ROOT project, CERN IT, Coffea). The AGC will function as an integration exercise and testbed for the latest developments, leading the analysis approaches into the HL-LHC era.

\subsubsection*{Existing work on the AGC}

During the first phase of IRIS-HEP, the AGC selected as a pathfinder analysis a relatively-simple physics analysis task based on a $t\bar{t}$ cross-section measurement with CMS Open Data. IRIS-HEP implemented analysis pipelines heavily based on the Python data science ecosystem (including and complemented with analysis tools developed by the Analysis Systems area). These made use of novel data delivery methods such as ServiceX and exploited new services in analysis facilities such as Coffea-Casa. Starting from centrally-provided Open Data samples at CERN, the AGC implements the common workflow items relevant to data analysis performed by physicists for this analysis: data extraction and filtering, object calibration and evaluation of systematic uncertainties, construction of observables, histogramming, construction of statistical models and inference, and visualizations to study and disseminate analysis details and results.

IRIS-HEP demonstrated the feasibility of its analysis systems approach at dedicated annual AGC workshops \cite{agctools1-indico, agctools2-indico} and through contributions to relevant HEP conferences, such as International Conference on High Energy Physics (ICHEP 2022) \cite{Held:2022RC} and International Workshop on Advanced Computing and Analysis Techniques in Physics Research (ACAT 2022)  \cite{agc-acat2022}.

Annual AGC workshops dedicated to showcasing tools and workflows related to the AGC analysis pipeline included:
\begin{itemize}
    \item \textbf{AGC Tools 2021 Workshop I}: The idea for the first AGC workshop was to introduce to end-users the various tools and services, as the components of an AGC analysis pipeline. These tools were developed in the Python ecosystem by and for the particle physics community as well and deployed on related cyberinfrastructure to be executed at scale.  The workshop agenda also included specific experiment sessions to collect early feedback about pipeline and deployment from CMS and ATLAS analysts.
    \item \textbf{AGC Tools 2022 Workshop II}: The agenda of the second AGC workshop featured detailed hands-on tutorials based on AGC tools as well as providing the possibility for participants to follow all tutorials on dedicated Analysis Facilities provided by SSL (Coffea-Casa analysis facilities at University of Nebraska-Lincoln and University of Chicago).
\end{itemize}

These forums allowed the team to discuss the physics analysis workflow vision developed in the AGC with the community and to demonstrate the pipelines in practice. Tests of pipelines at scale at various analysis facilities  revealed performance bottlenecks and usability issues which have subsequently been  addressed. Different analysis pipeline implementations have been contributed, including a version using ROOT’s RDataFrame, allowing the comparison of possible approaches available to users for physics analysis.

\subsubsection*{Next Phase of the AGC}

We have found the annual workshops showing increasing scale, complexity, and realism of HL-LHC analysis to be a useful structure for accomplishing the AGC's goals and plan to continue these in the future. This subsection provides an overview for how the AGC plans to integate with various areas internally followed by the plans and goals for the next phase of IRIS-HEP.

Given the wide variety of physics analyses within the LHC community, for the next phase of IRIS-HEP we have selected two HL-LHC analyses as flagship examples to complement the current pathfinder $t\bar{t}$ cross-section analysis.

\begin{itemize}
    \item \textbf{Top quark mass}: The top quark mass measurement in the $t\overline{t} \rightarrow {\rm lepton+jets}$ channel is expected to require an extremely large analysis dataset due to the significant backgrounds in this channel.  This will serve as a {\bf high-volume} analysis that pushes the system's limits in terms of event rate.
    \item \textbf{Di-Higgs search}: The di-Higgs analysis program at the LHC probes the Higgs self-coupling, which is believed to be {\em just} within the reach of the HL-LHC dataset. Maximizing sensitivity to the trilinear Higgs coupling implies the use of the latest available ML and statistics techniques; the software developed by the AS area is meant to enable these approaches. This will serve as a {\bf high-complexity} analysis example to bring together these tools.
\end{itemize}

\paragraph{\textbf{Connections to Analysis Systems}}

Unsurprisingly, there are many deep connections between the AGC and the Analysis Systems (AS) area; indeed, the AS team is the primary provider of tooling for the AGC exercises.  Community interactions during AGC workshops proved to be a useful way to understand user experiences for analysis work.  Based on these experiences and the remaining incomplete items from the overall vision, topics of work planned for the next phase of IRIS-HEP include:

\begin{itemize}

\item \textbf{Handling Systematic Uncertainties}: The handling of systematic uncertainties in an analysis pipeline was outlined as the top pain point for end-users in the Analysis Ecosystem Workshop II workshop report \cite{Stewart:AnalysisEcosystem}. The AGC analysis tasks capture all relevant types of systematic uncertainties and the associated metadata handling to probe the associated user experience. The top quark mass measurement in particular is an ideal environment for this: dominant contributions limiting the sensitivity come from sources of systematic uncertainty, and a rich set of different types of uncertainties needs to be evaluated.

\item \textbf{ML Services}: Integration of machine learning into the AGC framework has been a frequently requested item in community interactions. ML is increasingly commonplace in physics analysis but users often have to split ML training into a separate pipeline. The AGC can be used to investigate workflows that streamline the user experience, including identifying  services facilities should provide to help with this.

\item \textbf{Differentiable Analysis}: A differentiable analysis pipeline would allow for end-to-end analysis optimization via gradient descent, increasing the overall physics reach of the analysis. To accomplish this goal, all the different parts of the analysis pipeline need to support evaluating and passing gradient information along. The AGC can be used to track the progress towards increasing the number of pieces that provide the required technical capability and to test the functionality in a realistic environment.

\end{itemize}

\paragraph{\textbf{Connections to DOMA}}
The DOMA team contributes important data delivery services (capabilities) and expertise on scaling up data rates on the facilities used by the AGC (capacity).  Topics that will be emphasized for the next phase include:

\begin{itemize}


\item \textbf{Data Delivery}: The data extraction and data delivery edge service, ServiceX, is providing data access functionality for the pipeline. It allows users to filter events, add additional columns to an event (for example, CMS NanoAOD) on-demand, and have the result placed at the facility.  The target workflow for the next phase will be to allow users to select columns from NanoAODs, using the desired selection to skim and extract a column from MiniAOD files centrally, and finally retrieving the new column for the selection and integrating it back into the existing NanoAOD workflow. An analogous capability is planned for ATLAS datasets as well.

\item \textbf{Data Management}: A critical capability for HL-LHC analysis will be to manipulate, augment, and subset data in-place without having to create a derived copy of the dataset (as is frequently done for the LHC). As discussed in Section \ref{sec:doma}, the next phase of the Skyhook project will be to allow joining together data from derived columns (in formats such as NanoAOD or PHYSLITE) used by analyzers with new generated columns or columns derived from the reference dataset (in formats such as MiniAOD or PHYS) as delivered by ServiceX.

\end{itemize}

\paragraph{\textbf{Connections to Facilities R\&D}}
The third key component for the success of AGC is the investigation of modern cyberinfrastructure approaches that can compose these services into a facility, and to allow duplicating it to several locations. The AGC will work closely with the Facilities R\&D area and other facilities, such as Fermilab and BNL, to provide a complex ``in-house" use case demonstrating the needs of a modern facility.

\begin{itemize}

\item {\bf Coffea-Casa Facility}: The prototype facility heavily used by the AGC is the Coffea-Casa analysis facility, developed as a cross-cutting project in IRIS-HEP that brings new, interactive paradigms for users from R\&D into production.  This facility is used as a testbed, offering the possibility for end-users to execute analysis at HL-LHC-scale data rates. It uses an approach that allows it to transform existing facilities (e.g., U.S. LHC T2 sites) into composable systems using Kubernetes as the enabling technology.

\item \textbf{Scale}: We will extend the AGC input datasets to 200 TB to enable testing the I/O performance of analysis pipelines at a scale that is interesting to the cyberinfrastructure (this is about 20x larger than what is commonly used today for end-stage analysis). One component of each workshop will be to achieve certain processing time targets with this input dataset. Between workshops, the AGC will perform periodic benchmarking exercises on various partner facilities to ensure that no regressions appear when executing the pipeline at scale and with larger computation complexity.

\item \textbf{ML Services}: The AGC will work to integrate emergent functionalities and services required by the analysis community such as making Machine Learning (ML) more readily accessible to users.  We aim to include a ML inference server (one of the existing examples is SONIC \cite{krupa2021gpu}), built on NVIDIA's Triton Inference server, as a way to have transparent access to high-speed, GPU-based inference as part of user applications. The environment will be extended to include tools like MLFlow to help tracking user experiments with AI training.  Where possible, the focus will be to integrate national-scale AI resources instead of only those internal to HEP facilities.
\end{itemize}

\paragraph{\textbf{AGC Targets}}

The AGC's annual workshop format allows us to have specific targets toward a full-scale HL-LHC analysis. The first AGC workshop of a possible next phase to IRIS-HEP would be in 2024; while there's an understanding of the volume and scale targets throughout the entire phase, it is more difficult to map out the precise ordering of new technologies this timescale. Table \ref{tab:agc} outlines the possible goals and targets for the AGC.

\begin{table}[ht]
    \centering
    \begin{tabular}{|c|p{0.8\linewidth}|}
        \hline
        \rowcolor{msdarkblue}
        \textcolor{white}{\bf Year} & \textcolor{white}{\bf Target} \\ \hline
\rowcolor{verylightgray}
             & $\bullet$ Define analysis tasks for the top quark mass and di-Higgs measurement. \\
\rowcolor{verylightgray}
2024 & $\bullet$ High-volume analysis done on dataset 20\% the scale needed for HL-LHC and completed within 1 hour. \\
\rowcolor{verylightgray}
& $\bullet$ Integrate ML inference service with AGC. \\
        \hline
\rowcolor{mslightblue}
2025 & $\bullet$ High-volume analysis done on dataset 40\% the scale needed for HL-LHC and completed within 1 hour. \\
\rowcolor{mslightblue}
& $\bullet$ Demonstrate AOD column extraction workflow \\
        \hline
\rowcolor{verylightgray}
        2026 & $\bullet$ High-volume analysis done on dataset 60\% the scale needed for HL-LHC and completed within 1 hour. \\
\rowcolor{verylightgray}
             & $\bullet$ Demonstrate fully differentiable analysis \\
        \hline
\rowcolor{mslightblue}
2027 & $\bullet$ High-volume analysis done on dataset 80\% the scale needed for HL-LHC and completed within 1 hour. \\
        \hline
\rowcolor{verylightgray}
        2028 & $\bullet$ High-volume analysis done on dataset 100\% the scale needed for HL-LHC and completed within 1 hour. \\
        \hline
    \end{tabular}
    \caption{Targets by year for the AGC in a five-year timespan.}
    \label{tab:agc}
\end{table}

\newpage

%% file: 302-data-grand-challenge.tex
\subsection{Data Grand Challenge}
\label{sec:domachallenge}

\begin{figure}[h]
\begin{tcolorbox}
The data challenges are the global community's mechanism to prepare the cyberinfrastructure for the HL-LHC data rates and a mechanism for IRIS-HEP to ready new technologies for the production infrastructure. 
\end{tcolorbox}
\end{figure}

HL-LHC has multiple challenges around data organization, management, and access.  Perhaps too often, the ‘data issue’ in the HL-LHC focuses on the raw volume of data; however, it’s insufficient to simply purchase a certain volume of hard drives.  The problems also include \textit{data velocity} – raw data must be moved from the detector to distributed sites, from simulation site to archival, and from buffers to HPC resources - and data management.  For coordination of the distributed cyberinfrastructure, the data velocity is the more vexing issue as the data must be moved over shared network resources (implying coordination and resource management with multiple providers) and the technology stacks select must be interoperable with a global set of facilities.

Considering the datasets involved, the velocity – the global aggregate terabits per second – is also a scaling issue.  The HL-LHC community will need to \textbf{sustain data rates 20 times larger} than the baseline set by LHC Run 3.

Simply put, the LHC community is not ready for the HL-LHC rates and the preparation is not solely a function of buying switches and fiber immediately before the accelerator turns on.  Instead, a robust R\&D program is needed in this area as outlined in Section \ref{sec:doma}, introducing new technologies into the software stack and demonstrating components’ readiness for HL-LHC data rates.

To help integrate the global R\&D efforts and show progress toward the ultimate HL-LHC goals, the LHC community has defined a Data Grand Challenge, a series of biennial exercises for data movement.  These exercises have global sustained data movement targets as outlined in Table \ref{tab:dgctargets}.  Each exercise has two targets – the “minimal scenario” and “flexible scenario”.  The minimal scenario is derived from the input parameters from the computing models – number of events expected per year, size of each event format, and the minimum number of times a dataset needs to be transferred or processing.  The flexible scenario is based on the experience of how LHC operates – data transfers are bursty and not flat over the year, we may need to move data to HPC resources to take advantage of large time-limited allocations, and data sometimes needs to be transferred for processing multiple times to account for software errors.

\begin{table}[ht]
\centering
\begin{tabular}{| c | r | r | r |}
\hline
\rowcolor{msdarkblue}
\textcolor{white}{\bf Year}    & \textcolor{white}{\bf Percent of}  & \textcolor{white}{\bf Minimal Scenario} & \textcolor{white}{\bf Flexible Scenario}  \\
\rowcolor{msdarkblue}
 & \textcolor{white}{\bf HL-LHC Scale} & \textcolor{white}{\bf Agg. Targets (Gbps)} & \textcolor{white}{\bf Agg. Targets (Gbps)} \\
\hline
\rowcolor{verylightgray}
2021	& 10\%	& 480	& 960 \\
\rowcolor{mslightblue}
2023	& 30\%	& 1,440	& 2,880 \\
\rowcolor{verylightgray}
2025	& 60\%	& 2,880	& 5,760 \\
\rowcolor{mslightblue}
2027	& 100\%	& 4,800	& 9,600 \\
\hline
\end{tabular}
\caption{The aggregate global targets for the biennial data challenges as presented to the WLCG Management Board in February 2021 \cite{Campana:DataChallengeOutline}.}
\label{tab:dgctargets}
\end{table}

\begin{table}[htbp]
\centering
\begin{tabular}{|c|r|r|}
\hline
\rowcolor{msdarkblue}
\textcolor{white}{\bf LHC Tier-1 Site}	& \textcolor{white}{\bf HL-LHC Network Needs (Gbps)} & \textcolor{white}{\bf HL-LHC Network Needs (Gbps)} \\
\rowcolor{msdarkblue}
 & \textcolor{white}{\bf Minimal Scenario}	& \textcolor{white}{\bf Flexible Scenario} \\
\hline
\rowcolor{verylightgray}
CA-TRIUMF	& 200	& 400 \\
\rowcolor{mslightblue}
DE-KIT	& 600	& 1,200 \\
\rowcolor{verylightgray}
ES-PIC	& 200	& 400 \\
\rowcolor{mslightblue}
FR-CCIN2P3	& 570	& 1,140 \\
\rowcolor{verylightgray}
IT-INFN-CNAF	& 60	& 1,380 \\
\rowcolor{mslightblue}
KR-KISTI-GSDC	& 50	& 100 \\
\rowcolor{verylightgray}
NDGF	& 140	& 280 \\
\rowcolor{mslightblue}
NL-T1	& 180	& 360 \\
\rowcolor{verylightgray}
NRC-KI-T1	& 120	& 240 \\
\rowcolor{mslightblue}
UK-T1-RAL	& 610	& 1,220 \\
\rowcolor{verylightgray}
RU-JINR-T1	& 200	& 400 \\
\rowcolor{mslightblue}
US-T1-BNL	& 450	& 900 \\
\rowcolor{verylightgray}
US-FNAL-CMS	& 800	& 1,600 \\
\rowcolor{mslightblue}
(Atlantic Link)	& 1,250	& 2,500 \\
\hline
\rowcolor{msdarkblue}
\textcolor{white}{\bf SUM}	& \textcolor{white}{\bf 4,810}	& \textcolor{white}{\bf 9,620} \\
\hline
\end{tabular}
\caption{The breakdown of expected data rates per site as presented to the WLCG Management Board in February 2021 \cite{Campana:DataChallengeOutline} for the purposes of planning the data challenge exercises.}
\label{tab:dgc-t1targets}
\end{table}

Each experiment’s draft computing model provides a rough partitioning of data to different computing sites for HL-LHC; this is reproduced in Table \ref{tab:dgc-t1targets}.  The global experiments’ plans do not forecast at the Tier-2 level but the U.S. LHC programs have also participated in a network capacity planning exercise with ESNet \cite{Zurawski:ESNetReq} which additionally concluded that a \textbf{nominal Tier-2 site will need 400Gbps capacity} for HL-LHC.

\subsubsection*{Existing work on the DGC}

The inaugural data challenge, DC21, had a target of 10\% HL-LHC scale.  This target is similar to the current production scale for LHC Run 3 meaning the challenge largely focused on (a) building a reusable infrastructure necessary for running this and future data challenges, (b) demonstrating the ability of the global community to collaborate on such a challenge, and (c) prove new technologies at production scale.

There were two important new technologies for DC21:
\begin{itemize}
    \item \textbf{HTTP-TPC}: A majority of transfers were made with WebDAV-based HTTP-TPC protocol.  HTTP-TPC is an agreed-upon interpretation of the WebDAV protocol that is built on industry-standard HTTPS.  After the data challenge, the experiments officially committed to using HTTP-TPC in production for LHC Run 3; this represents the first upgrade in transport protocols in two decades and the only change in protocol since the start off LHC Run 1.  Using HTTPS as a base reflects evolutions in the wider industry, and keeps the LHC community on a \textbf{sustainable path for the future}.
    \item \textbf{SRM/HTTP}: The tape / archive management protocol used by the WLCG is Storage Resource Management (SRM) protocol.  SRM utilizes a proprietary TLS-like transport layer with SOAP as the RPC layer; neither technology is widely used in modern environments.  Further, SRM is based upon a model of remote space management that has been discarded by the WLCG – only the small subset of the protocol (tape management) is used.  The LHC cyberinfrastructure is in the middle of replacing this dead-end technology; the first step was during DC21 where, instead of layering SRM on top of GridFTP, HTTP-TPC was used.
\end{itemize}

DC21 successfully executed all its goals.  The exercise was held in October 2021, performed using the same Rucio / FTS software stack used in production by ATLAS and CMS, and hit the target data rates for both the minimal and flexible scenarios.\\

\begin{figure}[htbp]
\centering
\includegraphics[width=0.70\textwidth]{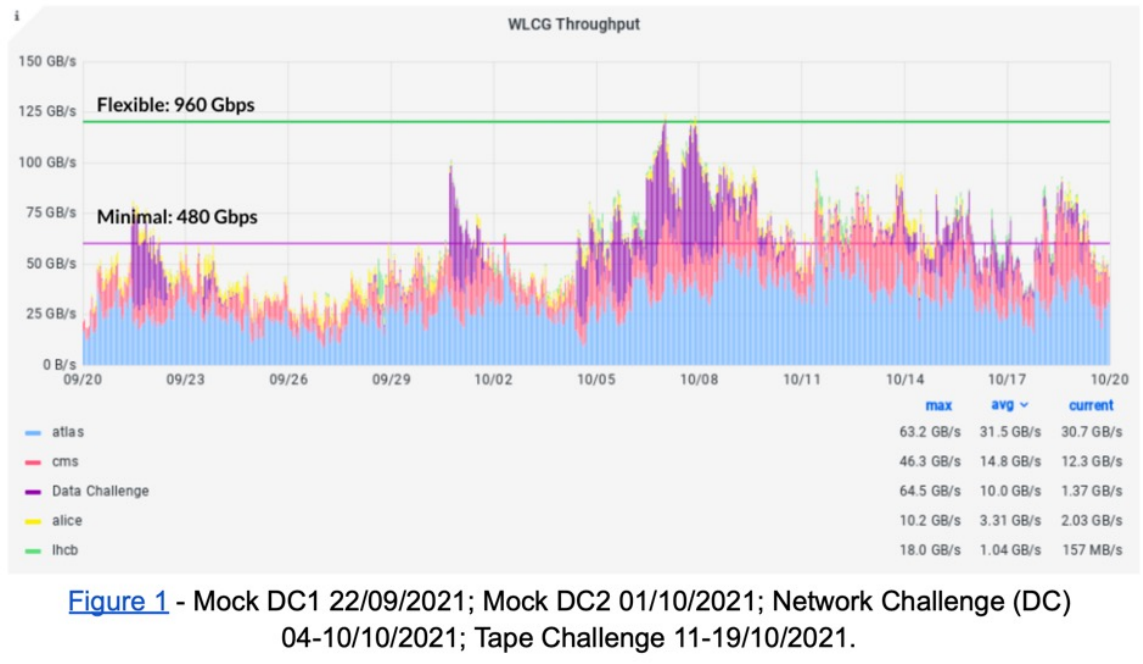}
\caption{The hourly average data rates achieved by successful transfers during DC21.  Figure is reproduced from the report at \cite{Forti:DC21}.}
\label{fig:}
\end{figure}

\subsubsection*{Next Phase of the DGC}

In the next phase of IRIS-HEP, we will participate in the next four data challenges.  Below, we outline the known plans for the next challenge - DC23 (due to scheduling conflicts, currently planned for March 2024) - and subsequent exercise. \\

\noindent \textbf{DC23}: DC21 validated the approach of executing coordinated data movement exercises for the Data Grand Challenge.  The next data challenge, DC23, is more difficult as it has both a robust ``new technology'' component and serious scale challenges, tripling the target data rates from 10\% to 30\% of HL-LHC scale (the largest relative increase in the DGC program).  Planned new technologies and capabilities for DC23 include:

\begin{itemize}
\item \textit{Dynamic engineered network paths}: Using technologies provided by ESNet, show the ability to migrate specific flows to dynamically engineered paths, associating the traffic type with a specific path enabling the ability to do differentiation in the future.
\item \textit{New authorization paradigm}: The U.S. LHC computing facilities have already transitioned from identity-mapping based authorization to capability-based, providing a fine-grained approach to authorization.  The equivalent change for data transfers is underway and expected to be a cornerstone for DC23.
\item \textit{Emphasis on IPv6}: IPv6 is becoming widely deployed across the WLCG and is a prerequisite for new network technologies.  Its header extensions are used for adding metadata to network flows, enabling improved monitoring of flows outside of engineered network paths.  We aim to have a majority of transfers use IPv6 during DC23.
\end{itemize}

Additionally, the U.S. LHC community is investigating regional mini-challenges to ensure that its facilities are prepared for the increased data rates.  For more information on the technologies involved, see Section \ref{sec:doma}. \\

\noindent \textbf{Post-DC23}: The precise set of technologies that will be ready to integrate into production in 2025 and 2027 are less clear.  Given DC21 and DC23 represent a generational change of transfer protocols and authorization approaches, the expected focus for DC25 and DC27 will be increasing the scale of the production system.  Software scaling tests, with some efforts starting already, will be critical to get the storage endpoints ready for these challenges; DC25’s target data rates will be an order of magnitude larger than today’s production traffic.  Post-DC23, production facilities will start the purchase ramp-up of hardware, allowing the community to rely more on production endpoints over testbed facilities.


%% file: 303-reco-grand-challenge.tex


%% file: 305-training-grand-challenge.tex
\subsection{Training Grand Challenge}
\label{sec:tgc}

The science of the HL-LHC will be highly software-enabled. 
Maximizing the physics output requires that
community members from all career levels have a solid base
of knowledge in order to leverage the software tools being
developed. Even for small group collaborations or single individuals developing a physics 
analysis, software literacy is fundamental. 
For this category of ``Cyberinfrastructure Users'', training
should include basic programming, version control, best practices 
of state-of-the-art software engineering, etc. A large subset of the 
community will then also be involved in developing software tools that will 
be used by their collaborators and perhaps contributing to the community
ecosystem (``Cyberinfrastructure Contributors''). These people will likely need more advanced 
skills and some of them will become ``Cyberinfrastructure Professionals'' within our
community. Section~\ref{sec:training} has detailed our
approach to scaling up software training activities to match the
large size of the HL-LHC workforce.

The Training Grand Challenge defines a series of goals to guide 
our efforts. We consider the following three areas of impact:

\begin{itemize}

\item
\textbf{Democratizing science with standardized prerequisites
(done):} A series of standardized training modules covering all
basic software prerequisites allows students to close any gaps in
their knowledge before starting experiment onboarding.

\item
\textbf{Scaling up training for intermediate and advanced topics:} 
Regular intermediate- to advanced-level courses and workshops need
to be organized and brought to the attention of both students and
their supervisors.  At the same time, all material should be
self-study friendly and easily discoverable in a way that does not
overwhelm students.  To that end, our training center should be
expanded to list all relevant and tested training resources across
the field.

\item
\textbf{Establishing career paths for developers:}
The career path toward cyberinfrastructure professionals or software-focused permanent positions
must be firmly established.  Specific workshops and mentoring
opportunities for students focusing on software engineering skills
need to be offered.
\end{itemize}

To achieve these goals in an efficient and sustainable fashion, we also define the following focus areas:

\begin{itemize}

\item
\textbf{Building a community of educators:}
We need to grow the community of cross-experiment educators that
has been created by the IRIS-HEP/HSF Training group. This includes
further improving the recognition and visibility of all contributors.
Regular talks at major conferences and publications in journals
such as the Journal of Open Source Education (JOSE) are important
parts of this strategy.  We should also expand our collaboration
with experiments and similar initiatives focusing on open-source
training, particularly the new initiatives centered around cyberinfrastructure for research.
\item 
\textbf{Training material sustainability:}
The community should firmly push to publish training material in
editable and community-maintainable formats with open-source
licenses. If traditional presentation software (PowerPoint, Keynote,
Impress) is used, this means publishing corresponding editable
files. However, the use of source file-based systems (e.g.,
presentations built from LaTeX files, Markdown files, or Jupyter
notebooks) should be encouraged.
Workshops should encourage educators to follow these guidelines
and, in particular, pay attention to licensing.
\end{itemize}

A series of milestones and an estimated timeline to achieve these
goals is shown in Table~\ref{tab:training-roadmap}.

\begin{table}[th]
\centering
\begin{tabular}{|c|p{13cm}|}
\hline
\rowcolor{msdarkblue}
\textcolor{white}{\bf Year} & \textcolor{white}{\bf Target} \\ \hline
\rowcolor{verylightgray}
2018 -- 2022& 
$\bullet$ Develop and teach standardized training modules covering essential software prerequisites (completed). \newline
$\bullet$ Establish and grow a training group that coordinates and sustains cross-experiment training efforts (completed).
\\ \hline
\rowcolor{mslightblue}
2023 & 
$\bullet$ Seed topical subgroups that create new intermediate/advanced training material based on the State of Training 2022 survey.\newline
$\bullet$ Rebuild and expand the Training Center to become a focal point of \emph{all} software training resources in HEP.\newline
$\bullet$ Strengthen collaborations with experiments and new organizations that support the career path of CI Professionals.
\\ \hline
\rowcolor{verylightgray}
2024 & The first intermediate-advanced training modules are being taught.\\ \hline
\rowcolor{mslightblue}
2025 & 
$\bullet$ 80\% of all cross-experiment software topics that apply to Ph.D. students should be covered by standardized training modules.\newline
$\bullet$ 90\% of HEP Ph.D. students should be aware of the material offered by the IRIS-HEP/HSF Training group.\newline
$\bullet$ 50\% of HEP Ph.D. students should participate in at least one intermediate/advanced training.
\\ \hline
\rowcolor{verylightgray}
2026 - 2028 & 
$\bullet$ Additional focus on workshops, networking opportunities, and discussion platforms for aspiring developers and CI professionals within HEP.\newline
$\bullet$ 20\% of Ph.D. students and postdocs should be enrolled in software-related communities or have attended advanced workshops.\newline
$\bullet$ 20\% of HEP Ph.D. students should teach or support a software workshop at one point during their Ph.D.
\\ \hline
\end{tabular}
\caption{Training activities and Training Grand Challenge goals to prepare the HL-LHC workforce.}
\label{tab:training-roadmap}
\end{table}

\tempnewpage

%% file: 320-broadening-participation.tex
\section{Outreach and Broadening Participation}
\label{sec:broadeningparticipation}

The participation of women and ethnic minorities is generally low
in the HEP world, and 
fractionally it is even lower in the HEP Software and Computing (S\&C) 
world.
We estimate that fewer than 10\% of people in HEP S\&C are women 
while (from LHC
experiment statistics) between 13\% and 20\%
of the LHC experiments' collaborators are women. 
In the U.S., 7.4\% of high-tech employment is black~\cite{GuardianSV}, 
while in HEP S\&C the fraction is negligibly small. Other measures
of HEP and HEP S\&C diversity in the U.S.\ are likely to diverge 
in similar fashion from society at large.
Looking forward, increasing the diversity of the HEP S\&C
workforce promises two types of benefits.
From first principles, the top 5\% of a larger pool should 
always be better than
the top 10\% of a pool half as large.
In addition, studies show~\cite{diversity1,diversity2,diversity3} that 
teams of people from diverse backgrounds are more innovative when 
crafting solutions to complex problems and can make better and 
more profitable  decisions.

Minimally, IRIS-HEP must be (and is!) sensitive to diversity in building 
its own team, but that is not enough.
Through its own directly-funded team IRIS-HEP alone cannot however 
significantly increase the fraction of under-represented populations
in the larger HEP community or even in HEP S\&C broadly. Even an NSF 
institute-class project like IRIS-HEP is too small a player by itself. 
What a project of this size can do, however, is help build the 
larger pipeline, and carefully align its efforts to be maximally 
effective by partnering with other institutions and projects actively 
working in the same direction. 

In practice this means focusing on aspects of the program which engage
people earlier in their academic careers. Most of the directly
funded IRIS-HEP team is faculty, staff, postdocs or graduate students. At 
the high school level, IRIS-HEP has piloted at UPR-Mayagüez in Puerto Rico 
outreach  activities with high school teachers around coding, machine learning and 
physics. By focusing on enabling teachers, we can multiply our impact.
This is being expanded nationally via partnership with the QuarkNet~\cite{quarknet} 
program which also engages high school teachers, including those connected 
with diverse groups of students. In particular IRIS-HEP has partnered 
post-COVID with QuarkNet in 2022 to develop a ``Coding Camp 2'' for high 
school teachers, which was piloted in summer 2022, at Fermilab. This adds to 
existing QuarkNet physics and software curriculum. In 2022-2023 we will
leverage QuarkNet's existing national network of connections and
local outreach programs at IRIS-HEP institutions to run 6 
additional Coding Camp workshops for teachers in various locations. 
The overall aim is to reach nearly 100 high school teachers per year to help
them develop basic software and physics-related training they
can use in their classrooms. This program will continue in a second phase 
for IRIS-HEP.

At the undergraduate level, IRIS-HEP has developed an extensive training
program (Section~\ref{sec:training}), including in particular a ``Fellows''
program which engages people in software-related physics research 
projects. This allows students to take general software skills they
have learned and apply them to developing research software to
solve problems in particle physics.
Originally the Fellows program was focused on graduate students,
however during COVID it evolved into a program that works primarily with 
undergraduate students. Each undergraduate Fellow works with a mentor
on a (typically summer) research software project relevant for particle
physics. Typically they work at a distance with their mentor, although
some of the Fellows have traveled for a short period to work with
their mentor in-person or to a conference or workshop.

Over the past couple of years IRIS-HEP has begun to align its training
(Section~\ref{sec:training} and Fellows mechanisms in ways that can
also contribute to broadening participation. As noted earlier, 
student knowledge of basic software skills is very uneven. Experience
with research software (and research activities in general) is less common
at smaller and non-R1 institutions.

\newpage

%% file: 350-funding-levels.tex
\section{Funding Scenarios}
\label{sec:funding}


In terms of scale, structure, and organization, the next phase of
IRIS-HEP has a clear baseline for comparisons, namely the currently-funded
institute.  As outlined in Section \ref{sec:role}, we envision that the
basic concepts and organizational outlines of the institute as
successful and thus would remain in place for the next phase.  The large
majority (approximately 90\%) of the institute's funding goes to
personnel; major non-personnel costs include the IRIS-HEP Fellows
program, support for workshop, broader outreach activities, and 
one-time hardware purchases
to bootstrap the Analysis Facilities activity.  Table \ref{fig:fte}
shows that for IRIS-HEP year 5, about 90\% of the personnel effort
are in technical areas (either R\&D or operations), 5\% to training
and outreach, and 5\% to project management.  While these precise
splits will differ each year of the next phase, the approximate
breakdowns are expected to remain.

In this strategic plan, three basic funding scenarios are considered:
\textbf{low}, providing \$4 million per year, \textbf{medium},
providing \$5 million per year (the baseline level, the same as the
current IRIS-HEP award), and \textbf{high}, providing \$6 million
per year.  In each scenario, we plan for five years of funding,
supporting IRIS-HEP from the end of the first phase through the end
of the research years running up to the HL-LHC.

In Section \ref{sec:focusareas}, each area has outlined a set of
expected activities.  The activities are further annotated to
indicate the associated funding scenarios.  For example, an activity
marked as \textit{medium scope (reduced activity), high} is expected
to be fully supported in the high funding scenario, have a significantly
reduced scope and scale in the medium funding scenario, and receive
no support in the low funding scenario.  Note the funding scenario
labels to not necessarily correspond to \textit{priority}; for
example, in the DOMA area, the ``Scaling the CI to HL-LHC data
rates" is highest importance but, due to the effort required, would
not be plausible in the low funding scenario (given this is key for
the HL-LHC program as a whole, the Operations programs potentially
would need to take on this activity).  We do not attempt to rank
the activities within each scenario; this level of detail is deferred
for an actual project proposal. \\

A summary of each scenario is as follows:

\begin{itemize}
    \item \textbf{Medium / Baseline Scenario}: In this scenario,
      which would most closely resemble the existing IRIS-HEP, 
      with a scope encompassing the same activities as it
      does today. The Translational AI area (Section \ref{sec:ai})
      would still be organized but it would be of modest size,
      approximately the same size as the aggregate exploratory AI
      R\&D distributed throughout the institute areas.  Areas would
      modestly evolve - for example, DOMA would still target adding
      the new XRootD activity at the cost of small reductions to other
      projects.  The second phase of IRIS-HEP would slowly evolve
      from its current level of research today to focus more on 
      development, deployment and sustainability towards the end of
      an 2nd 5-year mandate for the institute.
    \item \textbf{High Scenario}: In the high funding scenario, the
      new Translational AI area would be fully funded and expected
      to make a significant impact in the field.  Other, more ambitious
      efforts in existing areas would be achievable -- for example,
      the Analysis Systems would be able to make a realistic push
      toward completing the technologies required for a fully
      differentiable analysis in time for HL-LHC.  The Reconstruction
      and Trigger Algorithms area would be able to directly support
      ML-based approaches to tracking and take larger risks in trying
      to close the resource gaps foreseen for HL-LHC.  IRIS-HEP's
      second phase would have a larger mix of research activities
      throughout the project as it could support both research and
      sustainability activities simultaneously.
    \item \textbf{Low Scenario}: In this scenario, a 20\% decrease
      to the budget from the current funding levels (a larger impact
      than 20\% given the cumulative effects of inflation over the
      last five years), IRIS-HEP would not start any new activities
      like XRootD or areas such as Translational AI.  To meet funding
      levels, high-importance activities like the data grand challenge
      or entire areas such as Facilities R\&D may need to be cut or
      combined into larger areas.  The institute would pivot almost
      immediately to focus on completing and sustaining activities.
      Potentially, the institute would not be able to support any of
      the LHCb-related R\&D activities.
\end{itemize}

\newpage

%% file: 810-workshop-list.tex
\section{Appendix - Workshop List}
\label{workshoplist}

\vskip 0.10in
One aspect of the Intellectual Hub role of IRIS-HEP is to help organize and coordinate community-wide meetings; this activity has been ongoing throughout the lifetime of the institute.  During calendar year 2022, meetings IRIS-HEP was involved in or helped defined community needs and requirements include (in chronological order):

\vskip 0.10in
\noindent{\bf Analysis Ecosystems II Workshop} \\
\noindent{\it Date}: 23--25 May 2022 \\
\noindent{\it Location}: Laboratoire de Physique des 2 infinis Irène Joliot-Curie (Orsay, France) \\
\noindent{\it URL}: \url{https://indico.cern.ch/event/1125222/}\\
\noindent{\it Summary report}: \url{https://zenodo.org/record/7003962}\\
\noindent{\it Description}: (From workshop website) It has been five years since the first Analysis Ecosystems Workshop organised by the HSF in 2017. Since that time many changes have happened, with the advent of new projects, tools, and data formats, intense activity and progress in established projects. Still, the challenge of efficient analysis for the HL-LHC era is not yet solved and so the HSF and IRIS-HEP, together with IJCLab, are organising the Second Analysis Ecosystems Workshop.

Topics for the workshop include, amongst others:

\begin{itemize}
\item Analysis Facilities
\item ML tools and differentiable computing workflows
\item “Real-time” trigger-level analysis
\item Analysis User Experience and Declarative Languages
\item Analysis on reduced formats or specialist inputs
\item Bookkeeping and systematics handling
\end{itemize}
\vskip 0.01in

\vskip 0.1in
\noindent{\bf Connecting the Dots 2022} \\
\noindent{\it Date}: 31 May -- 2 June 2022 \\
\noindent{\it Location}: Princeton (Princeton, NJ) \\
\noindent{\it URL}: \url{https://indico.cern.ch/event/1103637/}\\
\noindent{\it Description}: (From workshop website) The Connecting The Dots workshop series brings together experts on track reconstruction and other problems involving pattern recognition in sparsely sampled data. While the main focus  is on High Energy Physics (HEP) detectors, the Connecting The Dots workshop is intended to be inclusive across other scientific disciplines wherever similar problems or solutions arise. 
 \vskip 0.01in

\vskip 0.1in
\noindent{\bf Snowmass Community Summer Study Workshop} \\
\noindent{\it Date}: 17 -- 26 July 2022 \\
\noindent{\it Location}: University of Washington (Seattle, WA) \\
\noindent{\it URL}: \url{http://seattlesnowmass2021.net/}\\
\noindent{\it Description}: (From workshop website) Snowmass21 is a yearlong study hosted by the Division of Particles and Fields (DPF) of the American Physical Society (APS), which takes place approximately every ten years. Its purpose is to define the most important questions for the field of particle physics and identify promising opportunities to address them. The study aims to be community-driven, inclusive, global, transparent, and interdisciplinary.

The Seattle meeting is the culmination of the various workshops and Town Hall meetings that have taken place during 2020, 2021, and 2022 as part of Snowmass21.
 \vskip 0.01in

\vskip 0.1in
\noindent{\bf Second MODE Workshop on Differentiable Programming for Experiment Design} \\
\noindent{\it Date}: 12 -- 16 Sept 2022 \\
\noindent{\it Location}:  OAC conference center (Kolymbari, Crete, Greece) \\
\noindent{\it URL}: \url{https://indico.cern.ch/event/1145124/}\\
\noindent{\it Description}: The MODE (Machine-learning Optimized Design of Experiments) Collaboration aims to advance the idea of using differentiable programming for astro and particle physics.  This workshop included invited keynote speakers, tutorials, and hackathons in the topic area and aims to coordinate the field's approach to differentiable programming.
 \vskip 0.01in

\vskip 0.1in
\noindent{\bf PyHEP 2022 Workshop} \\
\noindent{\it Date}: 12 -- 16 September 2022 \\
\noindent{\it Location}:  (Virtual) \\
\noindent{\it URL}: \url{https://indico.cern.ch/event/1150631}\\
\noindent{\it Description}: (From workshop website) The PyHEP workshops are a series of workshops initiated and supported by the HEP Software Foundation (HSF) with the aim to provide an environment to discuss and promote the usage of Python in the HEP community at large.
 \vskip 0.01in

\vskip 0.1in
\noindent{\bf IRIS-HEP Institute Retreat} \\
\noindent{\it Date}: 12 -- 14 October 2022 \\
\noindent{\it Location}:  Princeton (Princeton, NJ) \\
\noindent{\it URL}: \url{https://indico.cern.ch/event/1196111/}\\
\noindent{\it Description}: (From workshop website) The IRIS-HEP software institute has recently completed its fourth year after being funded by the National Science Foundation. The goals of the retreat were:

\begin{itemize}
\item Checkpoint the status of the IRIS-HEP efforts to date and specific plans and achievable goals for the next year (Year 5 of the project)
\item Clarify the gaps between where we are now and what will be needed for the HL-LHC startup
\item Elaborate a vision for what IRIS-HEP could do with an additional 5 year program of work
\end{itemize}
 \vskip 0.01in

\vskip 0.1in
\noindent{\bf A Coordinated Ecosystem for HL-LHC Computing R\&D} \\
\noindent{\it Date}: 7 -- 9 November 2022 \\
\noindent{\it Location}: University of Notre Dame's Keough School of Global Affairs (Washington, DC) \\
\noindent{\it URL}: \url{https://indico.cern.ch/event/1203733}\\
\noindent{\it Description}: (From workshop website) The research and development efforts required to address the HEP challenges for the HL-LHC are challenging. The current LHC physics program is enabled by an elaborate software and computing ecosystem. The planned major hardware upgrades for the HL-LHC, and its planned physics program, will require significant evolution of this ecosystem. Major advances in software performance, adaptability, sustainability, workforce development and training that take full advantage of future data \& compute platforms and leverage developments from outside of HEP are needed to succeed. Maintaining a coherent R\&D effort in software and computing is required to achieve the physics goals of that era.
 \vskip 0.01in

\vskip 0.1in
\noindent{\bf Software Citation and Recognition in HEP} \\
\noindent{\it Date}: 22 -- 23 November 2022 \\
\noindent{\it Location}: (Virtual) \\
\noindent{\it URL}: \url{https://indico.cern.ch/event/1211229}\\
\noindent{\it Description}: (From workshop website) This meeting aims to provide a community discussion around ways in which HEP experiments handle citation of software and recognition for software efforts that enable physics results disseminated to the public.
 \vskip 0.01in
 
\tempnewpage

%% file: 820-institute-organization.tex
\section{Appendix - Institute Organizational Structure and Evolutionary Process}
\label{sec:orggov}



\begin{figure}[ht]
\begin{center}
\includegraphics[width=0.9\textwidth]{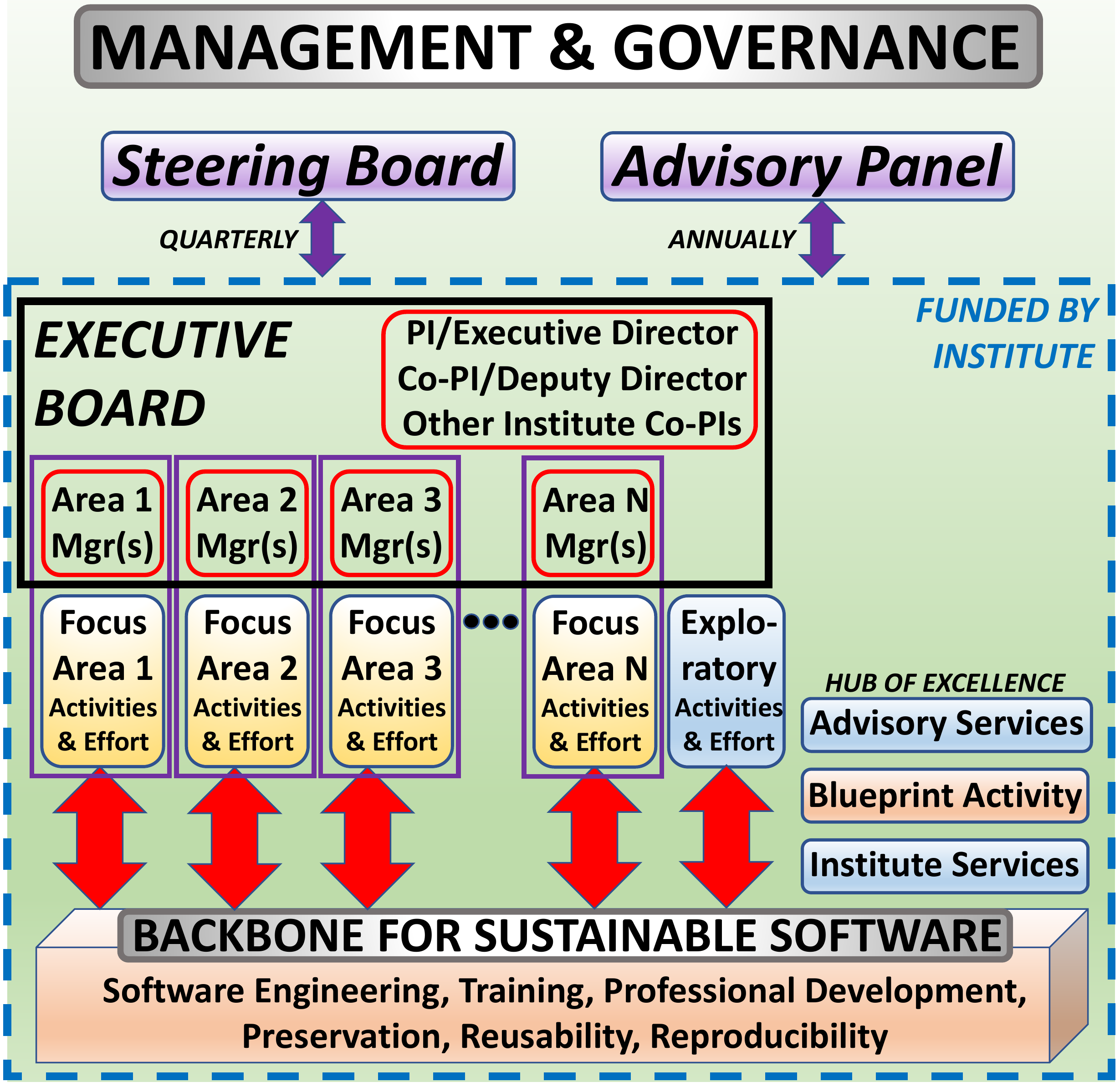}
\caption{Baseline Model for Institute Management and Governance.}
\label{fig:orgmgmt}
\end{center}
\end{figure}

The Institute's baseline organizational, management, and steering model is in 
Figure~\ref{fig:orgmgmt} and largely reflects the approach of IRIS-HEP.
The specific choices may evolve in an eventual implementation phase depending
on funding levels and project participants but the basic framework here is expected
to be unchanged.

\noindent The main elements in this organizational structure and their
roles within the Institute are:

\vskip 0.10in
\noindent{\bf PI/co-PIs:} The PI/co-PIs on an eventual Institute
implementation proposal will have project responsibilities as defined
by NSF.

\vskip 0.10in
\noindent{\bf Focus Areas:} A number of Focus Areas will be defined
for the Institute at any given point in time. These areas will
represent the main priorities of the Institute in terms of activities
aimed at developing the software infrastructure to achieve the mission
of the Institute. The initial set of strategic area are
described in Section~\ref{sec:focusareas}. The number
and size of the areas ultimately included in the
implementation will depend on the resources available
to achieve the goals. The areas could evolve over the course of
the Institute, but it is expected to be typically between six and
eight. Each focus area within an Institute will have a written set of
goals for the year and corresponding Institute resources as laid out in the project execution plan. The active
focus areas will be reviewed
once per year as part of the annual planning process and decisions will be
taken on updating the list of areas and their yearly goals (with input from the Steering Board).

\vskip 0.10in
\noindent{\bf Area Manager(s):} Each Area Manager will manage the
day to day activities within a focus area. Special care will be taken to identify a deputy for each area to ensure continuity when a manager takes on a new position or role. An appropriate mix of HEP, Computer
Science and representation from different experiments will be a
goal when selecting the area managers.

\vskip 0.10in
\noindent{\bf Executive Board:} The Executive Board will manage the
day to day activities of the Institute. It will consist of the PI,
co-PIs, and the area managers. A weekly meeting will be
used to manage the general activities of the Institute and make
shorter term plans. Liaisons from other organizations such as the U.S. LHC Ops programs
would be invited as an ``observer'' to
weekly Executive Board meetings in order to facilitate transparency
and collaboration (e.g.\ on shared services or resources).

\vskip 0.10in
\noindent{\bf Steering Board:} A Steering Board will be defined to
meet with the executive board approximately quarterly to review the
large scale priorities, strategy, and focus areas of the Institute.  The steering
board will consist of two representatives for each participating
experiment, representatives of the U.S. LHC Operations programs, plus 
other representatives including CERN and FNAL. Members of the
Steering Board will be proposed by their respective organizations
and accepted by the Executive Director in consultation with the
Executive Board.

\vskip 0.10in
\noindent{\bf Executive Director:} An Executive Director will manage
the overall activities of the Institute and its interactions with
external entities. In general, day-to-day decisions will be taken
by achieving consensus in the Executive Board and strategy and
priority decisions based on advice and recommendations by the
Steering and Executive Boards. In cases where consensus cannot be
reached, the Executive Director will take a final decision. A Deputy
Director will be included in the Institute organization, to assume
duties of the Executive Director during periods of unavailability to
ensure continuity of Institute operations.

\vskip 0.10in
\noindent{\bf Advisory Panel:} An Advisory Panel will be convened
to conduct an internal review of the project and future plans as needed. The
members of the panel will be selected by the PI/co-PIs with input
from the Steering Board. The panel will include experts not otherwise
involved with the Institute in the areas of physics, computational
physics, sustainable software development and cyberinfrastructure.

\tempnewpage

%% file: 840-glossary.tex
\section{Appendix - Glossary of Acronyms}
\label{glossary}

\begin{description}[leftmargin=0pt]

\item[ABC] Approximate Bayesian Computation

\item[ACAT] A workshop series on Advanced Computing and Analysis Techniques in HEP.

\item[ALICE] A Large Ion Collider Experiment, an experiment at the LHC at CERN.

\item[ALPGEN] An event generator designed for the generation of Standard Model
processes in hadronic collisions, with emphasis on final states with
large jet multiplicities. It is based on the exact LO evaluation of
partonic matrix elements, as well as top quark and gauge boson decays
with helicity correlations.

\item[AOD] Analysis Object Data is a summary of the reconstructed event and
contains sufficient information for common physics analyses.

\item[ATLAS] A Toroidal LHC ApparatuS, an experiment at the LHC at CERN.

\item[BaBar] A large HEP experiment which ran at SLAC from 1999 through 2008.

\item[BSM] Physics beyond the Standard Model (BSM) refers to the theoretical
developments needed to explain the deficiencies of the Standard Model
(SM), such as the
\href{https://en.wikipedia.org/wiki/Origin_of_mass}{origin of mass}, the
\href{https://en.wikipedia.org/wiki/Strong_CP_problem}{strong CP
problem},
\href{https://en.wikipedia.org/wiki/Neutrino_oscillation}{neutrino
oscillations},
\href{https://en.wikipedia.org/wiki/Baryon_asymmetry}{matter--antimatter
asymmetry}, and the nature of
\href{https://en.wikipedia.org/wiki/Dark_matter}{dark matter} and
\href{https://en.wikipedia.org/wiki/Dark_energy}{dark energy}.

\item[CDN] Content Delivery Network

\item[CERN] The European Laboratory for Particle Physics, the host laboratory
for the LHC (and eventually HL-LHC) accelerators and the ALICE, ATLAS, CMS
and LHCb experiments.

\item[CHEP] An international conference series on Computing in High Energy and Nuclear 
Physics.

\item[CI] Cyberinfrastructure, referring to the people, software, and hardware necessary that provide powerful and advanced capabilities 

\item[CMS] Compact Muon Solenoid, an experiment at the LHC at CERN.

\item[CMSSW] Application software for the CMS experiment including the processing framework itself and components relevant for
event reconstruction, high-level trigger, analysis, hardware trigger emulation, simulation, 
and visualization workflows.

\item[CMSDAS] The CMS Data Analysis School

\item[CoDaS-HEP] The COmputational and DAta Science in HEP school.




\item[CP] Charge and Parity conjugation symmetry

\item[CPV] CP violation

\item[CS] Computer Science

\item[CRSG] Computing Resources Scrutiny Group, a WLCG committee
in charge of scrutinizing and assessing LHC experiment yearly
resource requests to prepare funding agency decisions.


\item[CTDR] Computing Technical Design Report, a document written by one of
the experiments to describe the experiment's technical blueprint for building 
the software and computing system

\item[CVMFS] The CERN Virtual Machine File System is a network file system
based on HTTP and optimised to deliver experiment software in a fast,
scalable, and reliable way through sophisticated caching strategies.

\item[CVS] Concurrent Versions System, a source code version control system

\item[CWP] The Community White Paper is the result of an organised effort to 
describe the community strategy and a roadmap for software and computing R\&D 
in HEP for the 2020s. This activity is organised under the umbrella of the HSF.

\item[DASPOS] the Data And Software Preservation for Open Science project

\item[Deep Learning] one class of Machine Learning algorithms, based on a
high number of neural network layers.

\item[DES] The Dark Energy Survey

\item[DIANA-HEP] the Data Intensive Analysis for High Energy Physics project,
funded by NSF as part of the SI2 program


\item[DOE] The Department of Energy

\item[DHTC] Distributed High Throughput Computing

\item[DOMA] Data Organization, Management and Access, a term for an integrated
view of all aspects of how a project interacts with and uses data.


\item[EFT] the Effective Field Theory, an extension of the Standard Model


\item[EYETS] Extended Year End Technical Stop, used to denote a period 
(typically several months) in the winter when small upgrades and maintenance
are performed on the CERN accelerator complex and detectors





\item[FNAL] Fermi National Accelerator Laboratory, also known as Fermilab,
the primary US High Energy Physics Laboratory, funded by the US Department
of Energy

\item[FPGA] Field Programmable Gate Array

\item[FTE] Full Time Equivalent

\item[FTS] File Transfer Service

\item[GAN] Generative Adversarial Networks are a class of
\href{https://en.wikipedia.org/wiki/Artificial_intelligence}{artificial
intelligence} algorithms used in
\href{https://en.wikipedia.org/wiki/Unsupervised_machine_learning}{unsupervised
machine learning}, implemented by a system of two
\href{https://en.wikipedia.org/wiki/Neural_network}{neural networks}
contesting with each other in a
\href{https://en.wikipedia.org/wiki/Zero-sum_game}{zero-sum game}
framework.

\item[GAUDI] An event processing application framework developed by CERN

\item[Geant4] A toolkit for the simulation of the passage of
particles through matter.

\item[GeantV] An R\&D project that aims to fully exploit the
parallelism, which is increasingly offered by the new generations
of CPUs, in the field of detector simulation.

\item[GPGPU] General-Purpose computing on Graphics Processing Units
is the use of a Graphics Processing Unit (GPU), which typically
handles computation only for computer graphics, to perform computation
in applications traditionally handled by the Central Processing
Unit (CPU). Programming for GPUs is typically more challenging, but
can offer significant gains in arithmetic throughput.

\item[HEP] High Energy Physics 

\item[HEP-CCE] the HEP Center for Computational Excellence, a DOE-funded
cross-cutting initiative to promote excellence in high performance
computing (HPC) including data-intensive applications, scientific
simulations, and data movement and storage

\item[HEPData] The Durham High Energy Physics Database is an open access
repository for scattering data from experimental particle physics.

\item[HEPiX] A series of twice-annual workshops which bring together IT
staff and HEP personnel involved in HEP computing


\item[HL-LHC] The High Luminosity Large Hadron Collider is a proposed
upgrade to the Large Hadron Collider to be made in 2026. The upgrade
aims at increasing the luminosity of the machine by a factor of 10,
up to $10^{35}\mathrm{cm}^{-2}\mathrm{s}^{-1}$, providing a better
chance to see rare processes and improving statistically marginal
measurements.

\item[HLT] High Level Trigger. Software trigger system generally using a large computing cluster located close 
to the detector. Events are processed in real-time (or within the latency defined by small buffers) 
and select those who must be stored for further processing offline.

\item[HPC] High Performance Computing.

\item[HS06] HEP-wide benchmark for measuring CPU performance based on the 
SPEC2006 benchmark (\href{https://www.spec.org}{{https://www.spec.org}}).

\item[HSF] The HEP Software Foundation facilitates coordination and common
efforts in high energy physics (HEP) software and computing
internationally.

\item[IgProf] The Ignominius Profiler, a tool for exploring the CPU and 
memory use performance of very large C++ applications like those used in HEP

\item[IML] The Inter-experimental LHC Machine Learning (IML) Working Group is
focused on the development of modern state-of-the art machine learning
methods, techniques and practices for high-energy physics problems.

\item[INFN] The Istituto Nazionale di Fisica Nucleare, the main funding agency 
and series of laboratories involved in High Energy Physics research in Italy


\item[JavaScript] A high-level, dynamic, weakly typed,
prototype-based, multi-paradigm, and interpreted programming language.
Alongside HTML and CSS, JavaScript is one of the three core technologies
of World Wide Web content production.

\item[Jupyter Notebook] This is a server-client application that allows
editing and running notebook documents via a web browser. Notebooks are
documents produced by the Jupyter Notebook App, which contain both
computer code (e.g., python) and rich text elements (paragraph,
equations, figures, links, etc...). Notebook documents are both
human-readable documents containing the analysis description and the
results (figures, tables, etc..) as well as executable documents which
can be run to perform data analysis.

\item[LEP] The Large Electron-Positron Collider, the original accelerator
which occupied the 27km circular tunnel at CERN now occupied by the Large
Hadron Collider

\item[LHC] Large Hadron Collider, the main particle accelerator at CERN.

\item[LHCb] Large Hadron Collider beauty, an experiment at the LHC at CERN



\item[LIGO] The Laser Interferometer Gravitational-Wave Observatory

\item[LS] Long Shutdown, used to denote a period (typically 1 or more years) 
in which the LHC is not producing data and the CERN accelerator complex and
detectors are being upgraded.

\item[LSST] The Large Synoptic Survey Telescope



\item[ML] Machine learning is a field of computer science that gives computers
the ability to learn without being explicitly programmed. It focuses on
prediction making through the use of computers and encompasses a lot of
algorithm classes (boosted decision trees, neural networks\ldots{}).


\item[MREFC] Major Research Equipment and Facilities Construction, an NSF
mechanism for large construction projects

\item[NAS] The National Academy of Sciences

\item[NCSA] National Center of Supercomputing Applications, at the University 
of Illinois at Urbana-Champaign

\item[NDN] Named Data Networking

\item[NSF] The National Science Foundation

\item[ONNX] Open Neural Network Exchange, an evolving open-source standard for exchanging AI models


\item[Openlab] CERN openlab is a public-private partnership that accelerates
the development of cutting-edge solutions for the worldwide LHC
community and wider scientific research.

\item[OSG] The Open Science Grid

\item[P5] The Particle Physics Project Prioritization Panel is a scientific
advisory panel tasked with recommending plans for U.S.\ investment in
particle physics research over the next ten years.

\item[PI] Principal Investigator




\item[QA] Quality Assurance

\item[QC] Quality Control

\item[QCD] Quantum Chromodynamics, the theory describing the strong
interaction between quarks and gluons.

\item[REANA] REusable ANAlyses, a system to preserve and instantiate analysis workflows 

\item[REU] Research Experience for Undergraduates, an NSF program to fund undergraduate participation in research projects

\item[RRB] Resources Review Board, a CERN committee made up of representatives of funding agencies participating in the LHC collaborations, the CERN management and the experiment's management.


\item[ROOT] A scientific software framework widely used in HEP data
processing applications.


\item[SciDAC] Scientific Discovery through Advanced Computing, a DOE program
to fund advanced R\&D on computing topics relevant to the DOE Office of
Science


\item[SDSC] San Diego Supercomputer Center, at the University of California
at San Diego

\item[SHERPA] Sherpa is a Monte Carlo event generator for the Simulation of
High-Energy Reactions of PArticles in lepton-lepton, lepton-photon,
photon-photon, lepton-hadron and hadron-hadron collisions.

\item[SIMD] Single instruction, multiple data (\textbf{SIMD}), describes
computers with multiple processing elements that perform the same
operation on multiple data points simultaneously.

\item[SI2] The Software Infrastructure for Sustained Innovation program at NSF

\item[SKA] The Square Kilometer Array

\item[SLAC] The Stanford Linear Accelerator Center, a laboratory funded by the
US Department of Energy

\item[SM] The Standard Model is the name given in the 1970s to a theory of
fundamental particles and how they interact. It is the currently
dominant theory explaining the elementary particles and their dynamics.

\item[SOW] Statement of Work, a mechanism used to define the expected activities and deliverables of individuals funded from a subaward with a multi-institutional project. The SOW is typically revised annually, along with the corresponding budgets.

\item[SSI] The Software Sustainability Institute, an organization in the UK dedicated to fostering better, and more sustainable, software for research.

\item[SWAN] Service for Web based ANalysis is a platform for interactive data
mining in the CERN cloud using the Jupyter notebook interface.


\item[TMVA] The Toolkit for Multivariate Data Analysis with ROOT is a
standalone project that provides a ROOT-integrated machine learning
environment for the processing and parallel evaluation of sophisticated
multivariate classification techniques.

\item[TPU] Tensor Processing Unit, an application-specific integrated circuit 
by Google designed for use with Machine Learning applications

\item[URSSI] the US Software Sustainability Institute, an \s2i2 conceptualization activity recommended for funding by NSF



\item[WAN] Wide Area Network


\item[WLCG] The Worldwide LHC Computing Grid project is a global collaboration
of more than 170 computing centres in 42 countries, linking up national
and international grid infrastructures. The mission of the WLCG project
is to provide global computing resources to store, distribute and
analyse data generated by the Large Hadron Collider (LHC) at CERN.


\item[x86\_64] 64-bit version of the x86 instruction set, which originated wiht the Intel 8086, but has
now been implemented on processors from a range of companies, including the Intel and AMD processors that 
make up the vast majority of computing resources used by HEP today.

\item[XRootD] Software framework that is a fully generic suite for
fast, low latency and scalable data access.

\end{description}

%% file: iris-hep-strategic-plan.bbl
\begin{thebibliography}{10}

\bibitem{iris-hep}
{IRIS-HEP website}.
\newblock \url{http://iris-hep.org}.

\bibitem{iris-hep-nsf-award}
{National Science Foundation Cooperative Agreement OAC-1836650}.
\newblock
  \url{https://www.nsf.gov/awardsearch/showAward?AWD_ID=1836650&HistoricalAwards=false}.

\bibitem{CWPDOC}
Johannes Albrecht et~al.
\newblock {A Roadmap for HEP Software and Computing R\&D for the 2020s}.
\newblock {\em Comput. Softw. Big Sci.}, 3(1):7, 2019.

\bibitem{S2I2HEPSP}
Peter Elmer, Mark Neubauer, and Michael~D. Sokoloff.
\newblock {Strategic Plan for a Scientific Software Innovation Institute (S2I2)
  for High Energy Physics}.
\newblock 2017.
\newblock [arXiv 1712.06592] \url{https://arxiv.org/abs/1712.06592}.

\bibitem{atlas2022reviewdoc}
ATLAS Collaboration.
\newblock {ATLAS Software and Computing HL-LHC Roadmap}.
\newblock Technical report, CERN, Geneva, 2022.

\bibitem{cms2022reviewdoc}
CMS~Offline Software and Computing.
\newblock {CMS Phase-2 Computing Model: Update Document}.
\newblock Technical report, CERN, Geneva, 2022.

\bibitem{LHCbCollaboration:2776420}
CERN~(Meyrin) LHCb~Collaboration.
\newblock {Framework TDR for the LHCb Upgrade II - Opportunities in flavour
  physics, and beyond, in the HL-LHC era}.
\newblock Technical report, CERN, Geneva, 2021.

\bibitem{S2I2HEP}
S2I2-HEP project webpage: \url{http://s2i2-hep.org}.

\bibitem{HIGG-2012-27}
{G. Aad {\em et al.} [ATLAS Collaboration]}.
\newblock {Observation of a new particle in the search for the Standard Model
  Higgs boson with the ATLAS detector at the LHC}.
\newblock {\em Phys. Lett. B}, 716:1, 2012.

\bibitem{Chatrchyan:2012xdj}
{S. Chatrchyan {\em et al.} [CMS Collaboration]}.
\newblock {Observation of a new boson at a mass of 125 GeV with the CMS
  experiment at the LHC}.
\newblock {\em Phys. Lett. B}, 716:30--61, 2012.

\bibitem{Isidori:2010kg}
Gino Isidori, Yosef Nir, and Gilad Perez.
\newblock {Flavor Physics Constraints for Physics Beyond the Standard Model}.
\newblock {\em Ann.Rev.Nucl.Part.Sci.}, 60:355, 2010.

\bibitem{p5Final}
{Particle Physics Project Prioritization Panel}.
\newblock {Building for Discovery: Strategic Plan for U.S. Particle Physics in
  the Global Context}.
\newblock \url{
  http://science.energy.gov/~/media/hep/hepap/pdf/May%202014/FINAL\_DRAFT2\_P5Report\_WEB\_052114.pdf}.

\bibitem{cmstp}
D~Contardo, M~Klute, J~Mans, L~Silvestris, and J~Butler.
\newblock {Technical Proposal for the Phase-II Upgrade of the CMS Detector}.
\newblock Technical report, Geneva, 2015.
\newblock Upgrade Project Leader Deputies: Lucia Silvestris (INFN-Bari), Jeremy
  Mans (University of Minnesota) Additional contacts: Lucia.Silvestris@cern.ch,
  Jeremy.Mans@cern.ch.

\bibitem{atlastp}
{ATLAS Phase-II Upgrade Scoping Document}.
\newblock Technical report, CERN, Geneva, 2015.

\bibitem{hl-lhc-yellowreport2019}
A.~Dainese, M.~Mangano, A.~B. Meyer, A.~Nisati, G.~Salam, and M.~Vesterinen.
\newblock {CERN Yellow Reports: Monographs, Vol 7 (2019): Physics of the
  HL-LHC, and Perspectives at the HE-LHC}.
\newblock Dec 2019.

\bibitem{2021lhccreview}
A~Boehnlein, C~Biscarat, A~Bressan, D~Britton, R~Bolton, F~Gaede, C~Grandi,
  F~Hernandez, T~Kuhr, G~Merino, F~Simon, and G~Watts.
\newblock {HL-LHC Software and Computing Review Panel, 2nd Report}.
\newblock Technical report, CERN, Geneva, 2022.

\bibitem{Campana:DataChallengeOutline}
S.~{Campana}.
\newblock {HL-LHC network needs and data transfer challenges}.
\newblock
  \url{https://indico.cern.ch/event/1004222/contributions/4241102/attachments/2194497/3745965/HL-LHC\%20network\%20challenges\%20-\%20V2.pdf}.

\bibitem{Zurawski:ESNetReq}
Jason Zurawski, Benjamin Brown, Dale Carder, Eric Colby, Eli Dart, Ken Miller,
  Abid Patwa, Kate Robinson, Lauren Rotman, and Andrew Wiedlea.
\newblock High energy physics network requirements review (final report,
  july-october 2020), Jun 2021.

\bibitem{sustainability-blueprint-indico}
{Workshop on Sustainable Software in HEP}.
\newblock \url{https://indico.cern.ch/event/930127/}.

\bibitem{sustainability-blueprint-note}
Daniel~S. Katz, Sudhir Malik, Mark~S. Neubauer, Graeme~A. Stewart, Kétévi~A.
  Assamagan, Erin~A. Becker, Neil~P. Chue~Hong, Ian~A. Cosden, Samuel Meehan,
  Edward J.~W. Moyse, Adrian~M. Price-Whelan, Elizabeth Sexton-Kennedy,
  Meirin~Oan Evans, Matthew Feickert, Clemens Lange, Kilian Lieret, Rob Quick,
  Arturo Sánchez~Pineda, and Christopher Tunnell.
\newblock Software sustainability \& high energy physics.
\newblock Technical report, 2020.

\bibitem{HLLHCTIMELINE}
{Based on schedule graphic from HL-LHC Project website (Dec 2017):}.
\newblock
  \url{https://project-hl-lhc-industry.web.cern.ch/content/project-schedule}.

\bibitem{Neubauer:2021xhi}
Mark~S. Neubauer et~al.
\newblock {Learning from the Pandemic: the Future of Meetings in HEP and
  Beyond}.
\newblock In {\em {IRIS-HEP Blueprint Workshop}}, 6 2021.

\bibitem{OACBlueprint}
{Transforming Science Through Cyberinfrastructure: NSF’s Blueprint for a
  National Cyberinfrastructure Ecosystem for Science and Engineering in the
  21st Century}.
\newblock \url{https://www.nsf.gov/cise/oac/vision/blueprint-2019/}.

\bibitem{Harris2020}
C.~R. Harris et~al.
\newblock Array programming with numpy.
\newblock {\em Nature}, 585(7825):357--362, Sep 2020.

\bibitem{ATLAS:2020pnm}
{ATLAS HL-LHC Computing Conceptual Design Report}.
\newblock 2020.

\bibitem{ATLAS:2802918}
ATLAS Collaboration.
\newblock {ATLAS Software and Computing HL-LHC Roadmap}.
\newblock Technical report, CERN, Geneva, 2022.

\bibitem{CMS:2023-HL-LHC-CDR}
{U.S. CMS Software and Computing: HL-LHC R\&D Strategic Planning Document}.
\newblock 2023.

\bibitem{reana_github}
{REANA reproducible analysis platform}.
\newblock \url{https://github.com/reanahub/reana}.

\bibitem{neos_software}
Nathan Simpson and Lukas Heinrich.
\newblock {neos: version 0.2.0}.
\newblock \url{https://github.com/gradhep/neos}, 1 2021.

\bibitem{Simpson:2022suz}
Nathan Simpson and Lukas Heinrich.
\newblock {neos: End-to-End-Optimised Summary Statistics for High Energy
  Physics}.
\newblock In {\em {20th International Workshop on Advanced Computing and
  Analysis Techniques in Physics Research}: {AI Decoded - Towards Sustainable,
  Diverse, Performant and Effective Scientific Computing}}, 3 2022.

\bibitem{Dorigo:2022gqm}
Tommaso Dorigo et~al.
\newblock {Toward the End-to-End Optimization of Particle Physics Instruments
  with Differentiable Programming: a White Paper}.
\newblock 3 2022.

\bibitem{Pivarski_Uproot_2017}
Jim Pivarski, Henry Schreiner, Angus Hollands, Pratyush Das, Kush Kothari,
  Aryan Roy, Jerry Ling, Nicholas Smith, Chris Burr, and Giordon Stark.
\newblock {Uproot}.
\newblock
  \href{https://doi.org/10.5281/zenodo.4340632}{10.5281/zenodo.4340632}.

\bibitem{BRUN199781}
Rene Brun and Fons Rademakers.
\newblock {ROOT — An object oriented data analysis framework}.
\newblock {\em Nuclear Instruments and Methods in Physics Research Section A:
  Accelerators, Spectrometers, Detectors and Associated Equipment},
  389(1):81--86, 1997.
\newblock New Computing Techniques in Physics Research V.

\bibitem{servicex}
B.~Galewsky, R.~Gardner, L.~Gray, M.~Neubauer, J.~Pivarski, M.~Proffitt,
  I.~Vukotic, G.~Watts, and M.~Weinberg.
\newblock {ServiceX A Distributed, Caching, Columnar Data Delivery Service}.
\newblock {\em EPJ Web Conf.}, 245:04043, 2020.

\bibitem{Awkward_Array_2018}
Jim Pivarski, Ianna Osborne, Ioana Ifrim, Henry Schreiner, Angus Hollands,
  Anish Biswas, Pratyush Das, Santam Roy~Choudhury, Nicholas Smith, and Manasvi
  Goyal.
\newblock {Awkward Array}.
\newblock
  \href{https://doi.org/10.5281/zenodo.4341376}{10.5281/zenodo.4341376}.

\bibitem{func_adl_github}
{func\_adl}.
\newblock \url{https://github.com/iris-hep/func_adl}.

\bibitem{coffea}
Lindsey Gray, Nicholas Smith, Benjamin Tovar, Andrzej Novak, Jayjeet
  Chakraborty, Peter Fackeldey, Nikolai Hartmann, Gordon Watts, Douglas Thain,
  Giordon Stark, BenGalewsky, Jonas Rübenach, Benjamin Fischer, Devin Taylor,
  MoAly98, Dmitry Kondratyev, Paul Gessinger, Yi-Mu~"Enoch" Chen, Joosep Pata,
  Anna Woodard, Andreas Albert, slehti, Zoe Surma, Alexx Perloff, Kevin Pedro,
  dnoonan08, Andrew Hennessee, Karol Krizka, kmohrman, and Lukas.
\newblock coffea.
\newblock
  \href{https://doi.org/10.5281/zenodo.3266454}{10.5281/zenodo.3266454}.

\bibitem{boost-histogram}
Henry Schreiner, Hans Dembinski, Aman Goel, Jay Gohil, Shuo Liu, Jonas Eschle,
  Chanchal Maji, Andrzej Novak, Chris Burr, Doug Davis, Kilian Lieret,
  Konstantin Gizdov, Kyle Cranmer, and Pierre Grimaud.
\newblock boost-histogram.
\newblock
  \href{https://doi.org/10.5281/zenodo.3492034}{10.5281/zenodo.3492034}.

\bibitem{cabinetry}
Alexander Held, Matthew Feickert, Henry Schreiner, Lars Henkelmann, Angus
  Hollands, and Nathan Simpson.
\newblock cabinetry.
\newblock
  \href{https://doi.org/10.5281/zenodo.4742752}{10.5281/zenodo.4742752}.

\bibitem{pyhf}
Lukas Heinrich, Matthew Feickert, and Giordon Stark.
\newblock {pyhf: v0.7.0}.
\newblock
  \href{https://doi.org/10.5281/zenodo.1169739}{10.5281/zenodo.1169739}.

\bibitem{pyhf_joss}
Lukas Heinrich, Matthew Feickert, Giordon Stark, and Kyle Cranmer.
\newblock pyhf: pure-python implementation of histfactory statistical models.
\newblock {\em Journal of Open Source Software}, 6(58):2823, 2021.

\bibitem{Rodrigues:2020syo}
Eduardo Rodrigues et~al.
\newblock {The Scikit HEP Project -- overview and prospects}.
\newblock {\em EPJ Web Conf.}, 245:06028, 2020.

\bibitem{dask-awkward_github}
{dask-awkward}.
\newblock \url{https://github.com/dask-contrib/dask-awkward}.

\bibitem{numfocus-affiliated-projects}
{NumFOCUS Affiliated Projects}.
\newblock \url{https://numfocus.org/sponsored-projects/affiliated-projects}.

\bibitem{HS3}
{High Energy Physics Statistics Serialization Standard (HS3)}.
\newblock
  \url{https://github.com/hep-statistics-serialization-standard/hep-fit-serialization}.

\bibitem{Schulz:2021BAT}
Oliver Schulz, Frederik Beaujean, Allen Caldwell, Cornelius Grunwald, Vasyl
  Hafych, Kevin Kr{\"o}ninger, Salvatore~La Cagnina, Lars R{\"o}hrig, and
  Lolian Shtembari.
\newblock Bat.jl: A julia-based tool for bayesian inference.
\newblock {\em SN Computer Science}, 2(3):210, Apr 2021.

\bibitem{recat_atlas_github}
{RECAST for ATLAS}.
\newblock \url{https://github.com/recast-hep/recast-atlas}.

\bibitem{Bocci:2020pmi}
Andrea Bocci, Vincenzo Innocente, Matti Kortelainen, Felice Pantaleo, and Marco
  Rovere.
\newblock {Heterogeneous Reconstruction of Tracks and Primary Vertices With the
  CMS Pixel Tracker}.
\newblock {\em Front. Big Data}, 3:601728, 2020.

\bibitem{IAIFI}
{NSF AI Institute for Artificial Intelligence and Fundamental Interactions
  (IAIFI)}.
\newblock \url{https://iaifi.org}.

\bibitem{A3D3}
{A3D3: Accelerated Artificial Intelligence Algorithms for Data-Driven
  Discovery}.
\newblock \url{https://a3d3.ai/}.

\bibitem{Shlomi:2020gdn}
Jonathan Shlomi, Peter Battaglia, and Jean-Roch Vlimant.
\newblock Graph neural networks in particle physics.
\newblock {\em Machine Learning: Science and Technology}, 2(2):021001, dec
  2020.

\bibitem{atlasftagGNN2022}
The~ATLAS Collaboration.
\newblock {Graph Neural Network Jet Flavour Tagging with the ATLAS Detector}.
\newblock 2022.

\bibitem{DELPHES2014}
J.~de~Favereau, C.~Delaere, P.~Demin, A.~Giammanco, V.~Lemaître, A.~Mertens,
  and M.~Selvaggi.
\newblock Delphes 3: a modular framework for fast simulation of a generic
  collider experiment, Feb 2014.

\bibitem{NRP}
{The National Research Platform}.
\newblock \url{https://nationalresearchplatform.org/}.

\bibitem{FABRIC}
{FABRIC Testbed}.
\newblock \url{https://fabric-testbed.net/}.

\bibitem{SENSE}
{SENSE: SDN for End-to-End Networked Science at the Exascale}.
\newblock \url{https://sense.es.net/}.

\bibitem{Proffitt:ADL}
Mason Proffitt, Ingo Müller, Dan Graur, Mat Adamec, Pieter David, Enrico
  Guiraud, and Sebastien Binet.
\newblock {iris-hep/adl-benchmarks-index: ADL Functionality Benchmarks Index
  v0.1}.
\newblock
  \href{https://doi.org/10.5281/zenodo.5131287}{10.5281/zenodo.5131287}, July
  2021.

\bibitem{Graur:VLDB}
Dan Graur, Ingo Müller, Mason Proffitt, Ghislain Fourny, Gordon~T. Watts, and
  Gustavo Alonso.
\newblock Evaluating query languages and systems for high-energy physics data,
  Oct 2021.

\bibitem{RUGS}
{Research End User Group}.
\newblock \url{https://community.cncf.io/research-end-user-group/}.

\bibitem{CNCF}
{Cloud Native Computing Foundation}.
\newblock \url{https://www.cncf.io/}.

\bibitem{hsftraining}
{HSF Training Working Group}.
\newblock \url{https://hepsoftwarefoundation.org/workinggroups/training.html}.

\bibitem{uscms-pursue}
Tulika Bose, Sudhir Malik, and Meenakshi Narain.
\newblock U.s. cms - pursue (program for undergraduate research summer
  experience), 2022.

\bibitem{hsf-curriculum}
{HSF Training Curriculum}.
\newblock \url{https://hepsoftwarefoundation.org/training/curriculum.html}.

\bibitem{Aaij:2019zbu}
Roel Aaij et~al.
\newblock {Allen: A high level trigger on GPUs for LHCb}.
\newblock {\em Comput. Softw. Big Sci.}, 4(1):7, 2020.

\bibitem{https://doi.org/10.48550/arxiv.1505.04597}
Olaf Ronneberger, Philipp Fischer, and Thomas Brox.
\newblock U-net: Convolutional networks for biomedical image segmentation,
  2015.

\bibitem{agctools1-indico}
{IRIS-HEP AGC Tools 2021 Workshop}.
\newblock \url{https://indico.cern.ch/event/1076231/}.

\bibitem{agctools2-indico}
{IRIS-HEP AGC Tools 2022 Workshop}.
\newblock \url{https://indico.cern.ch/event/1126109/}.

\bibitem{Held:2022RC}
Alexander Held and Oksana Shadura.
\newblock {The IRIS-HEP Analysis Grand Challenge}.
\newblock {\em PoS}, ICHEP2022:235, 2022.

\bibitem{agc-acat2022}
{First performance measurements with the Analysis Grand Challenge }.
\newblock \url{https://indico.cern.ch/event/1106990/contributions/4998188/}.

\bibitem{Stewart:AnalysisEcosystem}
Graeme~A Stewart, Peter Elmer, Giulio Eulisse, Loukas Gouskos, Stephan
  Hageboeck, Allison~Reinsvold Hall, Lukas Heinrich, Alexander Held, Michel
  Jouvin, Teng~Jian Khoo, Paul Laycock, Pere Mato~Vila, Jim Pivarski, Jonas
  Rembser, Eduardo Rodrigues, Jana Schaarschmidt, Elizabeth Sexton-Kennedy,
  Oksana Shadura, Nathan Simpson, Nicola Skidmore, Michael Sokoloff, Gordon
  Watts, Aly Mohamed, Jackson Burzynski, Bryan Cardwell, Daniel~C Craik, Tomas
  Dado, Antonio Delgado~Peris, Caterina Doglioni, Engin Eren, Martin~B Eriksen,
  Jonas Eschle, Conor Fitzpatrick, José Flix~Molina, Sean Gasiorowski, Aman
  Goel, Enrico Guiraud, Kanhaiya Gupta, Michael Hernández~Villanueva, José~M
  Hernández, Julius Hrivnac, Kilian Lieret, Luke Kreczko, Nils Krumnack,
  Thomas Kuhr, Baidyanath Kundu, Eric Lancon, Johannes Lange, Nicholas~J
  Manganelli, Andrzej Novak, Antonio Perez-Calero~Yzquierdo, Mason Proffitt,
  Grigori Rybkin, Henry~F Schreiner, Markus Schulz, Andrea Sciabà, Sezen
  Sekmen, Tal van Daalen, Tibor Simko, Jaydip Singh, Nicholas Smith, Giles~C
  Strong, Gokhan Unel, Vassil Vassilev, Mark Waterlaat, Efe Yazgan, Ayanabha
  Das, and Ben Galewsky.
\newblock {HSF IRIS-HEP Second Analysis Ecosystem Workshop Report}.
\newblock
  \href{https://doi.org/10.48550/arXiv.2212.04889}{10.48550/arXiv.2212.04889},
  12 2022.

\bibitem{krupa2021gpu}
Jeffrey Krupa, Kelvin Lin, Maria~Acosta Flechas, Jack Dinsmore, Javier Duarte,
  Philip Harris, Scott Hauck, Burt Holzman, Shih-Chieh Hsu, Thomas Klijnsma,
  et~al.
\newblock {GPU coprocessors as a service for deep learning inference in high
  energy physics}.
\newblock {\em Machine Learning: Science and Technology}, 2(3):035005, 2021.

\bibitem{Forti:DC21}
Forti Alessandra, Di~Maria Riccardo, Dona Rizart, Garrido Borja, Lassnig Mario,
  Dimitrios Christidis, Ellis Katy, Paspalaki Garyfalia, and Gomez Felipe.
\newblock {WLCG Network Data Challenges 2021: wrap-up and recommendations}, Dec
  2021.

\bibitem{GuardianSV}
{Segregated Valley: the ugly truth about Google and diversity in tech:}.
\newblock
  \url{https://www.theguardian.com/technology/2017/aug/07/silicon-valley-google-diversity-black-women-workers}.

\bibitem{diversity1}
Max Nathan and Neil Lee.
\newblock {Cultural Diversity, Innovation, and Entrepreneurship: Firm-level
  Evidence from London}.
\newblock {\em Economic Geography}, 89(4):367--394, 2013.

\bibitem{diversity2}
Cristina Díaz-García, Angela González-Moreno, and Francisco~Jose
  Sáez-Martínez.
\newblock {Gender diversity within R\&D teams: Its impact on radicalness of
  innovation}.
\newblock {\em Innovation}, 15(2):149--160, 2013.

\bibitem{diversity3}
Sheen~S. Levine, Evan~P. Apfelbaum, Mark Bernard, Valerie~L. Bartelt, Edward~J.
  Zajac, and David Stark.
\newblock Ethnic diversity deflates price bubbles.
\newblock {\em Proceedings of the National Academy of Sciences},
  111(52):18524--18529, 2014.

\bibitem{quarknet}
{QuarkNet website}.
\newblock \url{https://quarknet.org}.

\end{thebibliography}
